\documentclass[journal]{IEEEtran}

\ifCLASSINFOpdf
\else
\fi

\usepackage{float}
\usepackage{amsmath}
\usepackage{amssymb}
\usepackage{graphicx}
\usepackage{esint}
\usepackage{subcaption} 
\usepackage[T1]{fontenc}
\usepackage[utf8]{inputenc}
\usepackage[edges]{forest}
\usetikzlibrary{shadows.blur}

\usepackage{graphicx}
\usepackage{lineno,hyperref}
\usepackage{amsfonts}%
\usepackage{amssymb}
\usepackage{epsfig}
\usepackage{algorithm}
\usepackage{algorithmic}
\usepackage{xcolor}
\usepackage{array}
\usepackage{subcaption}
\usepackage{adjustbox}
\usepackage{makecell}
\hypersetup{draft}
\usepackage{multicol, blindtext}
\usepackage{tabularx}
\usepackage{ltablex}
\usepackage{longtable}
\usepackage{xltabular}

\begin{document}

\title{Survey on Machine Learning for Traffic-Driven Service Provisioning in Optical Networks}

\author{Tania Panayiotou, Maria Michalopoulou, Georgios Ellinas
\noindent \thanks{The authors are with the Department of Electrical and Computer Engineering and the KIOS Research and Innovation Center of Excellence, University of Cyprus, Nicosia, 1678, Cyprus. E-mail:
    {\tt\small \{panayiotou.tania, michalopoulou.maria, gellinas\}@ucy.ac.cy}}%
}


\maketitle

\begin{abstract}
The unprecedented growth of the global Internet traffic, coupled with the large spatio-temporal fluctuations that create, to some extent, predictable tidal traffic conditions, are motivating the evolution from reactive to proactive and eventually towards adaptive optical networks. In these networks, traffic-driven service provisioning can address the problem of network over-provisioning and better adapt to traffic variations, while keeping the quality-of-service at the required levels. Such an approach will reduce network resource over-provisioning and thus reduce the total network cost. This survey provides a comprehensive review of the state of the art on machine learning (ML)-based techniques at the optical layer for traffic-driven service provisioning. The evolution of service provisioning in optical networks is initially presented, followed by an overview of the ML techniques utilized for traffic-driven service provisioning. ML-aided service provisioning approaches are presented in detail, including predictive and prescriptive service provisioning frameworks in proactive and adaptive networks. For all techniques outlined, a discussion on their limitations, research challenges, and potential opportunities is also presented. 

\end{abstract}


\IEEEpeerreviewmaketitle

\section{Introduction}
The annual global Internet traffic is increasing at an unprecedented rate and is estimated to reach $4.8$ ZB/year by the end of 2022~\cite{cisco20} (essentially increasing threefold over the last five years).  Further, with the advent of new mobile technologies (5G currently and 6G in the future), a large number of services and applications have emerged [e.g., the Internet-of-Things (IoT), connected autonomous vehicles, augmented/virtual reality] and others are planned (e.g., holographic telepresence and immersive communications, as well as applications and services that are artificial intelligence (AI)-inspired such as the interaction 
on human-digital-physical worlds and the Internet of Senses~\cite{xiao2020selflearning}). It is forecasted that 6G-based networks will support in excess of 125 billion of connected devices by year 2030, extensively utilizing AI to improve the network's performance, but also to be able to offer AI-as-a-Service in a federated network architecture~\cite{5G-PPP}. This growth in Internet traffic is combined with large spatio-temporal fluctuations that create tidal traffic conditions, due primarily to the daily movement of population~\cite{7444562}.

In general, in telecommunications networks it is possible to design different virtual network topologies (VNTs) on top of the same infrastructure to support various services, and applications with diverse quality-of-service (QoS) requirements. In today's networks static topologies  are  commonly  designed  to  cope  with  the  traffic  forecast, with the connections permanently established and modified only in exceptional cases, such as failures. In this approach, the virtual topology establishes more connections than necessary, to avoid losses if equipment fails or traffic peaks occur. Thus, such an approach leads to resource over-provisioning, significantly increasing the total network cost. This is the case, as it is possible that the VNT that has been designed at a given moment of the network's lifecycle no longer satisfies the network's objectives/requirements previously set. Nevertheless, modern networks can handle traffic variations by resorting to traffic-driven service provisioning, i.e., some of the active connections are reconfigured taking into consideration both the current network state (i.e., virtual topology composition and resource availability) and the current traffic demand. 

Utilizing traffic-driven service provisioning the network can better adapt to traffic variations, while at the same time providing the required QoS/quality-of-transmission (QoT) and ensuring that the network resources are utilized efficiently (essentially reducing the total network cost). Clearly, to  enable dynamic (automated) traffic-driven service provisioning requires significant amounts of traffic data, driving the need for real-time data monitoring, collection, and analysis functionalities in the network. While such an approach has been widely used at the IP layer, this also holds true for the optical transport networks where virtual network topologies can also be created (i.e., set of all lightpaths that belong to a service, set of all lightpaths with the same QoS, etc.). Further, similar to the introduction of the ``AI everywhere'' concept in 6G networks, the introduction of AI/machine learning (ML)-based techniques at the optical layer constitutes a crucial component for the efficient realization of the aforementioned traffic-driven provisioning functionality. 

\subsection{Positioning of the Survey}
Several surveys on ML-based techniques for optical networks appear in the literature, including~\cite{Furdek:21} and~\cite{Musumeci:19} that focus on ML models for threat detection and failure management in optical networks, respectively. In particular, the work in~\cite{Furdek:21} focuses on security against attacks in optical networks, presenting  ML-based approaches for security diagnostics and required architectures and functionalities to enable the automated management of optical network security. In a similar vein, the work in~\cite{Musumeci:19} examines the state of the art concerning ML-based approaches for failure prediction, detection, localization, and identification, in an effort to present a framework for automating fault management in optical networks. Further, surveys on ML techniques for QoT estimation in optical networks are presented in~\cite{Pointurier:21,9799746,ZHANG2022102804}. Indicatively, in~\cite{Pointurier:21}, the author focuses on the reasons why inaccuracies appear in QoT estimation, on a classification of ML-based QoT estimation techniques, as well as on ML-aided optical performance monitoring.

Moreover, a more general survey on AI methods in optical networks is presented in~\cite{MATA201843}. In this work,  the authors initially address the use of AI-based approaches in problems related to physical (optical) layer optical transmission, including performance monitoring and QoT estimation. Subsequently, they consider ML-based techniques for addressing optical network control and management issues, including network planning, connection establishment, network reconfiguration, intra-datacenter networking, software defined networking, optical burst switching, etc. 

Finally, additional (general) surveys on ML for optical network automation include the work in~\cite{8527529} that, similar to~\cite{MATA201843}, presents an overview on the application of machine learning techniques in optical networks (with a focus on ML rather than AI), as well as the work in~\cite{rafique} that is a tutorial focusing on the major concepts and applications of ML in optical networking. In particular, the work in~\cite{8527529} focuses on applications of ML at the transmission and network layers and surveys works in the state of the art related to QoT estimation, optical power monitoring, optical amplifier control, non-linearity mitigation, traffic prediction and virtual topology design, path computation, fault management, etc. Similarly, the work in~\cite{rafique} introduces the various aspects of ML, data management issues, as well as network management architectures and use cases utilizing ML for network automation.

Specifically, all surveys currently available in the literature that focus on ML-based techniques applied in optical networks for a variety of control and management functionalities are listed in Table~\ref{surveys}, with a brief description of their main focus. As shown from Table~\ref{surveys} and discussed above, some of the surveys listed focus on fault management, security diagnostics, and QoT estimation, while others are more general, encompassing a number of control and management functionalities in optical networks, but without examining in depth any particular functionality.  

\begin{table}[htbp]
\begin{center}
\footnotesize
\caption{Current state of the art regarding surveys on ML-based control and management functionalities in optical networks.}
\label{surveys}
    \begin{tabular}{ | l | p{6.5cm} |}
    \hline
    {\bf Survey} & {\bf Topic Covered} \\ \hline
    \cite{Furdek:21} &  ML-based models for security diagnostics in optical networks \\ \hline
    \cite{Musumeci:19}  & ML-aided failure management (failure prediction, detection, localization, and identification) in optical networks \\ \hline
    \cite{Pointurier:21,9799746,ZHANG2022102804} & ML techniques for QoT estimation in optical networks  \\ \hline
    \cite{MATA201843} & General survey on AI/ML-based control and management in optical networks. \newline AI-based approaches for performance monitoring and QoT estimation. \newline ML-based techniques for network planning, connection establishment, network reconfiguration, intra-datacenter networking, software defined networking, optical burst switching, etc. \\ \hline
    \cite{8527529} & General survey on the applications of ML at the transmission and network layers (includes QoT estimation, optical power monitoring, optical amplifier control, non-linearity mitigation, traffic prediction and virtual topology (re)design, path computation, fault management, etc.) \\ \hline
    \cite{rafique}  &  General tutorial on ML-based network automation (includes the various aspects of ML, data management issues, and network management architectures and use cases)\\ \hline
\end{tabular}
\end{center}
\end{table}

Additional surveys in the literature concerning related topics for optical networks include the work in~\cite{920841} that discussed reconfiguration issues in traffic adaptive wavelength division multiplexed (WDM) networks, without however addressing any ML-aided approaches, surveys on optical performance monitoring~\cite{7273738},~\cite{8770528}, focusing on current and future technologies as well as on the benefits of data analytics, and surveys on routing and wavelength/spectrum assignment in WDM or flex-grid optical networks~\cite{6083231,Zang00areview,7105364}. Finally, there is a vast survey literature on ML techniques, as well as the utilization of ML-based techniques to address several networking problems. All these surveys are used in this work as background material in order to provide the information on the techniques/methodologies that are utilized in addressing the problem at hand. 

Given the aforementioned efforts in the literature that attempted to review ML-based techniques for a number of optical network operations, such as QoT estimation, fault management, network planning, and network security, the main contribution of this survey is to present a comprehensive survey of current solutions, as well as identify limitations and research challenges, related to ML-based techniques for traffic-driven service provisioning in optical networks, focusing on the service provisioning evolution, as well as the ML approaches that can enable this evolution. Various resource allocation and service provisioning frameworks are subsequently presented and extensively analyzed, explaining the state of the art as well as the challenges in traffic-driven service (re)provisioning.

\begin{figure*}[ht]
\begin{center}
\includegraphics[scale=0.35]{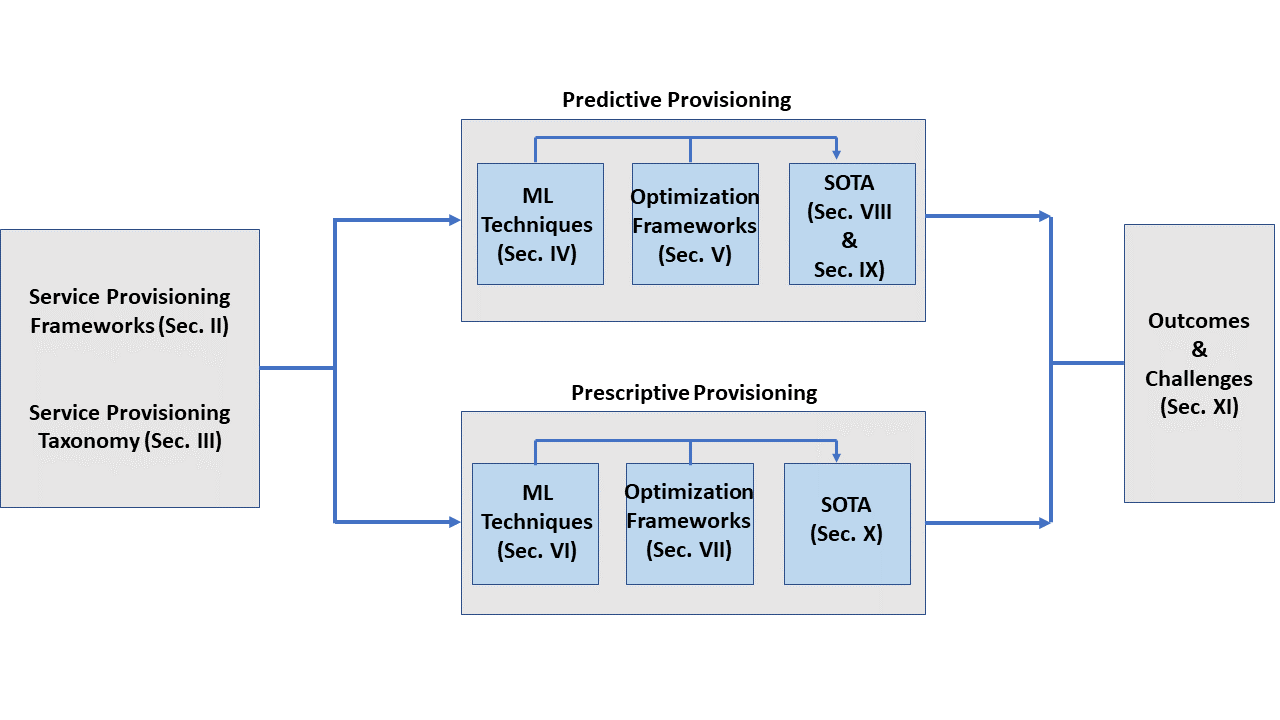}
\caption{Structural organisation of the survey.}
\label{organization}
\end{center}
\end{figure*} 

Specifically, the already rich literature on ML-aided traffic-driven service (re)provisioning is categorized into predictive and prescriptive frameworks for proactive and adaptive networks, with each category relying on a different ML subfield. The novel categorization serves to highlight the capabilities and limitations of each approach, providing also qualitative comparisons when appropriate. For each category, the ML background is described with an orientation to the specific use case surveyed in this work. To the best of our knowledge this is the first survey that appears in the literature that provides an in-depth analysis of the state of the art on ML-aided traffic-driven service provisioning in optical networks. This survey complements very well the general surveys on AI/ML-based control and management functionalities in optical networks, with its added value with respect to other surveys lying in the fact that it provides an in-depth examination of one of these functionalities, namely \emph{traffic-driven service provisioning}. 

Finally, the reader should note that other service provisioning techniques that are not traffic-driven, namely static service provisioning within static/reactive networks (with a rich literature as shown in the taxonomy presented in Section~\ref{taxonomy_section}), are out of scope of this work and they are not included in this survey. Specifically, even though recently RL has been applied for static service provisioning~\cite{9713679,9766412,9895124}, in this case the application of RL mainly concerns addressing the computational complexity of the underlying resource allocation problem. Hence, in this case RL constitutes a heuristic approach that does not include the notion of time or the notion of a reconfigurable programmable infrastructure; that is, it is not concerned with network (re)optimization as traffic varies over time. Hence, such works, even though important, are out of the scope of this survey and are not further discussed.

The rest of the paper is structured as follows. Section~\ref{evolution} discusses the evolution of service provisioning in optical networks and Section~\ref{taxonomy_section} discusses the taxonomy of the existing service provisioning approaches. Following the taxonomy, Sections~\ref{pred_ra} and~\ref{RA} present ML techniques for predictive reconfiguration and associated resource allocation frameworks, while Sections~\ref{pres_ra} and~\ref{prescriptive} present ML techniques for prescriptive reconfiguration and associated resource allocation frameworks. The literature review of predictive and prescriptive service provisioning for both proactive and adaptive networks is presented in Sections~\ref{surv_proactive}, \ref{traffic_prediction},  and~\ref{surv_prescr}, summarizing the existing state of the art and providing the main outcomes and research challenges for each provisioning approach. Section~\ref{conclusion} summarizes the work and highlights future research avenues. 

Figure~\ref{organization} presents graphically the structural organization of the survey. Following the service provisioning frameworks and taxonomy in Sections~\ref{evolution} and~\ref{taxonomy_section} the survey deals with predictive and prescriptive provisioning techniques within proactive/adaptive networks (Sections~\ref{pred_ra},~\ref{RA},~\ref{traffic_prediction}~\ref{surv_proactive}, and~\ref{pres_ra},~\ref{prescriptive},~\ref{surv_prescr}, respectively). For each of the two provisioning schemes, ML techniques, optimization frameworks (e.g., based on exact or heuristic ML-aided algorithms), and the state-of-the-art (SOTA) is presented in the sections as illustrated in the organization figure below. Finally, the discussions that follow each section drive the outcomes of the survey as well as various future research challenges as presented in Section~\ref{conclusion}. 

\section{Evolution of Service Provisioning}\label{evolution}
A reconfigurable optical network capable of adapting, as closely as possible, to the time-varying traffic demand conditions has always been on the wish list of network operators~\cite{ciena18adaptive}. In practice, however, traditional optical network architectures were not built to support dynamic operations, mainly due to the initial absence of tunable, flexible, and software programmable optical components. In fact, optical transport networks have been typically configured statically and engineered using worst-case, end-of-life conditions. The rapid and unpredictable growth in capacity needs, driven by the emergence of new types of networks (e.g., 5G, IoT, network slicing), applications (e.g., virtual reality, augmented reality) and services (e.g., cloud services, high-quality video, industrial automation, cloud robotics, vehicular communications), has led to several technological breakthroughs, gradually transforming the statically configured optical networks to more dynamic and reconfigurable infrastructures. Amongst the most notable technological advances, enabling agility and reconfigurability of optical transport networks, is the advent of the flexible grid and variable bit-rate coherent optics that can be externally programmed and controlled through software; that is, software-defined networking (SDN) is implemented in optical transport infrastructures~\cite{7503119} consisting of adaptive hardware components that can be intelligently programmed and controlled by analyzing real-time data monitored and collected by network telemetry services. In the heart of intelligently controlling an optical network to self-configure and self-optimize to a continuously changing network environment, is the application of AI and ML. 

\begin{figure}[h]
\begin{center}
\includegraphics[scale=0.32]{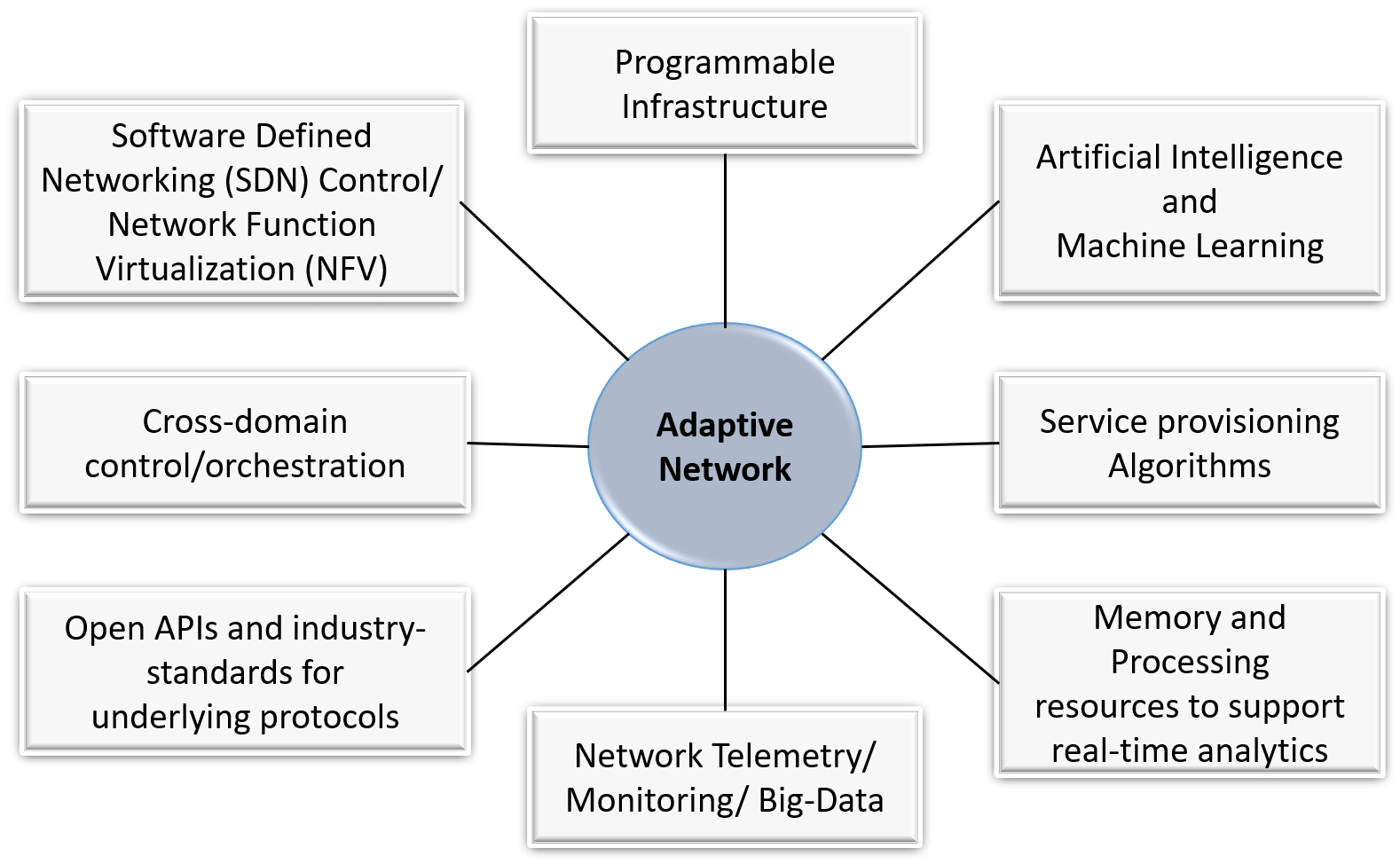}
\caption{Building blocks of adaptive optical networks.}
\label{f1}
\end{center}
\end{figure}

At present, the main building blocks of an adaptive optical network, such as more-programmable optics, SDN- and network function virtualization (NFV)-based software control and automation, cross-domain orchestration, monitoring and network telemetry, AI/ML-based intelligence, and processing resources to support AI/ML analysis (Fig.~\ref{f1}), are now available. Nevertheless, each one of these blocks is currently at different stages of maturity and commercial adoption. Thus, considerable technological advances are still required before truly adaptive optical networks can be realized. This survey focuses on a review of the state-of-the-art regarding the existing AI/ML-aided service (re)provisioning frameworks developed to best-fit time-varying traffic demands, as these have evolved over the years. It is worth mentioning that, the evolution towards adaptive optical networks is related to a number of other use-cases, such as self-healing operations upon failures and attacks to increase network reliability, and self-optimization of system margin utilization in QoT estimation models (i.e., to capture signal variability and degradation). Surveys related to these use cases can be found in~\cite{Furdek:21, Musumeci:19, Pointurier:21}. In this survey, however, the focus is on the use case of traffic-driven service (re)provisioning for which the underlying ML-aided resource allocation schemes developed are extensively reviewed. To the best of our knowledge, this is the first survey in the literature that is dedicated to this important use case.             

\begin{table*}[hbpt]
\center
\caption{Service provisioning frameworks.}
\footnotesize
\renewcommand{\arraystretch}{1.5}
\begin{tabular}{p{3cm}|p{4.5cm}|p{4.5cm}|p{4.5cm}}

  & {\bf Reactive} &{\bf Proactive} & {\bf Adaptive} \\ \hline
 
{\bf Infrastructure} &	Static & More dynamic with external software control & Dynamic and programmable with embedded software control \\ \hline

{\bf Analytics} &  Descriptive & Predictive/Prescriptive & Predictive/Prescriptive \\ \hline

{\bf Level of Automation} &  Low & Partial & 	High \\ \hline

{\bf ML- Aided} & 	No	& Yes & Yes \\ \hline

{\bf Description} &
$\bullet$ Services are provisioned to meet end-of-life traffic needs (i.e., years) \newline  
$\bullet$ Reprovisioning is initiated when utilization of reserved resources reaches a predefined threshold \newline
$\bullet$ Resource allocation algorithms are employed to a-priori reserve resources to meet updated traffic needs &  

$\bullet$  Services are provisioned to meet future, short-term, traffic needs (i.e., hours, weeks) \newline
$\bullet$  (Re)provisioning is initiated periodically (or if required) \newline
$\bullet$  ML-aided predictive or prescriptive resource allocation algorithms are employed to a-priori reserve resources towards meeting services' short-term traffic needs \newline
$\bullet$  Network environment is periodically monitored to allow predictive/prescriptive ML model adaptation to the time-varying traffic needs
 & 

$\bullet$ Services are provisioned on the fly according to their reported traffic needs while also accounting for the future traffic behavior (i.e., to improve network efficiency, congestion, fragmentation, etc.)\newline  
$\bullet$ ML-aided predictive or prescriptive resource allocation algorithms are employed to dynamically provision services towards meeting services' current and future traffic needs  \newline
$\bullet$ Network environment is continuously monitored (in real-time) to allow predictive/prescriptive ML model adaptation to the time-varying traffic needs 
\end{tabular}
\label{f2}
 \end{table*}

In general, traffic-driven service provisioning frameworks have evolved over the years, following the technological advances towards the realization of adaptive networks. Specifically, the evolution has gradually transformed these frameworks from reactive, to proactive, towards more adaptive service provisioning operations (Table~\ref{f2}). Each transformation step increases the levels of network automation, programmability, intelligence, and dynamicity of service provisioning operations, with the aim of more efficiently utilizing the available network resources, while ensuring that pre-defined QoS targets of existing and future services and applications are met. Table~\ref{f2} summarizes the main characteristics of each service provisioning framework with a brief description on how variations in traffic-demand are captured to initiate the service reprovisioning mechanisms. As described in Table~\ref{f2}, reactive networks are mainly static and reprovisioning is initiated when predefined network performance metrics are violated (e.g., link utilization reaches predefined thresholds). Intelligence, through the application of AI/ML for traffic-driven service (re)provisioning purposes is present only in proactive and adaptive frameworks. Even though several works exist in the literature that apply AI/ML methods (e.g., RL, genetic algorithms) for service provisioning purposes in static networks~\cite{9713679,9766412,9895124}, this is mainly to address computational complexity of resource allocation problems when the problem size makes the problem intractable (i.e., AI/ML is not concerned with traffic analysis and therefore such approaches are not reviewed in this survey).

From the service provisioning and resource allocation perspective, proactive networks mainly differ from adaptive networks on how resources are reserved (i.e., a-priori or on the fly). In the proactive case, resources for each service are reserved a-priori to meet their short-term traffic needs (e.g., for an hour). In the adaptive case, resources for each service are reserved on the fly, to meet their reported traffic needs, and these resources are released when not needed. In both proactive and adaptive cases, ML-aided predictive or prescriptive resource allocation algorithms are applied. 

Given this, a first qualitative comparison between the reactive, proactive, and adaptive networks is presented in Table~\ref{f3}, in order to highlight the main reasons behind each evolution step. Briefly, Table~\ref{f3} indicates that each evolution step is mainly driven by the need to more efficiently utilize the available network resources (e.g., spectrum) as capacity needs continuously increase. Specifically, according to Table~\ref{f3}, each evolution step increases the dynamicity of the underlying network infrastructure (i.e., the reconfigurability capabilities), allowing service (re)provisioning to more closely follow the traffic demand variations over time. As such, spectrum and energy savings are increased as we move from reactive, to proactive, towards adaptive networks. Of course, the price of the increased dynamicity is the control and management overhead associated not only with the dynamic service provisioning operations, but also with the application of ML that requires continuous monitoring, model adaptation, and on-line decision making. Furthermore, violations on predefined QoS requirements may be more evident with each evolution step (e.g., due to the more frequent reconfigurations causing disruptions). Nevertheless, such undesired side-effects can be controlled to desired levels through the ML-aided service provisioning algorithms developed.

\begin{table*}[hbtp]
\center
\caption{Qualitative comparison of service provisioning in reactive, proactive, and adaptive networks.}
\footnotesize
\renewcommand{\arraystretch}{1.5}
\begin{tabular}{p{4cm}|p{2cm}|p{5cm}|p{5cm}}

  & {\bf Reactive} &{\bf Proactive} & {\bf Adaptive} \\ \hline
 
{\bf Service over-provisioning Level} \newline (i.e., in spectrum utilization) &	High & High to Low \newline (as planning intervals
reduce) & Low\\ \hline

{\bf Service Reconfiguration Frequency} &	Low &	Low to High \newline (as planning intervals reduce) & High\\ \hline

{\bf QoS Guarantees Level} 
\newline (service availability, bit-rate guarantees, provisioning latency guarantees)
 &	High & High to Low \newline (depending on the traffic prediction accuracy, speed, and whether reconfigurations can be seamless) & Low \newline (in the absence of in-advance resource reservation, fast and seamless reconfigurations) \\ \hline

{\bf Network Throughput 
Level} \newline (service admission) &	Low	 & Low to High \newline (as planning intervals
reduce) &	High  \\ \hline

{\bf Energy Consumption Level} \newline (virtual - physical layers)	& High  &	High to Low \newline (as planning intervals reduce) &	Low \\ \hline

{\bf Control and Management Overhead} &	Low &	Low to High \newline (as planning intervals
reduce) &	High \\ \hline 
\end{tabular}
\label{f3}
 \end{table*}
 
Machine learning is by and large categorized into three major subfields; supervised learning (SL), unsupervised learning (UL), and reinforcement learning (RL)~\cite{Bishop06}. Even though all subfields have been used in the literature for traffic analysis, in general SL and RL are the subfields mainly utilized for traffic-driven service provisioning. Specifically:
\begin{itemize}  
\item \emph{Predictive service provisioning} is based on SL and specifically on {\it predictive analytics} providing insights on what will happen in the future, subsequently driving resource allocation decisions. An analytical description is given in Table~\ref{f2}, where it is shown that predictive service provisioning applies to both proactive and adaptive networks.
\item \emph{Prescriptive provisioning} is based on RL and specifically on {\it prescriptive analytics} to help with decision making by providing actionable advice. An analytical description is given in Table~\ref{f2}, where it is shown that prescriptive service provisioning applies to both proactive and adaptive networks.
\end{itemize}

In essence, in predictive provisioning, actionable advice is derived by a resource allocation algorithm that utilizes as inputs the traffic predictions (outputs) of trained ML models. Hence, when coordination of the various services is required (e.g., to meet predefined QoS targets in congested networks), this is mainly handled by the resource allocation algorithm, by exploiting the traffic predictions. In prescriptive schemes, actionable advice can be directly derived by reinforced resource allocation policies that are learned by trial and error (i.e., without the use of labeled data) to coordinate, if required, the various services towards meeting targeted QoS requirements. 

It is worth noting, that network traffic analysis is also associated with UL, most commonly for anomaly and attack detection/identification, and the characterization of traffic trends for capacity network planning.
Indicatively, UL was applied to perform clustering with the aim of identifying groups of IP traffic flows, mobile base-stations, or applications (e.g., Facebook, Twitter, etc.) with similar underlying profiles (trends), to reveal their underlying spatio-temporal similarities~\cite{7917576}. This information can be used for understanding the reasons behind peak traffic demands during a time interval (e.g., a day) and the main sources causing network congestion, in order  to improve network capacity planning. Alternatively, it can be used to reveal groups of services or applications with similar QoS requirements and traffic trends to guide resource allocation and network slicing decisions. Hence, clustering can be used in conjunction with predictive and prescriptive service provisioning (e.g., with each service cluster being subject to different QoS constraints), to better guide the resource allocation decisions. Note, however, that the outputs of UL (e.g., clustering) are static, as they do not capture future traffic trends, therefore they do not provide traffic predictions for future time points.     

Finally, it is worth mentioning that, for adaptive networks, several efficient resource allocation algorithms, initially developed for dynamic networks, usually constitute building blocks of the overall service provisioning algorithms, with ML enhancing their efficiency (e.g., service admission rate of the system). Such algorithms are usually referred to as \emph{dynamic service provisioning algorithms}. The same holds for proactive networks, where several efficient resource allocation algorithms, initially developed for static networks (i.e., to a-priori reserve resources over a fixed planning interval), usually constitute building blocks of the overall service provisioning algorithms, with ML providing an effective means of accurately estimating the traffic needs over the planning intervals of interest. In this survey, to differentiate between the various service provisioning schemes in the diverse types of networks, we provide the following additional definitions:
\begin{itemize}  
\item \emph{Dynamic service provisioning} applies to \emph{dynamic networks} where services are provisioned on the fly according to their reported traffic needs. In dynamic networks, unlike adaptive networks, ML-aided programmability through embedded software control is not present.
\item \emph{Static service provisioning} applies to \emph{static} or \emph{reactive networks} where services are provisioned to meet end-of-life traffic needs. In reactive networks, unlike proactive networks, ML-aided programmability though software is not present. Hence, in reactive networks descriptive analytics are utilized for service re-provisioning purposes (Table~\ref{f2}). In static networks, unlike proactive networks where the infrastructure is partially dynamic, the infrastructure is static.   
\end{itemize}
 
In this survey, however, as emphasized above, the focus is to review the literature related to the ML-aided proactive and adaptive networks, and more specifically to the traffic-driven ML-aided resource allocation algorithms for optical networks. As in both proactive and adaptive networks predictive and prescriptive resource allocation algorithms are applied, with each type related to different ML subfields, the survey is partitioned according to these ML subfields (see Fig.~\ref{organization}). Specifically, predictive schemes for both proactive and adaptive networks relate to SL and are described in Section~\ref{pred_ra}, while prescriptive schemes relate to RL are described in Section~\ref{pres_ra}. This categorization is opted for so as to be able to describe each ML framework according to the particular use case this survey is focused on, rather than as general ML subfields applicable to any use case, discussing for each their limitations and research challenges.  

\section{Taxonomy of Service Provisioning Approaches}
\label{taxonomy_section}

\begin{figure*}
\centering
 \scriptsize
\begin{forest}
  forked edges,
  for tree={align=center, anchor=children},
  before packing={where n children=7{calign child=3, calign=child edge}{}},
  before typesetting nodes={where content={}{coordinate}{}},
  where level<=1{line width=0.5pt}{line width=0.5pt},
[Taxonomy of Service Provisioning Approaches, rounded corners, top color=gray!40, bottom color=gray!40, edge+={darkgray, line width=0.5pt}, draw=darkgray 
    [Static \\ Networks , rounded corners, top color=white, bottom color=white, edge+={darkgray, line width=0.5pt}, draw=darkgray
     [Static \\ Provisioning, rounded corners, top color=white, bottom color=white, edge+={darkgray, line width=0.5pt}, draw=darkgray 
       [End-of-Life \\ Off-line \\ Optimization
         [(M)ILP/ \\ Heuristics [\cite{6083231,Zang00areview} \\ \cite{7105364,KLINKOWSKI201858}]]
        ]
      ]
    ]
    [Reactive \\  Networks, rounded corners, top color=white, bottom color=white, edge+={darkgray, line width=0.5pt}, draw=darkgray
     [Static \\ Provisioning, rounded corners, top color=white, bottom color=white, edge+={darkgray, line width=0.5pt}, draw=darkgray 
       [Rule-Based \\ Off-line  \\ (Re)Optimization
          [(M)ILP/ \\ Heuristics [\cite{328952, 594381} \\ \cite{1194820, 891340} \\ \cite{920841,7308092} \\ \cite{5272256,Yin:12} \\ \cite{CASTRO20122869,6524878} \\ \cite{6532668}]]       
       ]
     ]
    ]
 [Proactive\\  Networks, rounded corners, top color=gray!20, bottom color=gray!20, edge+={darkgray, line width=0.5pt}, draw=darkgray
     [Predictive \\ Provisioning, rounded corners, top color=gray!5, bottom color=gray!5, edge+={darkgray, line width=0.5pt}, draw=darkgray 
       [Short-Term \\ Traffic \\ Prediction 
         [ Supervised \\ Learning \\{\color{gray} (Section~\ref{pred_ra})}  
            [ \cite{8501527,9204965}\\ \cite{8047676,7830260} \\ \cite{8473978,8430520} \\ \cite{8386186,8696381} \\ \cite{Balanici:21,doi:10.1002/dac.4516} \\ \cite{9748600,8845132} \\ \cite{9782838,MARYAM202313} \\{\color{gray} (Section~\ref{traffic_prediction}) } ]         
         ]       
       ]
       [Traffic-Driven \\  Off-line  \\ (Re)optimization 
          [(M)ILP/ \\ Heuristics \\ {\color{gray} (Section~\ref{proactive_pred})}
          [\cite{ZHANG2011171, 5540249,6515884} \\ \cite{6384633,Alvizu2017EnergyED,7444562} \\ \cite{8047676,7830260,s11227-020-03493-7} \\ \cite{6381741,6831425,8473978} \\ \cite{8501527,8430520,8686111}\\ \cite{9322381,9347908,9522187} \\ \cite{9748600,9782838,MARYAM202313} \\ {\color{gray}(Section~\ref{surv_proactive_opti})}]       
          ]
       ]
     ]
     [Prescriptive \\ Provisioning, rounded corners, top color=gray!5, bottom color=gray!5, edge+={darkgray, line width=0.5pt}, draw=darkgray 
       [Short-Term \\ Resource Allocation \\ Policies
       [Reinforcement \\  Learning \\ {\color{gray} (Section~\ref{pres_ra})}
       [\cite{8396130,8751369} \\ \cite{8620207,9256655} \\ {\color{gray} (Section~\ref{presc_proact_survey})} ]]]
       [Policy-Driven \\  Off-line  \\  (Re)Optimization 
       [(M)ILP/ \\ Heuristics \\ {\color{gray} (Section~\ref{optimization_prescr_proact})}[\cite{8396130,8751369} \\ \cite{8620207} \\ {\color{gray} (Section~\ref{presc_proact_survey})}]]     
       ] 
     ]
   ]
   [Dynamic \\ Networks, rounded corners, top color=white, bottom color=white, edge+={darkgray, line width=0.5pt}, draw=darkgray
     [Dynamic \\ Provisioning, rounded corners, top color=white, bottom color=white, edge+={darkgray, line width=0.5pt}, draw=darkgray 
     [Rule-Based \\ On-Line \\ Optimization 
     [Heuristics [\cite{6083231,Zang00areview} \\ \cite{7105364,KLINKOWSKI201858} \\ \cite{7490359,7293303}]]]]
   ]
   [Adaptive \\ Networks, rounded corners, top color=gray!20, bottom color=gray!20, edge+={darkgray, line width=0.5pt}, draw=darkgray
     [Predictive \\ Provisioning, rounded corners, top color=gray!5, bottom color=gray!5, edge+={darkgray, line width=0.5pt}, draw=darkgray 
     [Short-Term \\ Traffic \\ Prediction
     [Supervised \\ Learning \\{\color{gray} (Section~\ref{pred_ra})} 
     [\cite{8501524,XIONG201999} \\ \cite{8737631,Zhao:18} \\ \cite{9203477,8436062} \\{\color{gray} (Section~\ref{traffic_prediction}) } ]]]
     [Traffic-Driven \\ On-Line \\ Optimization
     [ Heuristics \\ {\color{gray} (Section~\ref{adaptive_pred})} [ \cite{8501524,9042293} \\ \cite{9203477,8436062} \\ \cite{XIONG201999,Zhao:18} \\ {\color{gray}(Section~\ref{surv_adaptive_opti})}]]]]
     [Prescriptive \\ Provisioning, rounded corners, top color=gray!5, bottom color=gray!5, edge+={darkgray, line width=0.5pt}, draw=darkgray 
     [Policy-Driven \\ On-Line \\ Optimization
     [Reinforcement \\ Learning \\ {\color{gray} (Section~\ref{pres_ra})}
         [Heuristics \\ {\color{gray} (Section~\ref{optimization_prescr_adapt})}
     [\cite{8738827,8847548,3229554} \\ \cite{Zhao:18,8485853,9083336} \\ \cite{9375634,9373585,SUN2021107891} \\  \cite{Zhao:21,Aibin2020MonteCT,9507559} \\  \cite{9609608,9748476,9279336} \\ \cite{9779080,KHOSHKHOLGHI2022109451} \\ {\color{gray} (Section~\ref{presc_adaptive_survey})}]]]] ]
   ]
 ]
\end{forest}
 \caption{Taxonomy of service provisioning approaches.}
 \label{taxonomy}
\end{figure*}
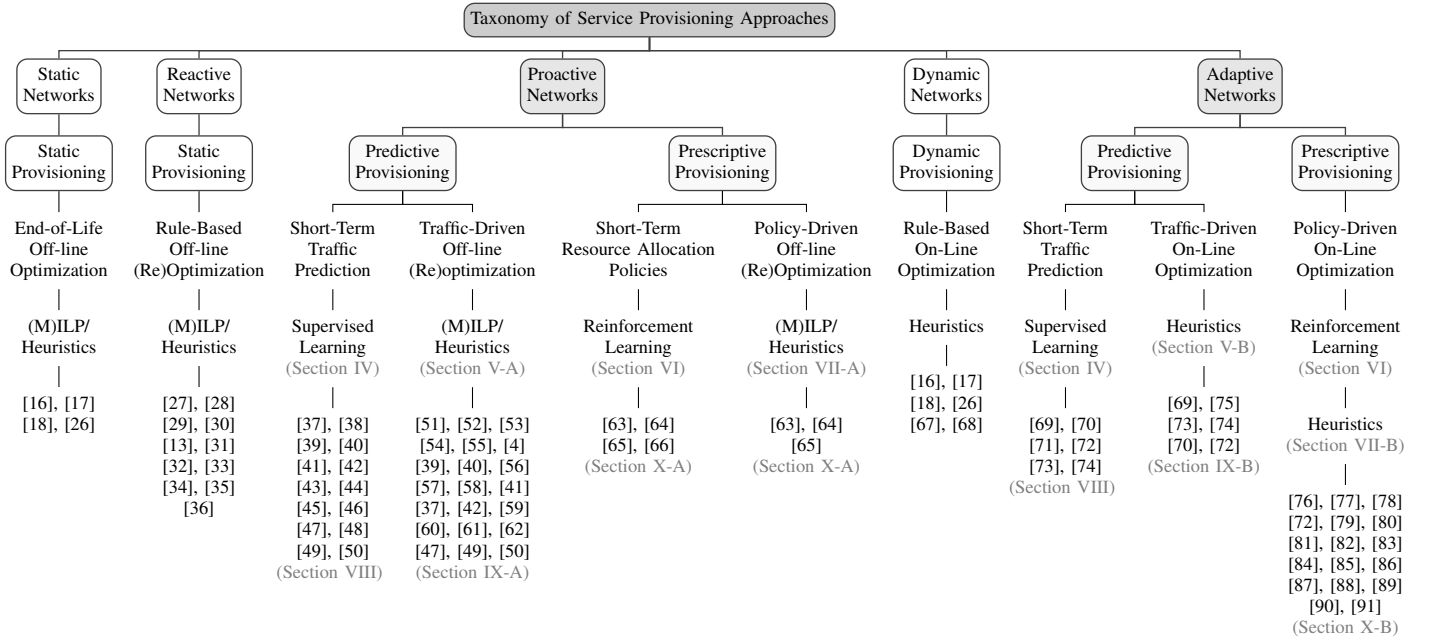

The taxonomy of the existing service provisioning approaches for static, reactive, proactive, dynamic, and adaptive networks is given in Fig.~\ref{taxonomy} to better guide this survey. As previously mentioned, in this survey we are concerned with the literature related with proactive and adaptive networks, where ML-aided predictive and prescriptive service provisioning is applied. However, in Fig.~\ref{taxonomy}, we also provide references of previous surveys and other publications related with service provisioning in static, reactive, and dynamic networks. This is mainly done to provide a complete picture to the reader regarding works on service provisioning problems and optimization approaches in all types of networks. 

It is worth mentioning that service provisioning in proactive networks is usually compared with service provisioning in static and reactive networks, whereas service provisioning in adaptive networks is usually compared with service provisioning in dynamic networks. Most of the times, for comparison purposes, baseline service provisioning schemes are opted for. Furthermore, such baseline service provisioning schemes, previously developed for static, reactive, and dynamic networks, usually constitute building blocks for the development of service provisioning approaches in proactive and adaptive networks. 

Indicatively, the reader is referred to~\cite{6083231, Zang00areview} for service provisioning in wavelength division multiplexed (WDM) networks, to~\cite{5948951} for service provisioning in waveband (WBS) switching networks, to~\cite{7105364} for service provisioning in elastic optical networks (EONs), and to~\cite{KLINKOWSKI201858} for service provisioning in space division multiplexed (SDM) optical networks. Note that in~\cite{6083231, Zang00areview,7105364,KLINKOWSKI201858}, heuristic and mixed integer linear algorithms (MILP) are presented/described to solve the underlying service provisioning problem in various types of networks and optical spectrum multiplexing technologies. 

Regarding the service provisioning schemes that this survey focuses on, these are further described and surveyed in the sections indicated in Fig.~\ref{taxonomy}. As such, Fig.~\ref{taxonomy} complements Fig.~\ref{organization} by providing also a brief overview of the structure of this survey, indicating how the sections that follow are related to each other. It should be noted that works exist in the literature that may deviate from the taxonomy in Fig.~\ref{taxonomy}. Such works are discussed in the section that best fits the methodology followed by the authors. Importantly,  the taxonomy in Fig.~\ref{taxonomy} is subject to extensions, in the near future, as the research community proposes novel ways on how to utilize/develop ML for optical networks, or even on how to build optical networks for ML.  

\section{Machine Learning for Predictive Provisioning}~\label{pred_ra} 
Predictive provisioning can in general be decomposed into two interdependent, but complementary, sub-problems (Fig.~\ref{taxonomy}); that is, the ML-aided short-term traffic prediction problem, and the resource optimization problem that is driven by the predictions. Hence, each sub-problem is described and discussed separately. In this section, we focus on the description of the traffic prediction sub-problem which is subsequently surveyed in Section~\ref{traffic_prediction}. 

The network traffic prediction problem is essentially a time-series forecasting problem, in which past observations are exploited to predict one or more future observations. It has been shown that network traffic is, to some extent, predictable~\cite{7917576,SANG2002329, 6381046}, thus network traffic prediction is plausible in the sense that it can yield a sufficiently accurate outcome for decision-making and network optimization procedures.

Over the previous years, statistical linear models have been extensively used to address time-series prediction problems, with the Auto Regressive Integrated Moving Average (ARIMA) model and its variants (i.e., Seasonal ARIMA, Fractional ARIMA, etc.)~\cite{DBLP:journals/corr/abs-1302-6613} being the most widely-used for network traffic prediction~\cite{1510713, 10.1007/978-3-540-31956-6_58, 10.1007/s12544-015-0170-8, 6887148, 7023486}. In general, ARIMA models are based on a linear combination of past values and/or errors. Despite their ease of use and interpretation, their pre-assumed linear form renders them inadequate in many practical situations and real-world problems~\cite{DEGOOIJER2006443}. In fact, from the late 1970s it became evident that such statistical models are often not satisfactory, as real time-series data are rarely purely linear~\cite{ZHANG2003159}. Therefore, although the ARIMA models exhibit an accurate forecasting ability for certain types of time-series, they are not the most suitable to describe characteristics of network data traffic, that exhibits a highly non-linear nature, long-range dependencies, self-similarity, and burstiness. 

Machine learning, that does not presume the linearity of a model, is considered today a powerful tool capable of more accurately capturing network traffic behavior. In fact, numerous studies comparing statistical linear and non-linear ML methods have verified that non-linear ML methods achieve higher traffic prediction accuracies~\cite{6047849, 1544219, 8530608, 9012669}. Among the various subfields of ML, SL is most commonly used for traffic analysis and prediction purposes. 

In general, SL techniques can be categorized in {\it classification} and {\it regression} techniques. For network traffic prediction, regression techniques are commonly applied, due to the continuous nature of the output, enabling flexibility in the predictions (i.e., traffic predictions are not restricted to predefined classes). Nevertheless, works exist in the literature that also address the traffic prediction problem as a classification problem. For an underlying network infrastructure that operates on fixed bit-rates [e.g., on wavelengths in WDM optical networks where the objective is to predict the number of wavelengths required to support future traffic needs], classification may be sufficient, especially for a small number of classes. However, for flexible infrastructures [e.g., elastic optical networks (EONs)], allowing the allocation of various spectrum sizes, classification restricts the predictions to the considered classes, possibly leading to improper resource allocation decisions. Furthermore, classification restricts the model to predicting a traffic demand within the predefined classes, eventually rendering the output of the model to be static; that is, classification does not sufficiently capture future traffic trends. Hence, Section~\ref{SL} focuses on describing the general traffic regression framework, surveying, however, papers that are addressing traffic prediction problems through classification.

\subsection{SL for Network Traffic Prediction}~\label{SL} 
Supervised learning is based on analyzing labeled data (patterns) to find a model that predicts the labels of unseen data. In general, a labeled dataset $D$ can be described by $D=(X, Y)$, 
where $X$ is the set of input patterns and $Y$ is the set of their output labels (ground truths). In a network traffic prediction problem, where the past values of traffic are used to predict one or more future values,
an input pattern is usually represented through a features' vector
\begin{equation*}
{\bf x}=[x_{t-1}, x_{t-2}, x_{t-3}, \cdots, x_{t-w}] \in X,
\end{equation*}
to predict the output vector
\begin{equation*}
{\bf y}=[{x_{t}}, {x_{t+1}}, \cdots, {x_{t+s}}]  \in  Y,
\end{equation*}
where $x_t$ denotes the traffic value at time slot $t$, $w$ denotes the number of previous time slots considered in the input patterns, and $s$ is the number of future time slots for which the model predicts the traffic values (i.e., model outputs). A time slot, $t$, is a time interval of a fixed duration defining the time scale of predictions. 

In the simplest model training and testing example, dataset $D$ is partitioned into the training and test datasets, with the training dataset used to optimize the unknown parameters of the ML model of choice, and the test dataset used to evaluate the model's accuracy. In the literature, various ML models are examined and compared for the network traffic prediction problem (e.g., DNNs, LSTMs, etc.), and these are analytically discussed in Section~\ref{model}. 

Specifically, irrespective of the underlying ML model of choice, the goal of the training phase is to obtain a model that accurately associates the input patterns to their corresponding labels. 
Hence, during training, an optimization algorithm (e.g., Adam~\cite{kingma2017adam}) measures the accuracy achieved so far, through a loss function, updating the model parameters until the loss is sufficiently minimized. In regression analysis, mean absolute error (MAE) and mean squared error (MSE) are the loss functions commonly applied, while other loss function are also possible [e.g., Mean Squared Logarithmic Error (MSLE)]. Subsequently, after training, the accuracy of the model is evaluated over the patterns of the test dataset. Common measures of accuracy are the MAE, MSE, the Root Mean Squared Error (RMSE), and the Mean Absolute Percentage Error (MAPE).   

\subsection{Feature Selection, Data Preprocessing, and Inference}
In the context of network traffic prediction, the term traffic is used in a broader meaning and refers to any quantity that characterizes the amount of data being transferred at a given point of time, such as, for instance, data rate (bits per second), number of packets, link load, number of on-going user sessions, etc. Moreover, the problem of network traffic prediction may be addressed at various network levels. For example, some authors consider the network traffic on a certain based station~\cite{8047676}, while others consider the aggregated traffic associated with a certain network hub or link~\cite{8845132,8737631,Zhao:18,8501527,9204965,8047676,7830260,8386186,8696381,Balanici:21,9748600}. Furthermore, many studies tackle the more complex problem of traffic demand (TD) matrix prediction~\cite{8473978}, where the traffic between all source-destination pairs in the network is considered.

Depending on the traffic prediction problem (e.g., prediction of TD matrix, node data rates, link load), the input features of the model/s must be appropriately selected and represented. In general, features refer to the inputs of the model, and, as previously described, these features at least correspond to the traffic values of several previous time steps, in order to predict the traffic value of the next time points or intervals. Of course, additional features may be also considered to better describe the underlying problem and increase performance accuracy. Such features may include the identity of the source/destination nodes (e.g., base stations, IP nodes), day of the week, or month of the year, etc. In general, the features considered may significantly affect the achievable accuracy of the model, and hence must be carefully selected to best describe the underlying problem. A recent study, demonstrating the importance of appropriately selecting the input features towards accurate traffic prediction, can be found in~\cite{9782850}. Section~\ref{traffic_prediction} provides more information on what features are commonly considered in conjunction with the traffic prediction problem examined.  
  
Nevertheless, irrespective of the features considered, a common approach for data preprocessing is to discretize time into intervals of fixed duration (time slots), and account for traffic on a per time slot basis (i.e., depending on the desired time scale of predictions). Since traffic may fluctuate within the slot duration (i.e., several traffic values may be reported for the same time slot), which is normally the case for large time slot durations (e.g., hours), traffic fluctuations are either averaged for each time slot, or, alternatively, the maximum traffic value is opted for in order to avoid service under-provisioning. As over time the total number of past traffic values grows continuously resulting in a prohibitive computational complexity, a sliding window is typically used to create a dataset consisting of patterns of a fixed size. 

Before training and testing, dataset scaling is also applied to improve the learning performance, in particular with regards to the training convergence time. After scaling, the dataset is split into the training and test datasets. Since in traffic prediction the interest is on traffic estimations of future time slots, the dataset must be split in such a way so that the test dataset includes patterns of earlier points in time compared to the training dataset. For the inference phase, the trained model can be applied to new inputs, as these are extracted from the evolving network. 

\subsection{Available Datasets}\label{avail_datasets}
In the literature, model training and testing is performed either according to synthetic datasets or according to real traffic traces. Synthetic datasets are commonly generated by event driven simulators (e.g., available in MATLAB, OMNeT++~\cite{804564}), by traffic demand models~\cite{1194820}, or by distributions describing the traffic behavior. 
Indicatively, Poisson distribution is used for simulating arrivals and departure times, while the log-normal distribution is also applied to simulate the holding times or traffic volumes.
There are, however, some real-world datasets utilized in the literature as well, with several of them  shown in Table~\ref{source_dataset} (publicly available or upon request). The most utilized datasets are from the Abilene and GEANT networks~\cite{Abilene_dataset, totem_project}. In particular the Abilene dataset provides traffic bit-rate information for every pair of nodes in the backbone network, over a $5$-minute time horizon, and for a six-month period (March-September 2004). 
In general, real-world datasets provide similar information, including traffic loads (i.e., in bit-rate) on the links of the network (i.e., link utilization), with time-scales of the traffic samples varying from seconds to minutes. In~\cite{CAIDA}, a variety of datasets can be found, including link loads in backbone networks that can be made available upon request. 

\begin{table}[htbp]
\begin{center}
\caption{Sources of real datasets.}
    \begin{tabular}{ | l | l |}
    \hline
    {\bf Real Datasets} & {\bf Source} \\ \hline
    Abilene network & \cite{Abilene_dataset} \\ \hline
    GEANT network &   \cite{totem_project} \\ \hline
    TELUS PureFibre & \cite{telus_fibre} \\ \hline
    Internet Service Providers (ISPs) & \cite{datamarket} \\ \hline
    CAIDA org. & \cite{CAIDA} \\ \hline
    Waikato network & \cite{Waikato} \\ \hline
    Telecom Italia & \cite{Titalia} \\ \hline
    Database of Cell Towers & \cite{ENAiKOON16} \\ \hline
    CRAWDAD Datasets & \cite{crawdaddatasets} \\ \hline
    SNDlib Datasets & \cite{OrlowskiPioroTomaszewskiWessaely2010} \\ \hline
    \end{tabular}
    \label{source_dataset}
    \end{center}
 \end{table}
 
\subsection{ML Models Applied for Network Traffic Prediction}\label{model}
While several ML models have been applied in the literature for network traffic prediction, this section briefly discusses the types of models that have been applied the most, or have shown promising performance indicators (i.e., concerning accuracy and computational complexity). A comprehensive list of the ML models applied for network traffic prediction is shown in Table~\ref{list_models}. Amongst them feedforward NNs [i.e., NNs, deep neural networks (DNNs)], recurrent neural networks and their variants [e.g., RNNs, LSTM, GRU, convolutional and/or graph NNs (e.g., (Deep) Convolutional Neural Network (DCNN/CNN), Graph Convolutional Network-Generative Adversarial Network (GCN-GAN)], as well and as Bayesian learning approaches [e.g., Gaussian Process Regression (GPR)] have been applied the most and are briefly discussed next. Other ML approaches that appear in the literature, such as Linear Regression, SVMs, Random Forests, and Boosted Decision Trees, are most of the times used as comparative approaches, with other types of models, specifically designed to model sequence data (i.e., LSTMs, GRUs, GPR) or to capture network (graph) dependencies (i.e., DCNN, GCN-GAN), outperforming such models. 

Finally, it should be emphasized that many of the existing works do not apply a single ML method, but develop {\it ensemble ML} algorithms; that is, they combine more than one ML learning techniques into one predictive model for obtaining better performance. Especially interesting are studies where NNs are applied in combination with linear regression methods in order to capture accurately both the non-linear and linear components of network traffic~\cite{8530608, 8352941}. Furthermore, a few works also connect deep learning models with {\it attention mechanisms}~\cite{LI2020102258} to enhance model performance. In general, attention implies a selection process in which certain features of the input vectors are more important than others (i.e., a NN with an attention mechanism learns how to focus on what is relevant and disregard the noise). Attention mechanisms have produced state-of-the-art results in machine translation and other natural language processing (NLP) tasks, when combined with neural word embeddings~\cite{vaswani2017attention, bahdanau2016neural}, and constitute one of the latest and most promising directions in deep learning.

\begin{table}[htbp]
\begin{center}
\footnotesize
\caption{ML models applied for network traffic prediction.\\
(\scriptsize{*Gaussian Processes Regression (GPR), Bayesian Estimation (BE), Linear Regression (LR), Elman NN (ENN), LSTM, Encoder-Decoder (ED), Gated Recurrent Units (GRU), Graph NN (GNN), Graph Convolutional Generative Adversarial NN (GCN-GAN), Nonlinear Autoregressive Neural Network (NANN), Deep Belief Network (DBN), Image Recognition Neural Network (IRNN), online Bayesian Moment Matching (oBMM)}).}
\label{list_models}
    \begin{tabular}{ | l | p{5cm} |}
    \hline
    {\bf ML model} & {\bf Ref.} \\ \hline
    NN/DNN & \cite{8047676, 7830260, Zhao:18, 8386186, 8436062, 8430520,8126198, 7785324, 2016220, 6047849, JIANG201175, 201902004, 8553653, 0191939,Huo2019,13006}  \\ \hline
     LSTM & \cite{8501524, 8386209,Balanici:21,9012669, 201902004, 8126198, 8553653,8406199, 8352941, azzouni2017long, LI2020102258}  \\ \hline
     ED-LSTM & \cite{9782838} \\ \hline
    ResNet-LSTM -Attention & \cite{LI2020102258}\\ \hline
    GRU & \cite{8473978,Balanici:21,8126198,LI2020102258, 9748600, MARYAM202313} \\ \hline  
     RNN & \cite{8126198, 8530608, 9163001}\\ \hline
     CNN & \cite{9163001} \\ \hline
      CNN-LSTM & \cite{LI2020102258} \\ \hline
     3DCNN & \cite{LI2020102258} \\ \hline
     DCRNN & \cite{8845132}\\ \hline
    IRNN & \cite{8126198} \\ \hline
    GCN-GAN & \cite{9203477}\\ \hline
    DBN & \cite{NIE201616} \\ \hline
    ENN &  \cite{XIONG201999} \\ \hline
    NANN & \cite{8696381} \\ \hline
    Linear Regression & \cite{doi:10.1002/dac.4516,7069393, 5693766, 8530608, 8352941, 9782850}\\ \hline
    Boosted Decisions Trees & \cite{8501527} \\ \hline
    Random Forests & \cite{8501527,8757058, 13006, 9782850} \\ \hline
    LightGBM & \cite{8757058}\\ \hline
    SVM & \cite{20030603,13006}\\ \hline
    GPR & \cite{8501527,7785324, 9204965, 8254808}\\ \hline
    Bayesian Estimation & \cite{2016220}\\ \hline
    oBMM & \cite{2016220}\\ \hline
    \end{tabular}
\end{center}
\end{table}

\subsubsection{FeedForward NNs}
As shown in Table~\ref{list_models}, feedforward NNs (denoted in this survey simply as NNs) and deep NNs (DNNs) have been applied in several works, with the DNNs shown to outperform NNs (i.e., one hidden layer models), most of the times.
It is worth mentioning, that classical feedforward NNs/DNNs are often used as benchmarks to more advanced deep learning methods, designed specifically to operate over sequential or spatial data (i.e., NNs/DNNs are more appropriate for mapping features to a more separable space). 

\subsubsection{Recurrent NNs}
Even though feedforward NNs have shown to outperform statistical models (e.g., ARIMA, FARIMA, etc.), recurrent NNs and their variants (e.g., LSTM, GRU) are the type of NNs that have gained considerable attention in the research community for network traffic prediction (see Table~\ref{list_models}). In general, a recurrent NN (RNN) is a type of artificial NN that contains a circular structure that feeds back the output of a unit as input to itself at the next iteration (i.e., the output is copied and reinjected to the recurrent network), in order to model temporal relationships (i.e., it keeps information about past inputs). Due to their internal memory, RNNs are in general considered suitable for dealing with sequential data (time series, natural language). Their advantage compared to feedforward NNs has been quantified, for example, in \cite{8126198, 8553653}. However, in ``original'' RNNs~\cite{PhysRevLett.59.2229}, issues have been reported in modeling long-term temporal relationships because the error vanishes due to the recurrent backpropagation process (i.e., due to the gradient vanishing problem it becomes difficult for a model to store long time-steps in its memory)~\cite{10.1142/S0218488598000094}.

In order to address this issue, authors in~\cite{10.1162/neco.1997.9.8.1735} proposed long short-term memory (LSTM) networks. In general, LSTMs were shown to learn much faster than RNNs and have been proven effective for modeling long-range temporal dependencies of input data. Subsequently, gated recurrent units (GRUs)~\cite{https://doi.org/10.48550/arxiv.1412.3555} have been developed in order to reduce the computational cost of LSTMs, with GRUs being one of the most popular LSTM variants. Specifically, GRUs have fewer parameters than LSTMs, and performances of these two architectures vary depending on the problem. Feedforward NNs and recurrent neural network variants (i.e., RNNs, LSTMs, and GRUs) are applied and compared in \cite{8126198}. 

\subsubsection{Encoder-Decoders with Recurrent NNs}
Amongst the most competitive neural sequence-to-sequence models are those having an encoder-decoder structure~\cite{cho2014learning, sutskever2014sequence}. These models are designed for sequence-to-sequence problems, rendering them capable of mapping an input sequence of past traffic observations into an output sequence of future traffic estimates (i.e., multi-step ahead traffic prediction), especially when the encoder-decoder components are designed with recurrent cells (e.g., GRUs, LSTMs). Although, encoder-decoder structures have been applied and tested for various other applications (e.g., machine translation, NLP, trajectory prediction), their capabilities were not greatly explored yet for network traffic prediction purposes in optical networks. As an example, and as shown in~\cite{9782838}, multi-step ahead prediction, may be an additional asset during the optimization phase of predictive provisioning (Fig.~\ref{taxonomy}) as the knowledge of how traffic is expected to vary over sequential future planning intervals may be used for minimizing undesired service disruptions.    

\subsubsection{Convolutional NNs}
While RNNs and their variants were designed to exploit sequential data, rendering them suitable for modeling dependencies from long time series (i.e., to capture information in the time domain), convolutional NNs (CNNs) and deep CNNs (DCNNs)~\cite{10.1145/3065386} were designed to exploit the spatial dependencies between the data (i.e., to capture information in the space domain); an important consideration for the network traffic prediction problem, especially when there is an interest in capturing the spatial dependencies between the traffic patterns/volumes of various source-destination pairs in the network (i.e., predicting TD matrices) or when there is an interest in predicting spatial dependencies of link traffic loads. 

Recently, some novel techniques, e.g., Deep Convolutional Recurrent Neural Networks (DCRNN) or DCNN combined with LSTM (DCNN-LSTM), have been designed to capture information in both the time and spatial domains of the input, taking advantage of the complementarity of RNNs and CNNs by integrating them into a unified architecture~\cite{7178838}; that is, such unified architectures are suitable for capturing the spatio-temporal dependencies in traffic patterns.

\subsubsection{Graph NNs}
Traditionally, ML algorithms do not have the ability to exploit the graph topology of a network. Recently, novel graph learning techniques have been designed for learning models of graph-structured data. Even though only a few works exploit the capabilities of such models for network traffic prediction (e.g., in~\cite{9203477}), such models are promising approaches for capturing and modeling the spatio-temporal dependencies of network traffic (i.e., predicting link loads or TD matrices), especially when combined with attention mechanisms~\cite{vel2018graph} and/or convolutions~\cite{10.1007/978-3-319-93417-4_38}.

In general, Graph NNs (GNNs) is a recent class of NNs designed to operate over graph-structured data, representing entities as vertices and relationships between them as edges, and can include data associated with either as features. Graph-based ML algorithms have been proposed to discover patterns (i.e., relations between graph elements) and answer questions about graph-structured data. They were introduced in~\cite{4700287} and numerous variants, with diverse learning tasks, have been developed since, including feature inference of vertices or edges and vertex- and edge-based prediction. A recent survey discussing the applications and challenges of graph learning (e.g., their limited capabilities on generalizing over non-stationary dynamic network states) can be found in~\cite{s40649-019-0069-y}.  

\subsubsection{Bayesian Learning}
Bayesian ML approaches are categorized under probabilistic ML approaches. Specifically, in Bayesian learning model parameters are treated as random variables (r.v.s) and parameter estimation is the computation of posterior distributions for these r.v.s based on the observed data (i.e., statistical inference)~\cite{theodoridis_book}. 

There are several advantages in Bayesian learning, with the most noteworthy being the fact that the trained models produce an output with a clear probabilistic interpretation, providing a measure of uncertainty for the obtained predictions. This is in contrast to other ML approaches (e.g., pure NN-based models, SVMs, etc.), which merely provide point predictions. 

The class of Gaussian Processes (GPs) is one of the most widely-used families of stochastic processes for modeling dependent data observed over time~\cite{3569}. Hence, they are useful for the case of sequential data, such as time series, and are thus adopted in a few works in the literature for network traffic purposes (see Table~\ref{list_models}). GPs (often denoted as GPRs when applied for regression) represent one of the most predominant Bayesian ML approaches and are based on a particularly effective method for placing a prior distribution over the space of regression functions. They have a small number of tunable parameters, can be trained on relatively small training sets, and allow us to capture non-linear and skewed artifacts, hence demonstrating considerable robustness to outliers and the ability to handle sparse data without becoming susceptible to overfitting~\cite{5966349}.
However, one of the limitations of Bayesian learning in general, is that algorithms (i.e., training) can become computationally intensive as the data size and/or dimensionality of the data grows.
 
\subsection{Further ML Models to Exploit}
Additional ML models, not yet exploited in the literature are discussed in this section. These ML models constitute promising approaches for solving the traffic prediction problem (i.e., were designed for sequence-to-sequence problems), with some of them considering in addition to the accuracy performance criteria, the improvement in training time as well as uncertainty estimation. 
    
\subsubsection{Transformers}
One of the latest breakthroughs based solely on attention mechanisms is the Transformers~\cite{vaswani2017attention}, a deep learning model that, like RNNs, is designed to handle sequential data. However, unlike RNNs, Transformers do not require processing of sequential data in order, allowing parallelization and requiring significantly less time to train compared to RNN and its variants. Currently, Transformers have become the model that is most commonly used to solve NLP  problems, compared to other older recurrent NN models such as LSTMs. Hence, Transformers, constitute a promising approach, especially in the presence of large datasets that require high training times. 

\subsubsection{Bayesian Deep Learning}
Bayesian Deep Learning (BDL) refers to the intersection of deep learning and Bayesian learning approaches, and has recently received great interest from the ML community. BDL approaches aim to combine the advantages of both Bayesian and deep learning, while coping with their limitations. Specifically, Bayesian learning models do not scale well for high dimensionality data but offer principled uncertainty estimates. On the other hand, deep learning models have been revolutionary for ML, especially due to their scaling capabilities over datasets of high dimensionality, but they lack uncertainty representation. Hence, the rise of BDL is mainly due to the capabilities of BDL models to offer principled uncertainty estimates from deep learning architectures (i.e., understanding what a model does not know) in a computationally efficient way. BDL models typically derive estimations of uncertainty by either placing probability distributions over model parameters (i.e., over the parameters of a NN), or by learning a direct mapping to probabilistic outputs. The reader is referred to a recent survey on BDL that appeared in~\cite{10.1145/3409383}. 

In general, BDL models constitute promising approaches for the traffic prediction problem, mainly due to their probabilistic estimates providing measures of uncertainty, potentially allowing dealing with undesired over- and under-provisioning effects related with model inaccuracies. In fact, the benefits of considering traffic prediction uncertainty during the optimization phase of predictive provisioning have been recently verified in~\cite{9748600}, indicating that by doing so significant spectrum savings are achieved (i.e., over-provisioning is mitigated) with a negligibly observed under-provisioning. However, this work was based on deep quantile regression which is not flexible as it concerns the uncertainty estimation as opposed to Bayesian frameworks~\cite{MaryamMed22}.   

Finally, it is worth mentioning that while several BDL models may be appropriate for network traffic prediction, Bayesian RNNs and their variants (e.g., Bayesian LSTMs)~\cite{fortunato2019bayesian} and Deep Gaussian Processes (DGPs)~\cite{damianou2013deep, salimbeni2017doubly, lee2018deep} seem to be the most promising models, mainly due to the fact that both RNNs/LSTMs and GPs have shown to perform well in network traffic prediction. Of course, another straightforward way of making DNNs ``Bayesian", is by using Monte Carlo dropout inference~\cite{gal2016dropout}. Monte Carlo dropout inference is based on approximating posterior
distributions over the weights of a DNN given the observed data. Unlike, however, BDLs and the general Bayesian modeling framework, Monte Carlo dropout provides
a simple and computational efficient way of approximating
posterior and predictive distributions, with the predictive distribution used for analyzing and measuring model uncertainty~\cite{MaryamMed22, MARYAM202313}.

\section{Optimization in Predictive Provisioning}~\label{RA}
Optimization frameworks for predictive provisioning can be categorized into two cases, depending on whether a partially dynamic or a dynamic infrastructure is considered (Fig.~\ref{taxonomy}). The former refers to proactive networks, where the optimization is performed off-line, while the latter refers to adaptive networks, where optimization is performed on-line. On this basis, predictions for each case concern different time scales and attributes.  

\begin{figure}[h!]
\begin{center}
\includegraphics[scale=0.38]{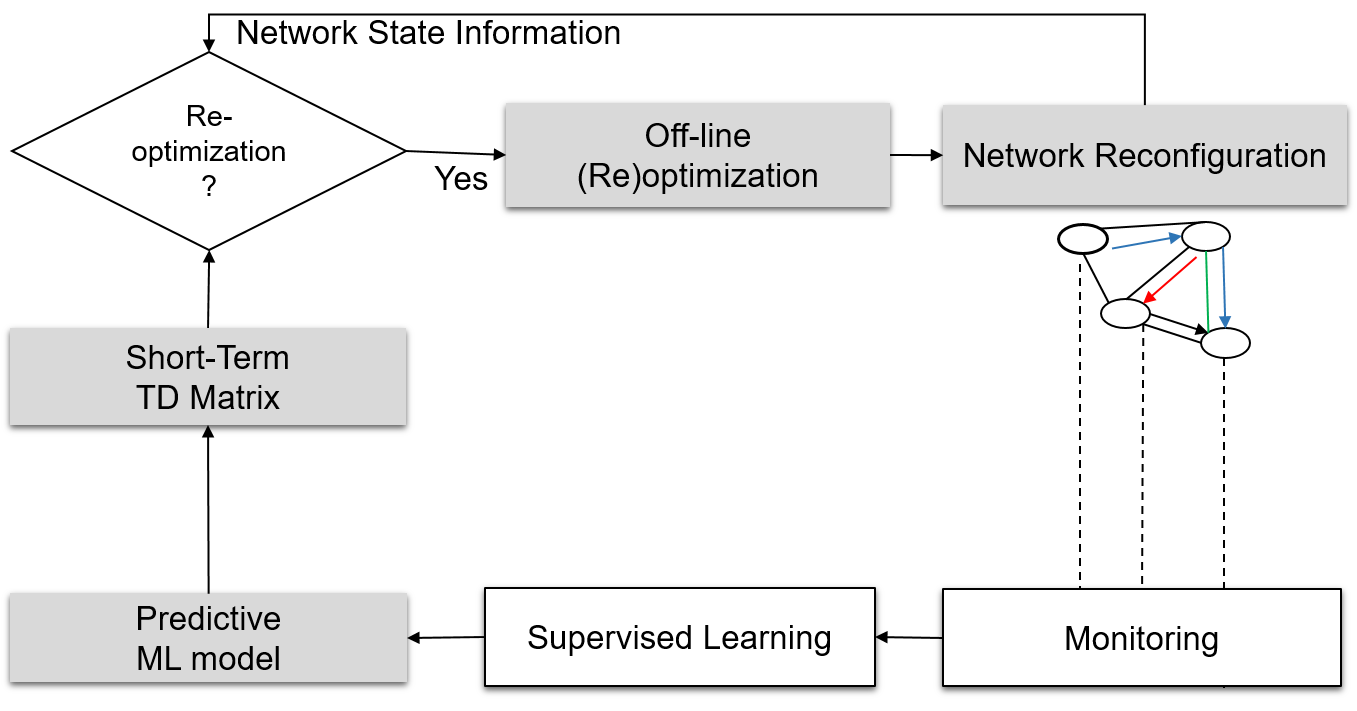}
\caption{Generic predictive provisioning in proactive networks.}
\label{proactive_fram1}
\end{center}
\end{figure}

\subsection{Proactive Networks}\label{proactive_pred}
In proactive networks (Fig.~\ref{proactive_fram1}), predictive provisioning concerns multi-period network re-optimization with the predictions usually concerning the TD matrix and with the time scales usually on the order of hours. 

Predictive provisioning mainly consists of four stages; that is, the TD matrix prediction stage, the reconfiguration decision stage, the resource allocation algorithm, and the reconfiguration stage that takes place dynamically at predefined time points. The purpose of the reconfiguration decision stage is to trigger a network re-optimization and subsequently a network reconfiguration only when it is really required, in order to avoid unnecessary service disruptions and computational overhead. Roughly, a network re-optimization is triggered when traffic demand predictions suggest that the current network configuration cannot accommodate the future traffic demand without violating predefined QoS targets (e.g., unserved traffic is expected to increase above the predetermined limits given the current network configuration and traffic predictions)~\cite{8430520}. Once a network reconfiguration decision is taken, the network optimization takes place during which a resource allocation algorithm is executed. 

In general, resource allocation algorithms implemented for static and reactive networks are extended for proactive networks, to account for various optimization objectives. Specifically, many works focus on exploiting the traffic demand predictions to minimize connection blocking and service disruptions between consecutive reconfigurations~\cite{7830260,s11227-020-03493-7,6381741,6831425, 9782838}, energy consumed by the active network components~\cite{7444562,8047676,7830260,ZHANG2011171, 5540249,6515884,6384633,Alvizu2017EnergyED}, and undesired over- and/or under-provisioning that is present in traditional static/reactive networks or arises due to ML-model uncertainties~\cite{8501527,8473978,8430520,9748600,8686111,9322381,9347908, MARYAM202313}.

Depending on the planning intervals and problem computational complexity, exact (i.e., ILP-based) or heuristics algorithms are utilized. For planning intervals allowing for optimally solving the problem, exact approaches are usually opted for; however, as planning intervals reduce, heuristic approaches become more appropriate. 

Predictive service provisioning in proactive networks, is commonly compared to static provisioning in static and reactive networks. It has been shown that proactive networks outperform both static and reactive networks in terms of energy consumption, over-provisioning, spectrum utilization, and connection blocking. A slight increase in unserved traffic (i.e., under-provisioning) was also observed, that was handled in a best-effort approach, by on-line provisioning the unpredicted amounts of traffic~\cite{8686111}. However, as best-effort approaches cannot provide QoS guarantees, margins can be considered along with the predicted traffic values to account for the unpredicted traffic~\cite{8473978,9748600}. Margins can be either based on the statistical error of the model~\cite{8473978} or on the estimation of the prediction uncertainty of the ML model~\cite{9748600}, with the latter approach best mitigating both over- and under-provisioning. A set of indicative works, summarizing the discussion above is shown in Table~\ref{proactive_surv_core}, with these works surveyed in Section~\ref{surv_proactive}.     

\subsection{Adaptive Networks}\label{adaptive_pred}
\begin{figure}[h!]
\begin{center}
\includegraphics[scale=0.38]{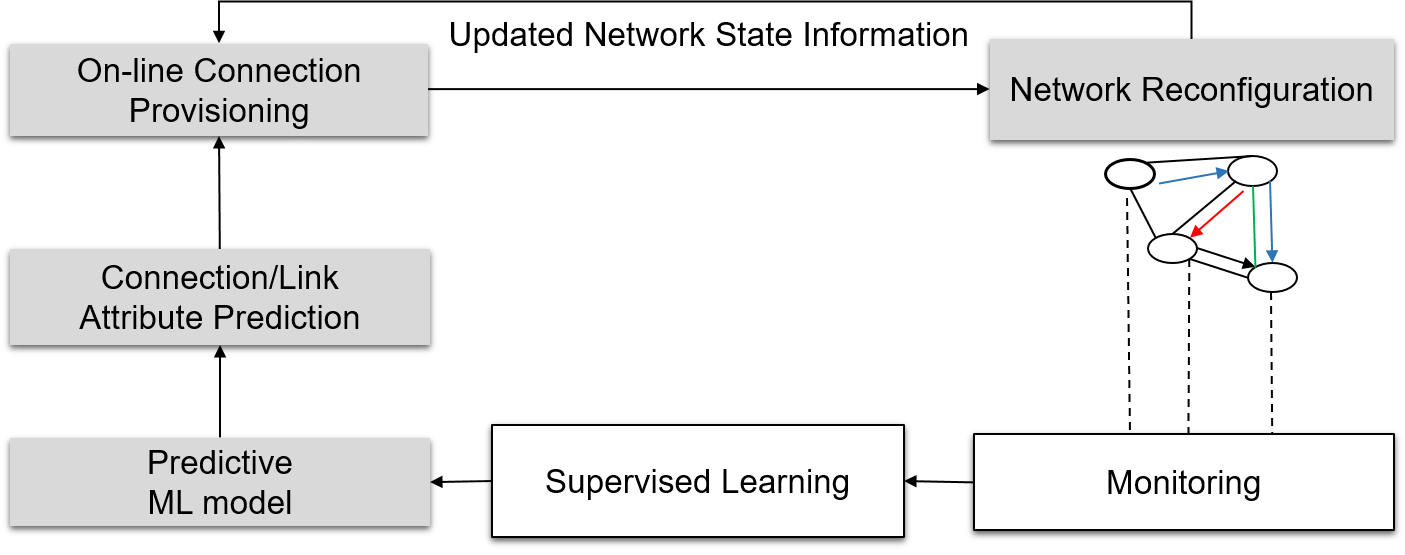}
\caption{Generic predictive provisioning in adaptive networks.}
\label{adaptive_fram2}
\end{center}
\end{figure}

In adaptive networks (Fig.~\ref{adaptive_fram2}), predictive provisioning concerns on-line optimization, for which heuristic algorithms are applied. Such heuristics are similar to those developed for conventional dynamic networks (Fig.~\ref{taxonomy}), but extended to take advantage of the predictions. The predictions are performed at short time scales (e.g., minutes or seconds) and usually refer to connection attributes such as the arrival or departure times of the connections and network link attributes such as the spectrum utilization. This information is used for prioritizing available network resources at decision time, by setting appropriate weights on the links of the temporal network states. The predictive weights guide the resource allocation decisions towards increasing the likelihood of accepting future connections into the network. As networks dynamically evolve between consecutive arrivals (i.e., when departures occur resources are released), the weights are continuously updated and sent to the on-line service provisioning algorithm to accommodate arriving requests. The most common objective is to reduce blocking probability by appropriately weighing the available resources.

Predictive provisioning in adaptive networks~\cite{8501524,XIONG201999,Zhao:18,9203477,8436062,9042293} is commonly compared to conventional dynamic provisioning. It has been shown that predictive provisioning outperforms conventional dynamic provisioning in terms of connection blocking, spectrum utilization, and fragmentation. However, this holds when the predictions are of sufficient accuracy. Otherwise, inappropriate resource allocation decisions may arise, with conventional dynamic provisioning, possibly outperforming predictive schemes~\cite{8501524}. A set of indicative works, summarizing the discussion above is shown in Table~\ref{proactive_surv_core}, with these works surveyed in Section~\ref{surv_proactive}.
     
\section{Machine Learning for Prescriptive Provisioning}~\label{pres_ra} 
While prescriptive provisioning, based on RL, has already gained significant attention for other types of networks, such as wireless networks and Internet-of-Things~\cite{8714026, 9403369}, it has only recently attracted an (increasing) interest by the optical networks research community (Table~\ref{list_models_rl}). This is mainly due to the deep learning advances that enable RL to scale to previously intractable problems arising in complex systems of high dimensionality~\cite{nature14539}. Unlike predictive provisioning, where SL is used to empower domain-specific resource allocation algorithms (i.e., traffic-driven optimization), prescriptive provisioning is empowered by RL to learn resource allocation policies (i.e., policy-driven optimization) by continuously exploring the network environment to improve over time through trial and error. Hence, unlike predictive provisioning, that requires designing rule-based heuristics or exact mathematical resource allocation algorithms, subsequently driven by the predictions, prescriptive approaches are model-free as they can learn optimal (or near-optimal) resource allocation policies (i.e., heuristics) from experience. 
In a network environment where traffic behavior and/or network topology and technology may change, prescriptive resource allocation policies can self-adapt, whereas predictive resource allocation algorithms may need to be redesigned.        

Even though RL has seen impressive advances over the last few years, especially in solving complex, challenging for AI, classical games (such as Go and StarCraft~\cite{nature24270,s41586-019-1724-z}), large amounts of training (i.e., in training experience and time) are required to successfully learn even simple games, hindering the practical applicability of RL, on real, non-stationary, and complex environments, such as telecommunication networks. While ML research focuses, amongst others, on advancing RL techniques to scale well to large applications in computational and time efficient ways, research on communication networks focuses on how to advance resource allocation algorithms by incorporating RL advances into well-known resource allocation schemes, in tractable ways. 

Hence, currently, proposed prescriptive resource allocation frameworks are rarely based purely on RL; rather, RL is applied to tackle sub-problems (e.g., only routing, only bandwidth allocation) supplemented by rule-based heuristics or exact algorithms~\cite{9083336,8644190}. To tackle the entire problem, the network environment is usually simplified to reduce problem dimensionality (e.g., considering a network with precomputed lightpaths~\cite{8847548} or precomputed Virtualized Network Functions (VNFs)~\cite{9609608}). By doing so, prescriptive schemes have demonstrated promising results, outperforming well-known rule-based benchmarks~\cite{8738827}, and in some cases, even predictive schemes~\cite{8751369}. This section describes the general field of RL and what approaches have been applied for prescriptive traffic engineering (TE) and/or service provisioning. 

\subsection{RL for TE and Service Provisioning}
In the general RL setup, an autonomous agent, that is controlled by a ML algorithm, interacts with its environment by taking an action at each time point of interest. When an action is taken, it causes the environment and the agent to transition to a new state, with the environment providing as a feedback a scalar reward to the agent. The aim of the agent is to learn a policy that maximizes the expected return. Hence, the best sequence of actions is determined by the rewards provided by the environment, which the agent uses to update its policy.  

In formal terms, RL can be described as a Markov Decision Process (MDP) over the tuple $\{\mathcal{S}, \mathcal{A}, \mathcal{T}, \mathcal{R}, b_0, \gamma\}$ where: 
\begin{itemize}
\item $\mathcal{S}$: set of states.
\item $\mathcal{A}$: set of actions.
\item $\mathcal{T}(s_{t+1}|s_t,a_t)$: distribution over next state $s_{t+1}$ that the agent may transition to from state $s_t$ after taking action $a_t$ (i.e., the transition dynamics).
\item $\mathcal{R}(s_t,a_t,s_{t+1})$: reward function specifying the immediate reward $r_{t+1}$ when action $a_t$ is taken at state $s_t$.
\item $b_0$: distribution of starting states.
\item $\gamma \in [0,1)$: discount factor that weighs the importance of current and future rewards [i.e., for $\gamma = 0$ the agent only cares about which action will yield the largest expected immediate reward and for $\gamma$ approaching $1$ the agent cares about maximizing the expected sum of future rewards (i.e., the cumulative discounted reward)]. Typically $\gamma$ is set above $0.9$ in applications where actions may have long-term consequences.
\end{itemize}

In general, at each time step $t$, the environment is at some state $s_t \in \mathcal{S}$. The agent takes an action $a_t \in \mathcal{A}$, and the environment transitions to state $s_{t+1}$ with probability $\mathcal{T}(s_{t+1}|s_t,a_t)$. At the same time, the agent receives a reward equal to $\mathcal{R}(s_t,a_t,s_{t+1})$. Then the process repeats. The aim of the agent is to learn an optimal policy $\pi^*(a|s)$ that maximizes the expected future discounted reward $\mathbb{E}[R]$, where $R= \sum_{t=0}^T \gamma^t r_{t+1}$, and $T$ is the number of time steps in an episode. Hence, an episode results in a sequence of states, actions, and rewards (i.e., a trajectory or rollout of the policy), and a policy $\pi$ is a mapping from states to a probability distribution over actions.

\subsection{RL Algorithms}
\begin{table*}[htbp]
\begin{center}
\footnotesize
\caption{RL Frameworks applied for traffic engineering and/or service provisioning.\\
(\scriptsize{*Deep Reinforcement Learning (DRL), Deep Q Learning (DQN), Asynchronous Advantage Actor-Critic (A3C), Advantage Actor-Critic (A2C), Deep Deterministic Policy Gradients (DDPG), Kronecker-Factored Trust Region (ACKTR), Trust Region Policy Optimization (TRPO), Proximal Policy Optimization (PPO), Path Consistency Learning (PCL), Actor Critic with Experience Replay (ACER), Soft Actor Critic (SAC), Twin Delayed Deep Deterministic Policy Gradient (TD3)}, actor-critic-based resource allocation (ACRA))}
\label{list_models_rl}
    \begin{tabular}{ | l | p{10cm} | p{3cm} |}
    \hline
    {\bf RL Method} & {\bf RL algorithm} & {\bf Ref.} \\ \hline
    Policy-based RL &  Learning Automata optimized with Policy Iteration & \cite{9256655}  \\ \hline
    Policy-based RL &  Monte Carlo Tree Search  & \cite{Aibin2020MonteCT}  \\ \hline
    Value-based RL &  Real time Dynamic Programming & \cite{8396130,8751369, 8620207}  \\ \hline
    Value-base RL &  Q-learning, Sample Average, Temporal Difference (TD) & \cite{9308056}  \\ \hline
    Value-based DRL & GNN optimized with DQN learning & \cite{almasan2020deep}  \\ \hline 
    Value-based DRL & DNN optimized with DQN learning & \cite{9279336} \\ \hline     
    Actor-Critic DRL &  DNNs optimized with A3C & \cite{8738827}  \\ \hline
    Actor-Critic DRL &   CNNs optimized with A3C  & \cite{3229554}  \\ \hline 
    Actor-Critic DRL  & CNNs optimized with A2C & \cite{Zhao:18, Zhao:21}  \\ \hline
    Actor-Critic DRL  & DNNs optimized with  A2C & \cite{9083336, KHOSHKHOLGHI2022109451}  \\ \hline
    Actor-Critic DRL & DNNs optimized with prioritized experience replay, DDPG & \cite{8485853}  \\ \hline
    Actor-Critic DRL &  GRUs optimized with ACKTR & \cite{SUN2021107891}  \\ \hline
    Actor-Critic DRL &  DNNs optimized with TRPO, PPO, DDPG, PCL, A3C, ACER & \cite{8847548}  \\ \hline
    Actor-Critic DRL &  DNNs optimized with Gradient Decent-DQN & \cite{9375634}  \\ \hline
    Actor-Critic DRL &   CNNs optimized according to ACRA & \cite{8644190}  \\ \hline
    Actor-Critic DRL &   CNNs optimized with TD3 & \cite{9748476}  \\ \hline
    Actor-Critic DRL &   ANNs and CNNs optimized with SAC & \cite{9507559}  \\ \hline 
    Actor-Critic DRL &   GCNs and RNNs optimized with A2C & \cite{9779080}  \\ \hline   
    Actor-Mimic with TL DRL &  DNNs & \cite{9373585}  \\ \hline
    \end{tabular}
\end{center}
\end{table*}
For approximating an optimal policy $\pi^*$, several RL algorithms have been developed, that can be categorized in three methods; that is, value-based, policy-based, and hybrid actor-critic methods. 

In value-based methods, a state-value function $V^{\pi}(s)$ estimates the total amount of reward an agent can expect to accumulate over the future (i.e., in the long-run), starting from a given state $s$ and following a policy $\pi$. Hence, value functions define a partial ordering over policies, from which an optimal policy is indirectly learned (e.g., Q-learning~\cite{BF00992698} and Deep Q Learning (DQN)~\cite{nature14236}).     

Policy-based methods (e.g., REINFORCE algorithms~\cite{Williams92}) do not need to maintain a value function; rather, they can directly learn an optimal policy. Typically, the policy is parameterized [i.e., $\pi(a|s,\theta)$] and its parameters
are updated to maximize the expected return using either gradient-based or gradient-free optimization~\cite{8186816}, with gradient-based methods being the method of choice for most DRL algorithms. 

Actor-critic methods, are amongst the most popular algorithms in the RL framework. In particular, DRL actor-critic methods, scaling to high-dimensional problems, are growing in popularity in both industry and academia. Actor-critic methods, combine the advantages of both value- and policy-based methods, while attempting to address their drawbacks. Specifically, they combine a value function (critic) with an explicit representation of the policy (actor). In this setting, the actor learns by using feedback from the critic to update its policy parameters so as to improve performance. Such methods, contrary to critic- and actor-only methods, may converge faster. For more information on value-based, policy-based, and actor-critic methods, RL advances, as well as multi-agent systems, the reader is referred to~\cite{Sutton18, 8103164}.   

According to Table~\ref{list_models_rl}, DRL actor-critic methods have attracted the most interest from the optical network research community for tackling various TE and service provisioning problems. In general, in a DRL framework critic and actor functions are parameterized with a NN (e.g., DNN, CNN, GRU, GNN, etc.), which must be, however, carefully chosen to capture the various spatio-temporal dependencies (e.g., Graph Convolutional Network (GCN)) within the network environment, especially when RL tackles the routing and/or spectrum allocation problems. Note, however, that when simpler problems are tackled (e.g., bandwidth estimation) RL frameworks with less computational and memory requirements may better apply (e.g. GRU).          

For training RL and/or DRL policies, several optimization algorithms have been exploited as indicated in Table~\ref{list_models_rl}. However, as demonstrated in~\cite{Henderson_2018}, a fair comparison amongst these algorithms is not trivial, as it depends on various factors (underlying problem, action and state representation, reward scaling, NN architecture, hyperparameter tuning etc.).

\subsection{State, Action, and Reward Representation}
Successfully addressing prescriptive TE and service provisioning to outperform baseline schemes does not merely depend on leveraging recent RL advances. Carefully designing reward, state, and actions spaces is even more important to allow the agent to capture the underlying problem and environment and to simplify the RL agent learning process in order to converge fast to a policy of desired performance. Even though DRL advances allow scaling to high dimensional state and action spaces, that were previously intractable, still DRL capabilities are not infinite. To address the RL's ``curse of dimensionality''~\cite{pmlr-v80-yang18d}, it is important to represent both states and actions as abstracted as possible (i.e., through auxiliary states and actions), especially in a large size network environment (i.e., in number of nodes, links, and available spectrum). This means that it may be more efficient to represent the state space at a higher level (e.g., at the route or even lightpath level) instead of at the physical layer level (e.g., at the node, link or spectrum level). Similarly, the action space can be also represented at a higher level (e.g., to represent lightpath, route, VNFs, or even baseline heuristic candidates) instead of physical layer decisions (e.g., link candidates to construct a route or spectrum block candidates to allocate). As discussed in Section~\ref{surv_prescr}, this trend is present in the related literature, shown to significantly accelerate policy convergence and efficiency. 

Reward representation has also a large impact on the learned policy, as this is the only signal the agent receives to understand how good or bad was an action taken, and accordingly update its policy. Reward representation may significantly affect policy direction and must be set appropriately to correctly evaluate the impact of each action with respect to the targeted network efficiency measure/s (i.e., the policy objective). In general, negative rewards are used to inform the agent that an action taken at a time point (i.e., at a state) was erroneous, and positive otherwise. Reward scaling and rescaling was also shown to have a large impact on the resulting policy, hence it must be properly considered~\cite{Henderson_2018}.            

\subsection{Performance Metrics}
Policy evaluation is commonly performed through the expected cumulative reward progress over the number of learning episodes. A good indicator of the learning progress is when the expected cumulative reward tends to increase over the learning episodes, until convergence. Of course, policy convergence does not necessarily mean that a good policy has been reached. To directly interpret the policy efficiency, commonly the policy is evaluated in terms of targeted performance measures (e.g., blocking probability) over the learning episodes, and compared against baseline service provisioning schemes. In particular, an indicative number of the latest traffic demand traces can be used for comparative purposes (e.g., after learning and by exploiting the learned policy). 
The number of learning episodes translates also to training time, which is another useful performance metric, especially in evaluating the practical limitations of the approach. 

It is worth mentioning, that traffic traces for training a service provisioning policy and for policy evaluation can be either found in openly available real datasets or can be synthetically generated as analytically described in Section~\ref{avail_datasets}. Specifically, in proactive networks the real datasets provided in Table~\ref{source_dataset} can be used as these provide traffic information per planning intervals (e.g., per minutes). However, in adaptive networks, traffic traces are usually synthetically generated from appropriate distributions as such information (e.g., arrival time, holding time of an arriving connection request) is not currently openly available.        

\section{Optimization in Prescriptive Provisioning}\label{prescriptive}
\begin{figure}[h!]
\begin{center}
\includegraphics[scale=0.38]{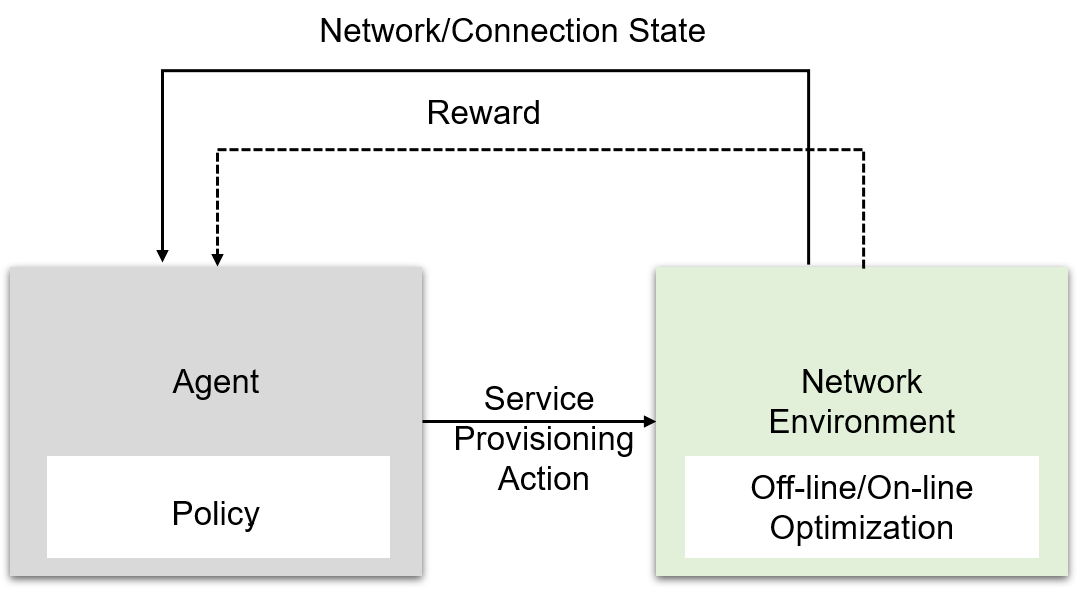}
\caption{Generic prescriptive provisioning for proactive and adaptive networks.}
\label{presc_fram}
\end{center}
\end{figure}

This section briefly describes how RL has been in general applied in the literature for prescriptive service provisioning in both proactive and adaptive networks. A generic provisioning scheme for both cases is illustrated in Fig.~\ref{presc_fram}. Both proactive and adaptive networks are illustrated under the same general scheme, as they mainly vary on how the optimization phase is implemented; off-line for proactive networks and on-line for adaptive networks. Depending on the case, the RL framework is differently formulated over the reward function, state, and action spaces.  

\subsection{Proactive Networks} \label{optimization_prescr_proact}
Acting in a prescriptive fashion in proactive networks means that the agent is able to provide actionable advise on how the network must be reoptimized off-line in order to meet predefined QoS targets over future planning intervals. The difference with the predictive case is that, instead of utilizing traffic demand predictions, it can provide actions dealing with network congestion and/or contention over the spectrum resources of the network links, while accounting for the service's (diverse) QoS requirements. As such, the agent policy can be learned for controlling over- and under-provisioning effects at desired levels. Of course, a policy may concern just finding actions that declare the traffic demand needs over future time points or intervals. In that case, the major difference with predictive schemes is that, unlike SL, labeled data are not needed. The agent is able to observe the traffic demand variations and learn a policy that eventually converges to the real traffic demand needs. Nevertheless, similar to the predictive case in proactive networks, the actions are used as inputs to the off-line optimization algorithm, and the network is reconfigured, if required, at the beginning of predefined planning intervals. Hence, similar off-line optimization algorithms can be applied in both cases.  

In the literature, only a few works exist for prescriptive provisioning in proactive networks~\cite{8620207,9256655}. In~\cite{9256655} the RL agent generated actions reflecting the traffic demand needs of future planning intervals and in~\cite{8620207}, a set of RL agents were trained in a cooperative fashion to deal with network congestion and the contention between connections by providing bandwidth reservation actions for each connection. An interesting outcome was that predictive and prescriptive schemes were compared, indicating the advantages of prescriptive approaches over predictive in terms of learning policies that better utilize network resources (i.e., mitigating over- and under-provisioning~\cite{8620207}). This is mainly due to the fact that agents are able to observe the impact of their actions, especially with regards to how traffic fluctuates over time, hence adjusting their policies accordingly. 

\subsection{Adaptive Networks} \label{optimization_prescr_adapt}
Acting in a prescriptive fashion in adaptive networks means that the agent is able to provide actionable advise on how a service request can be provisioned on-line to improve targeted network performance metrics (e.g., blocking probability, spectrum utilization, cost, latency). Hence, prescriptive provisioning in adaptive networks concerns learning heuristics/policies to on-line provision the arriving request, with the policies capturing both the traffic trends (e.g., arrival rates, bit-rates, holding times) and the long-term impact of agent actions to the system; that is, capturing how an action taken at a current time point affects actions for future arriving requests and consequently system performance.   

Prescriptive provisioning in adaptive networks has only recently received considerable attention from the optical networking community~\cite{Zhao:18,8738827,8847548,3229554,8485853,9083336,9375634,9373585,SUN2021107891,Zhao:21,9507559,9609608,9779080}. Even though diverse objectives are encountered (i.e., reward functions), most works focus on optimizing blocking probability and network spectrum utilization by on-line provisioning arriving connection requests through the learned RL policy. Furthermore, different optical multiplexing technologies (e.g., WDM, EON) have been considered, with different representations of space and action spaces. Recently, few works have targeted the VNF service chaining (SC) problem~\cite{9609608,KHOSHKHOLGHI2022109451}, that aims to finding an on-line VNF-SC policy that meets the services' SLA agreements with reduced cost, latency, and spectrum utilization.     
While RL can be utilized to jointly learn routing and spectrum allocation policies [e.g., routing modulation and spectrum allocation (RMSA) policy in~\cite{8738827}], most of the times RL is used to separately learn policies (e.g., on routing, spectrum allocation~\cite{8847548,9083336} or even VNF-SC~\cite{9609608,KHOSHKHOLGHI2022109451}), subsequently driving the rest of the decisions. This is done to reduce problem dimensionality, consequently allowing the RL to converge to a good policy requiring less experience (i.e., number of learning episodes). This is a critical consideration in non-stationary network environments where a policy may be rendered obsolete upon significant networks changes (i.e., leading to inappropriate actions before adapting to the new conditions). In general, however, existing works clearly demonstrate the potential of prescriptive provisioning in adaptive networks to outperform conventional dynamic provisioning (Fig.~\ref{taxonomy}). All relevant works are surveyed in Section~\ref{surv_prescr}.

\section{Predictive Provisioning: Survey on ML}~\label{traffic_prediction}
The survey of works dealing with traffic prediction is partitioned in three categories, depending on the targeted traffic prediction feature (i.e., connection bit-rate, connection arrival/holding time, network link load). Table~\ref{prediction_surv_core} provides information for a set of these works, including the network traffic feature considered, the ML methods applied, ML model inputs and outputs, time scale of predictions, performance evaluation metrics, and their key findings. This set of papers was mainly chosen to briefly summarize the state of the art. Note that we have chosen to separately survey works dealing with the traffic prediction sub-problem and/or the traffic driven optimization sub-problem (surveyed in Section~\ref{surv_proactive}) of the overall predictive service provisioning framework, since these two sub-problems are complementary to each other and independent. Further, many works in the literature address only either one of the sub-problems (e.g., only the traffic-driven optimization sub-problem, assuming, for example, that a perfect traffic prediction model is available). Hence, by doing, so we aim at revealing the main outcomes and challenges that arise separately from each sub-problem, without, however, excluding their interdependencies (e.g., how ML model inaccuracies affect the performance of traffic-driven optimization). 
 
\begin{table*}[htbp]
\footnotesize
\caption{Traffic prediction in optical networks.}
\begin{tabular} {p{0.5cm}|p{1.7cm}|p{1cm}|p{2.3cm}|p{2.5cm}|p{0.8cm}|p{0.9cm}|p{0.8cm}|p{3.4cm}}
{\bf Ref. }& {\bf Network \newline Traffic} & { \bf ML \newline Method} & { \bf Input $\bf {x}$} & {\bf Output  $\bf{y}$} & {\bf Time \newline Scale} & { \bf Dataset}  & {\bf Acc. \newline Metric} & { \bf Key \newline Findings} \\ \hline

\cite{8473978} & IP connection bit-rate & GRU & bit-rate TD matrices of $6$ previous slots & bit-rate TD matrix of next slot & Hourly & Real &  MAE & MAE $\leq 7.4$  \\ \hline

\cite{8501527} & IP connection bit-rate & GP, Boosted Decision Trees, Random Forests, Penalized Linear & bit-rate of $w$ previous slots  & bit-rate of next slot & Hours & Real &  MAE & GP outperforms all in Accuracy; \newline Accuracy is affected by $w$ and time scale selections \\ \hline

\cite{7830260} & IP connection bit-rate & DNN  & bit-rate of $w$ previous slots & bit-rate of next slot & Hourly & Synthetic &  AIC & Selection of $w$ affects Accuracy  \\ \hline

\cite{8047676} & IP connection/\newline Base Station bit-rate & NN  & avg. bit-rate of previous day; \newline bit-rate of the same hour of the previous day; \newline  bit-rate of the same hour and same day of previous week; \newline hour of the day; \newline day of the week; \newline flag for holidays, weekends; & bit-rate of next slot & Hourly & Real &  MAE, MAPE, RMSE & Higher Acc. for aggregated IP traffic compared to the Acc. at the base station level    \\ \hline

\cite{9204965} & IP connection bit-rate   \newline Virtual link load bit-rate & GPR Ensemble, LSTM, SVM & bit-rate of previous slots &  bit-rate of next slot & Minutes & Real &  NMSE  & GPR Ensemble model outperforms LSTM and SVM (on average by $12\%$ in Accuracy) \\ \hline 

\cite{9203477} & Physical link load (Erlangs) & GCN-GAN, LSTM  & link load matrix of previous slot & link load matrix of next slot & Hourly & Real &  MSE & MSE $\leq 0.01$; \newline GCN-GAN outperforms LSTM  \\ \hline

\cite{8845132} & Virtual link load bit-rate & DCRNN, CNN, LSTM, DNN & link load matrix of $10$ previous slots &  link load matrix of next slot & Hourly & Real &  MAPE, MAE, RMSE & DCRNN outperforms all in Acc.; \newline DCRNN better captures congestion information \\ \hline 

\cite{Balanici:21} & DC connection bit-rate & LSTM, GRU  & bit-rate of previous slots &  bit-rate of next $s$ slots & Seconds & Real &  MSE & LSTM outperforms GRU; \newline For $s>60$ Acc. dramatically drops \\ \hline

\cite{8501524} & DC connection holding time   & LSTM  & holding times of previous connections & holding time of next arrival & Seconds & Synthetic
 &  RMSE & $0 < RMSE < 20$; \newline Unexpected traffic decreases Accuracy  \\ \hline
 
 \cite{XIONG201999} & IP connection bit-rate, arrival time, holding time & ENN  & bit-rate, arrival time, holding time of previous slots & bit-rate, arrival time, holding time of next slots & Minutes & Real &  MSPE & MSPE $\leq 5 \%$ per prediction \\ \hline
 
\cite{9748600} & IP connection bit-rate & GRU  & bit-rate of previous 6 slots & bit-rate of next slot & 30 minutes & Real &  Quantile Error, MSE & Quantile Error $\leq 0.09 \%$; \newline MSE $\leq 0.11 \%$; \newline $9\%$ of the quantile predictions fall below their true value; \newline $45\%$ of the MSE predictions fall below their true value; \newline Quantiles mitigate potential under-provisioning    \\ \hline

\end{tabular}
\label{prediction_surv_core}
\end{table*}

\subsection{Bit-Rate Prediction}
Several of the works address the bit-rate prediction problem at the general IP connection level (i.e., bit-rate demand between IP nodes)~\cite{8501527,9204965,8047676,7830260,8473978,8386186,9748600,MARYAM202313}, or at the IP connection level of data center (DC) networks~\cite{8430520, 8386186, 8696381,Balanici:21,9782838}. Predictions may either refer to the entire TD matrix (i.e., by training a single ML model)~\cite{8473978,9782838}, or to the traffic demands of individual IP connections (i.e., by training a ML model for each connection) to subsequently form the TD matrix~\cite{8501527,9204965,8047676,7830260,8386186,8696381,Balanici:21,9748600}. The TD matrices can then be used for predictive provisioning, most commonly in proactive networks.

Specifically, in~\cite{8047676}, the TD matrix is predicted according to several ANNs (one for each base station in a mobile metro-core network). Two cases were examined, with the first examining the traffic prediction problem at the IP connection level (i.e., the base station traffic demand is aggregated to the metro network), and with the second examining the traffic prediction problem at the base station level. Real datasets were utilized for training the ANN models~\cite{ENAiKOON16} to perform on-hour predictions, demonstrating that aggregated base station traffic can be more accurately predicted as it appears to be more regular and stationary compared to the traffic at the base station level.

In general, ANNs and DNNs were applied in several works in the literature~\cite{7830260,8430520,8386186,8436062}, in which the TD matrix was formed by training several ML models (one for each source-destination pair). In~\cite{8430520, 8386186} it was shown that DNNs outperform ANNs in the prediction accuracy of DC traffic. A DNN was applied in~\cite{8436062} to predict multi-domain traffic in EONs (i.e., across metro networks, DCs, and other facilities) under various traffic loads, showing that prediction accuracy improves as traffic load increases. This trend, was also observed in~\cite{doi:10.1002/dac.4516}. In~\cite{7830260}, the performance of DNNs was tested under different hyperparameters, including the number of hidden layers and the number of previous bit-rate observations (i.e., the $w$ parameter). It was shown that a certain number of hidden layers and of previous observations resulted in a better accuracy, clearly demonstrating the importance of both hyperparameter optimization (i.e., tuning) and feature selection towards deriving models of sufficient accuracy.

The traffic prediction problem in electro-optical DCs was also examined in~\cite{8696381,Balanici:21}, enabling the transfer of heavy traffic streams in the optical domain and the short-lived, bursty data flows in the electrical domain. In particular, in~\cite{8696381} a NANN was applied with the objective of accurately predicting the heavy traffic streams of electro-optical DCs. The NANN was trained according to synthetic datasets generated by a chaotic model, preserving the long-range dependencies of DC traffic and it achieved a high accuracy, especially for deeper NN configurations, indicating, similarly to~\cite{7830260}, the importance of hyperparameter tuning. 

That work was extended in~\cite{Balanici:21}, in which real datasets were utilized to train LSTMs and GRUs. Both models were trained and compared according to different number of hidden LSTM/GRU layers. LSTM outperformed GRU, while a single hidden layer was shown to lead to models of higher accuracies. Additionally, LSTMs/GRUs were trained as single- and multi-step ahead regression models, with single-output regression shown to slightly outperform multi-step ahead regression in model accuracy. Furthermore, for multi-step ahead regression, it was shown that predictions up to certain number of future time slots (i.e., in these experiments up to $s=60$ seconds) could be obtained with sufficient accuracy. However when the number of future time slots increased beyond this value, model accuracy dropped dramatically, demonstrating the limitations of multi-step ahead regression that must be carefully considered during model training and inference. The multi-step ahead prediction problem was also examined in~\cite{9782838}, where an ED-LSTM model, specifically designed for sequence-to-sequence problems, was used to predict the TD matrices, of several steps ahead, in an optical DC and high-performance computing network.  

Recurrent DNNs with GRUs were also adopted in~\cite{8473978} aiming to directly predict the TD matrices of a backbone network. As, however, GRUs cannot directly process matrices, each TD matrix was reshaped into a vector. The GRU DNN was trained according to the real TD matrices of the Abilene network~\cite{Abilene_dataset} to perform on-hour predictions, with the GRU DNN shown to achieve a sufficient prediction accuracy. In general, directly predicting the TD matrix, is significantly more complex than the case of traffic prediction on a single network link or node, since not only time correlations, but also spatial correlations are present~\cite{8352941, 9163001}. It is worth mentioning, however, that GRUs are appropriate for capturing only the time correlations, hence the application of more advanced ML models (e.g., combination of GRUs and CNNs or graph DNNs) is expected to be more appropriate for this objective. 

GRUs were also applied in~\cite{9748600}, where several GRUs were trained to subsequently form the TD matrix. In that work, however, the GRUs were trained to optimize, instead of a least squares loss function (e.g., the MSE), the asymmetrically weighted sum of absolute errors; that is, in that work a deep quantile regression framework was adopted aiming to capture traffic prediction uncertainty over future network planning intervals. Several quantile GRU models were trained, considering various certainty levels, with all demonstrating sufficient inference accuracy. The performance of the quantile GRU models was compared against the performance of a least squares GRU model with respect to the percentage of the test patterns with predicted values underestimating the true traffic demand. This percentage was $9\%$ for the quantile GRU model, whereas it was $45\%$ for the least squares GRU model; an indicator that the quantile GRU models, capturing ML model uncertainty, significantly reduce undesired under-provisioning of the connections. This is one of the few works in the literature that aims to address ML model uncertainty in a principled way. Note, however, that the main limitation of this framework is that several quantile DNN models need to be trained, as the appropriate certainty level cannot be known a-priori~\cite{MARYAM2022108992}. Hence, this framework, unlike the general Bayesian inference framework, lacks in flexibility~\cite{MaryamMed22}. This limitation of of quantile regression was subsequently addressed in~\cite{MARYAM202313} where Monte Carlo dropout inference was opted instead.            
 
Bayesian inference frameworks were leveraged in~\cite{8501527,9204965}; however, the estimation of ML model uncertainty was not specifically examined. In particular, in~\cite{8501527,9204965} GPRs were trained to approximate Gaussian predictive distributions from which only their mean value was considered during inference. In~\cite{8501527} the GPR was trained according to input features that were optimized with respect to the number of previous time slots. Specifically, partial autocorrelation of the data was applied to choose an appropriate number of previous time slots for different prediction time scales. The GPR was trained and tested according to real datasets~\cite{citeulike:105953} and it was observed that the appropriate number of previous time slots depends on the time scale of predictions (i.e., hourly, weekly predictions require a different number of previous observations). The GPR performance was compared against penalized linear models, boosted decisions trees, and random forests, with GPR outperforming all in prediction accuracy.

Finally, in \cite{9204965}, an ensemble of GPR learners was proposed, taking into account the accuracy of individual learners and the diversity among their outcomes. Specifically, each learner found an optimal accuracy-diversity balance, so that the ensemble prediction error was minimized. A divide-and-conquer approach was utilized for executing the accuracy-diversity optimization, in the sense that each learner contributed to the optimization process by considering a small portion of the training samples in a region of the feature space. This renders the approach appropriate for large datasets. The proposed model was compared with several well-known time-series prediction algorithms (e.g., LSTM, ARIMA, SVR, LASSO etc.) using real traffic datasets from backbone networks~\cite{Abilene_dataset,CAIDA,Waikato}, demonstrating that GPR ensemble learning outperformed, in most of the cases, all other models considered. 

\subsection{Connection Arrival, Holding Time Prediction}
Another set of works addressed the arrival and/or holding time prediction of connections~\cite{8501524,XIONG201999}. Such predictions are usually used for on-line optimization in adaptive networks. Specifically, in~\cite{8501524} an LSTM was applied to predict holding times of future DC connection requests and its performance was evaluated for various traffic classes (i.e., short-, medium-, and long-lived connections), considering also the occurrence of unexpected traffic events. Traffic classes and unexpected events were synthetically generated. It was shown that LSTM achieves a higher accuracy under periodic or known circumstances, and that the accuracy is negatively affected by unexpected traffic events. Training an LSTM model according to various traffic profiles, instead of training it according to a single traffic profile, was shown to improve accuracy. Further, in~\cite{XIONG201999} an ENN was applied to predict traffic demand in multi-core (MC)-EONs, utilizing a real dataset~\cite{CAIDA} that spanned a period of two weeks. The ENN was trained to predict, apart from the bit-rate demand of future connection requests, their arrival and holding times as well. 

\subsection{Link Load Prediction}
A third set of works addressed link load prediction~\cite{8845132,8737631,Zhao:18}. Even though only~\cite{Zhao:18} utilized these type of predictions for predictive provisioning in adaptive networks, this information can be utilized for predictive provisioning in proactive networks as well, depending, of course on the time scale of predictions. In both types of networks this information can be utilized to address congestion, and in proactive networks it can be utilized, for example, for triggering re-configuration decisions upon expected congestion events. 

Specifically, in~\cite{Zhao:18} the authors applied an ANN in IP-over-WDM networks to predict link utilization for future time slots. Link utilization was measured in number of utilized wavelengths, and the model was formulated as a classifier over training inputs declaring edge-optical node connectivity, current point in time, and number of available wavelengths in the past few hours. The classification model was shown to predict, with sufficient accuracy, the trends of dynamically changing traffic. More recently, the authors in~\cite{8845132,8737631} applied graph learning methods to capture the spatio-temporal topological dependencies/relations of link loads in the network environment. 

In general, graph learning methods, are considered more appropriate for link load prediction purposes, compared to simpler ANN/DNN models, or even models that are solely based on RNN structures (e.g., LSTM); an outcome that was verified in~\cite{9203477}. The same holds for CNN models, that are capable of capturing the spatio-temporal dependencies, when combined with RNN structures (i.e., capturing both the spatial and temporal dependencies); an outcome that was verified in~\cite{li2018diffusion}.

Specifically, the authors in~\cite{9203477} applied a graph-based GCN-GAN~\cite{8737631} model to predict the link load matrix of an EON. In that work, the GCN-GAN processed historical graph states, describing temporal network loads, to predict future graph states. A graph state was formed by connecting the EON nodes through weighted links, with the weights representing the temporal link loads (in Erlangs) between the network nodes. GCN-GANs were trained for several traffic profiles (i.e., plateau, single-burst, and double-burst traffic profiles), utilizing synthetic and real datasets~\cite{telus_fibre} and their performance was compared to an LSTM, demonstrating that they can achieve higher performance accuracy.   

The combination of CNN and RNN structures was considered in~\cite{8845132} where the authors applied DCRNNs~\cite{li2018diffusion} to predict traffic loads (in Mbps) on the links of a real backbone network~\cite{Abilene_dataset}. Additionally, they evaluated the ability of the proposed method to predict the volume of traffic and congestion events. An extensive experimental study with a real dataset was presented which verified that the proposed method achieves better accuracy than other baseline NN methods (i.e., CNN, LSTM, DNN).
Overall, the authors in~\cite{9203477,li2018diffusion} verified the importance of appropriately representing and learning link load dependencies, especially when one is interested in finding link load prediction models guiding network management decisions (e.g., avoiding congestion, load balancing, etc.). 

\subsection{Main Outcomes and Research Challenges}~\label{chal1}
Several ML models have been already exploited for network traffic prediction, outperforming traditional statistical techniques. Most notably, RNNs and their variants are amongst the most promising approaches for capturing the temporal traffic dependencies, while CNNs and GNNs are capable of capturing the spatial dependencies. The combination of various NN structures (e.g., CNNs combined with RNNs) can be also considered for capturing both, spatial and temporal dependencies. Overall, the literature verified that depending on the underlying prediction problem (e.g., bit-rate prediction, link load prediction), the model must be carefully chosen. Ensemble learning mechanisms were also shown to be promising techniques towards further improving performance accuracy.

\subsubsection{Uncertainty Representation} Recent advances in ML, such as attention mechanisms and DBL models still remain unexploited and are interesting new avenues for research. The first mechanism can be considered to improve processing time, while the latter can be considered as a way of representing ML model uncertainty in a principled way. Even though the benefits of ML model uncertainty representation were quantified in~\cite{9748600, MARYAM202313}, this area remains greatly unexplored, especially as it concerns the representation of uncertainty in a way that is more flexible implementation-wise. 

\subsubsection{Multi-Step Ahead Prediction} As the vast majority of related works deals with the single-output prediction problem (i.e., predicting the traffic value only for the next time slot), the multi-step ahead prediction problem remains greatly unexplored. For this problem, encoder-decoder structures can be leveraged, while multi-step ahead prediction can be especially useful for predictive provisioning in proactive networks where the knowledge of how traffic varies over several planning interval ahead of time can be leveraged, amongst others, to minimize service disruption and/or energy consumption, especially of data center networks and the Internet network topology that mainly represent the overall energy consumption of cloud computing~\cite{6129370}.

\subsubsection{Disaggregation of Traffic}As the prediction of disaggregated traffic has shown to be more challenging that the prediction of aggregated traffic~\cite{8047676}, in the vast majority of existing works traffic is represented in an aggregated manner. However, the disaggregation of traffic may be essential to account for services with diverse SLAs (e.g., in the VNF-SC provisioning problem). Hence, more advances are expected towards this direction, most notably by leveraging recent ML advances, appropriate feature selection (i.e., traffic representation), time scales of prediction, and hyperparameter tuning, which were all shown to be essential towards finding a predictive model of sufficient accuracy. Of particular importance would be the development of methods that automatically tune time scales and hyperparameters to effectively deal with ML model adaptation, necessary to capture non-stationary traffic.   
 
\subsubsection{ML Lifecycle} The ML model adaption directly relates to the ML lifecycle, which is defined as a cyclical process that involves several phases (data preprocessing, model selection, training/testing, and inference phase). To initiate the first phase of the ML lifecycle requires continuously monitoring the performance accuracy of the model (i.e., to determine model degradation), towards model adaptation to the non-stationary traffic. While identification of the ML lifecycle is often mentioned in the literature, research efforts considering the unique characteristics of network traffic are currently missing. In general, and in practice, developing methods that are able to automatically and continuously monitor and maintain deployed model/s is a necessity. Additionally, since the degrading performance of a model also implies that the model becomes obsolete, systematic ways need to be developed that efficiently select a model~\cite{candela2020model}. Transfer learning (TL) can be an integral part of such methods to reduce the training dataset required for model adaptation. 

\subsubsection{Training at the Edge} Finally, in the related literature both centralized and decentralized training paradigms are considered. The first is considered for the prediction of TD matrices, whereas the second is considered for training several ML models at the edge to subsequently form the TD matrix. As with the advent of multi-access edge computing (MEC) architectures and with the emergence of edge-to-cloud continuum topologies~\cite{dell22}, supporting 5G and future 6G application and services, training paradigms are shifted to the edge, an interesting future direction is the consideration of federated learning (FL) paradigms. FL enables on the one hand training at the edge, without the exchange of data between the central controller and the edges, and on the other hand the creation of a centralized ML model. The ultimate advantage of FL is, apart from the lower overhead incurred by the exchange of data (e.g., in latency, bandwidth), the privacy that it offers by locally processing data susceptible to adversarial attacks. A tutorial of FL and a comprehensive survey on the issues regarding the MEC-enabled FL implementation can be found in~\cite{9060868}.

\subsubsection{Human-Centric ML} As ML needs to be also understood and trusted by humans (i.e., the operators in predictive service provisioning) to enable ML uptake, more human-centric ML-aided approaches need to be developed, leveraging or developing, for example, explainable supervised ML methods~\cite{Burkart_2021}. Furthermore, human-centric ML-aided approaches need to be developed that consider the fact that predictive service provisioning can never be fully autonomous, since human intelligence, providing expert domain knowledge, will always be needed. Towards this direction, human-in-the-loop (HITL) ML approaches~\cite{Monarch21} are expected to be developed leveraging both human and machine intelligence to create ML models.

An additional interesting direction, towards a more human-centric ML is identified in the FL paradigm where fairness issues may arise with respect to the accuracies achieved at each individual edge (i.e., fair federated learning~\cite{https://doi.org/10.48550/arxiv.1905.10497}), due to the presence of imbalanced and/or heterogeneous data at the edges. In the predictive service provisioning problem, unfair learning may result to erroneous service provisioning decisions, ultimately violated predefined QoS requirements. Hence, fair learning is expected to be of a particular importance towards trusted service provisioning decisions.   

\section{Predictive Provisioning: Survey on Optimization}~\label{surv_proactive}
In this section works focusing on predictive service provisioning are surveyed, with Table~\ref{proactive_surv_core} briefly describing their main components (i.e., network technology, algorithmic approach, comparisons, evaluation metrics, and key findings). Even though this table is not exhaustive, these works were chosen as they summarize the main outcomes in the literature, performance evaluation metrics considered, as well as the most common arising trade-offs and challenges. In the section that follows, these works are categorized according to their optimization objective. 
 
\begin{table*}[htbp]
\caption{Predictive service optimization in proactive and adaptive networks.\\
(\scriptsize{*Blocking Probability (BP), Spectrum Utilization (SU), over-provisioning (OP), under-provisioning (UP), Energy Consumption (EC), Regenerator Utilization (RU), Transponder Utilization (TU), Fragmentation (FRG), Crosstalk (XT), Packet Loss Rate (PLR), Multi-Core (MC), Routing and Spectrum Allocation (RSA), Spectrum Allocation (SA), Routing and Wavelength Assignment (RWA), Empirical Margin (EM), Quantile Margin (QM)).}}
\footnotesize
\begin{tabular} {p{0.5cm}|p{1.5cm}|p{1.5cm}|p{2cm}|p{3.5cm}|p{1cm}|p{5cm}}
{\bf Ref. }& {\bf Network} &{ \bf Problem} & { \bf Algorithm} & {\bf Comparisons} & {\bf Eval. \newline Metric}  & {\bf Key \newline Findings} \\ \hline 

\cite{8473978} &  WDM & Off-line RWA & Heuristic & Proactive, \newline Static & SU, OP, UP & Proactive outperform Static in SU and OP, \newline SU reduced $60\%$, \newline OP reduced $20\%$, \newline UP increased $3\%$ \\ \hline

\cite{8501527} &  Multi-layer \newline IP-WDM & Off-line RWA & Heuristic & Multi-layer Proactive, \newline  Multi-layer Static, \newline Virtual Static - Physical Proactive & SU,  \newline RU &  Proactive Multi-layer outperforms all \\ \hline

\cite{8047676} &   WDM & Off-line RWA & MILP, \newline Metaheuristic & Proactive, \newline Static &  EC & Proactive outperforms Static, \newline Energy Savings up to $31\%$ \\ \hline

\cite{7830260} &   WDM & Off-line RWA & MILP, Heuristic &  Proactive, \newline Static, \newline Reactive & BP, \newline TU &  Proactive outperforms both in Transponder Savings \\ \hline

\cite{6887148} &   WDM & Off-line RWA & Metaheuristic &  Proactive, \newline Static, \newline Reactive & PLR, OPEX &  Proactive outperforms both in OPEX, \newline Proactive slightly increases PLR\\ \hline 

\cite{7444562} &   WDM & Off-line RWA & MILP &  Proactive & EC, BP &  Trade-off: Reduction in EC is at the expense of BP\\ \hline 

\cite{8430520} & DC-EON & Off-line RWA  & Heuristic & Proactive, \newline  Reactive & BP, SU &  Proactive outperforms Reactive \\ \hline

\cite{8436062} &  EON & On-line RMSA & Heuristic &  Adaptive, Dynamic & BP &  Adaptive outperforms Dynamic  \\ \hline

\cite{8686111} &  EON & Off-line \newline RSA & ILP, \newline Heuristics  &  Proactive for different \newline prediction errors & BP, SU &  Wrong prediction error estimation leads to inappropriate resource allocations \\ \hline

\cite{9347908} &  EON & Off-line \newline SA & ILP &  Proactive with greedy SA, \newline Proactive with QoS-fair SA & BP, OP, UP, SU &  Proactive with fair QoS-fair SA outperforms Proactive with Greedy SA \\ \hline

\cite{XIONG201999} & MC-EON & On-line RSCA  & Heuristic &  Adaptive, \newline Dynamic & BP, SU, XT, FRG & Adaptive outperforms Dynamic  \\ \hline

\cite{8501524} & DC-MC-EON & On-line RSCA  & Heuristic &  Adaptive, \newline Dynamic & BP &  Adaptive outperforms Dynamic when predictions are sufficiently accurate \\ \hline

\cite{9748600} & EON & Off-line RSA  & Heuristic &  Static with EM, \newline Proactive with EM, \newline Proactive with QM & OP, UP &  Proactive with QM outperforms all, \newline Indicates the importance of appropriate ML-uncertainty representation \\ \hline

\end{tabular}
\label{proactive_surv_core}
\end{table*}

\subsection{Proactive Networks} \label{surv_proactive_opti}
\subsubsection{Optimizing Energy Consumption} Predictive provisioning aiming at reducing energy consumption is largely examined in the literature, mainly for proactive networks~\cite{7444562, 8047676,7830260,ZHANG2011171, 5540249, 6515884, 6384633, Alvizu2017EnergyED}. In general, these works suggest reconfiguring the network according to the traffic demand variations by shutting down unutilized resources during the low demand hours, instead of following a static/reactive configuration approach where the over-provisioned network resources are always on. All works indicate the potential of significant energy savings in proactive networks, as opposed to the static and reactive networks, while also demonstrating the arising trade-offs between energy savings and connection blocking/service disruptions.

In~\cite{ZHANG2011171}, a proactive multi-layer network planning approach was proposed for an IP-over-WDM network given realistic multi-period traffic variations of tidal traffic. A MILP algorithm was developed to off-line minimize the energy consumption incurred by the IP routers and optical cross-connects by shutting down idle linecards and chassis of routers, based on time-of-the-day network traffic variation. Service disruptions caused by the resulting reconfigurations were also considered through unconstrained and constrained reconfiguration schemes, with the results indicating the trade-off between traffic disruptions and energy savings (i.e., the unconstrained approach achieves higher energy savings but increases service disruptions).

In addition,~\cite{Alvizu2017EnergyED} examined proactive network optimization in mobile metro-core WDM networks, with the tidal traffic predictions being considered to off-line re-optimize the network, aiming to reduce energy consumption. Off-line (re)optimizations were followed by on-line network operations, to dynamically handle the unpredicted traffic variations. Heuristic and metaheuristic algorithms were proposed for both the off-line and on-line phases, considering also 1+1 protection of the aggregated mobile traffic. A real traffic dataset was utilized for evaluating the proposed framework, with the results indicating a significant reduction in energy consumption as opposed to the benchmark static network.

In~\cite{7444562}, authors proposed a proactive RWA algorithm to off-line re-optimize VNT to follow tidal traffic variations. The problem was formulated as a MILP to minimize both energy consumption and blocking probability, subject to different weights of importance for the two conflicting objectives. Traffic grooming techniques were also considered to further improve both energy consumption and connection blocking. In that work, the VNT performance was evaluated by simulating dynamic connection requests within the pre-defined planning intervals, showing that proactive VNT reconfigurations result in significant energy savings at the expense of connection blocking. Nevertheless, it was shown that the arising trade-offs could be controlled by appropriately weighing the importance of the diverse objectives in the MILP algorithm. 

In a similar vein, the authors in~\cite{8047676} exploited ANN traffic predictions for hourly reconfiguring the VNT of a WDM network. Predictions were used for identifying the near-optimal number of necessary (re)configurations during the day and for off-line VNT re-optimization. A simulated annealing-based heuristic was developed to reduce the number of (re)configurations. The VNT optimization problem was solved according to a MILP having as an objective the minimization of energy consumption (i.e., the number of transponders utilized). Subsequently, the VNT was used to guide the on-line routing decisions near-optimally. Significant energy savings were reported as opposed to the benchmark static network planning approach (i.e., up to $31\%$).

Finally, the authors in~\cite{7830260} utilized DNN traffic predictions to hourly trigger VNT reconfigurations in WDM networks. The objective was to reduce blocking probability and the number of utilized transponders (i.e., energy consumption), with the VNT reconfigurations triggered when the traffic predictions reached a predefined VNT utilization threshold. Both MILP and heuristic algorithms were developed to off-line re-optimize the VNT. The proactive scheme was compared to both conventional static and reactive networks, where, in the reactive network VNT reconfigurations were triggered when link utilization reached a predefined threshold (i.e., $90\%$ link utilization). It was shown that the proactive scheme resulted in significant energy savings, with a slight increase in connection blocking, as opposed to benchmark schemes. 

\subsubsection{Optimizing Service Disruptions}
Apart from the set of works leveraging traffic predictions towards energy savings, several works focused on examining the impact of constrained and unconstrained reconfigurations on service disruptions (i.e., traffic loss), OPEX, unserved traffic, and connection blocking~\cite{s11227-020-03493-7,6381741,6831425,6887148, 9782838}. 
In general, it was shown that network re-optimization that is subject to reconfiguration constraints reduces service disruptions and OPEX, but at the expense of unserved traffic and connection blocking.

Specifically, the authors in~\cite{6381741} examined predictive provisioning according to various RSA schemes, exploiting various grades of flexibility for the SA sub-problem, including fixed, semi-elastic, and elastic SA policies for adjusting the allocated spectrum to the time-varying traffic. 
All SA schemes were examined and compared through off-line ILP-based RSA algorithms, considering as inputs predictive TD matrices, and with each algorithm considering only one of the aforementioned schemes; that is, fixed (static network), semi-elastic, and elastic. It was shown that elastic SA, allowing of complete connection re-allocation (i.e., provides the maximum degree of flexibility), results in the lowest unserved traffic with semi-elastic SA outperforming conventional static SA.  

Similar to~\cite{6381741}, the authors in~\cite{6831425} examined predictive provisioning under two RSA schemes, with each scheme providing different degrees of flexibility for the SA sub-problem. In the first scheme, the objective was to reduce both the number of service disruptions and connection blocking, following an elastic SA approach (i.e., expanding/reducing the allocated spectrum and if that fails then a complete reallocation is followed). In the second scheme, the objective was to reduce connection blocking following a complete reallocation approach. For the first scheme, two heuristics were proposed, namely the Planning ahead Spectrum allocation (PAS) and the Predicative Planning ahead Spectrum allocation (PPAS). PAS considered the traffic demands at reconfiguration time, while PPAS additionally considered the traffic demand predictions of the next time slot. It was shown that PPAS and PAS outperformed the complete re-allocation scheme in the number of service disruptions, but at the expense of increased connection blocking. Importantly, PPAS, considering for the traffic predictions of the next time slot, outperformed PAS in service disruptions. As also elaborated in~\cite{s11227-020-03493-7}, service disruptions can be further improved by considering during network re-optimization the traffic predictions of several time slots ahead. Such consideration increase, however, the problem's computational complexity. 

In~\cite{6887148}, ARIMA predictions were used for reconfiguring hourly the VNT of a WDM network. For proactive VNT re-optimization a genetic RWA algorithm was implemented to minimize VNT OPEX  (i.e., cost of transponders, erbium-doped fiber amplifiers, and IP/MPLS nodes), while ensuring that the QoT requirements of the unestablished lightpaths will be met. Furthermore, a scheme was proposed to mitigate packet loss rate resulting from service disruptions during VNT reconfigurations. This predictive scheme was compared to conventional static and reactive networks and it is shown that while it significantly reduces OPEX, it slightly increases packet loss rate (i.e., due to service disruptions). 

Finally, in~\cite{9782838}, ED-LSTM predictions, providing information of how the TD matrices vary over several steps ahead, were used to reduce reconfiguration disruptions in an optical DC and high-performance computing network. To achieve this, an unsupervised learning algorithm was used to cluster, over time, predicted TD matrices. Hence, for the similar TD matrices (belonging to a cluster), a common connectivity graph was created, while reconfigurations were only triggered when a predicted TD matrix was clustered differently from the previous one. It was shown that the knowledge of how traffic varies over several steps ahead can be utilized to reduce service disruptions, but at the expense of an increased packet latency. 

\subsubsection{Optimizing Spectrum Utilization in Multi-Layer Networks}
The authors in~\cite{8501527} leveraged GP traffic predictions to off-line re-optimize an IP-over-WDM network (i.e., multi-layer optimization). Several heuristics were developed, with each allowing re-optimization at different layers of the network. Specifically, heuristics were proposed for re-optimizing both virtual and physical layers, and for re-optimizing only the physical layer (i.e., virtual is static and physical is proactive). The proposed schemes were compared with an approach where both virtual and physical layers are static, showing that the predictive techniques result in spectrum utilization and regenerator cost savings.
Importantly, it was shown that coordinated multi-layer optimization is more beneficial, as compared to the case where a single network layer is considered during the network (re)optimization phase.

\subsubsection{Optimizing QoS with ML-Uncertainty Considerations}
Some works proposed techniques towards improving over-provisioning (i.e., spectrum utilization) and under-provisioning (i.e., QoS) by accounting for the inaccuracies (i.e., uncertainty) in predictions~\cite{8473978,8686111,9347908}. These works, ultimately demonstrated the importance of appropriately representing ML model uncertainty during predictive provisioning.

The consideration of ML model uncertainty was first considered in~\cite{8473978}.  
Specifically, in that work, the authors leveraged RNN-GRU predictions to reconfigure hourly a WDM network by off-line solving the RMSA problem. 
To account for ML model inaccuracies, an empirical margin was considered in the TD matrices (i.e., predictions were increased by $30\%$), targeting mainly to mitigate under-provisioning arising by the inaccuracies. It was shown that predictive provisioning exhibits significant spectrum savings ($66\%$) and leads to lower over-provisioning ($20\%$), with a slight increase in under-provisioning ($3\%$), as opposed to the conventional static case. However, in that work, ML model uncertainty was addressed through a margin, that is rather myopic, largely ignoring the fact that diverse input patterns are
subject to different levels of uncertainty. Hence, even though under-provisioning was mitigated, further possible improvements in over-provisioning were neglected.     

The importance of appropriately representing ML model uncertainty during predictive provisioning was examined in~\cite{8686111}. Specifically, in that work, the authors examined off-line re-optimization in EONs, considering apart from the TD matrix predictions the prediction errors as well. Off-line re-optimization was followed by on-line network operations to allocate additional resources when needed (i.e., when more resources than the reserved were requested) to further improve connection blocking. 
The proposed framework was examined for different prediction errors and it was shown that inaccurate errors result in inappropriate resource reservation, consequently increasing connection blocking. If, however, the errors are appropriately measured,
to only slightly deviate from the true traffic demand, these errors can be effectively handled by the on-line resource allocation phase. 
Note, however, that the aforementioned work did not provide a mathematical framework capable of appropriately estimating errors in predictions (i.e., ML model uncertainty).

A mathematical framework was recently presented in~\cite{9347908}, where ML model uncertainty was appropriately captured through the approximation of quantile GRU models, while considering real traffic traces. The quantile GRU predictions were used for off-line re-optimizing an EON every $30$ minutes. The utilization of quantile GRU predictions was compared with the utilization of conventional GRU predictions increased by an empirical, myopic, margin. It was shown that quantile predictions, capturing the fact that each input pattern is subject to different levels of uncertainty, results in significant spectrum savings (i.e., up to $82\%$). The slight amounts of unpredicted  traffic observed, were successfully handled on-line by significantly reducing the operational overhead (i.e., $72\%$) required, when compared to the case where prediction uncertainty was completely ignored.  

\subsubsection{Optimizing QoS Fairness in Congested Networks} 
Fairness with respect to the achievable QoS of services in predictive provisioning is considered in~\cite{9322381, 9347908, 9522187}. Fairness becomes an issue when greedy spectrum allocation (SA) techniques are used in congested networks, where services content for the available spectrum resources. Specifically, in greedy SAs, highly uneven QoS guarantees arise, as some services may be highly over-provisioned, while others may be entirely blocked. In order to resolve this issue, the authors in~\cite{9322381} assumed that predictive traffic distributions are a-priori modeled, allowing the exploration of several combinations of possible SAs to obtain the one achieving fair QoS guarantees (i.e., by appropriately degrading the achievable bit-rate of some connections). An $\alpha$-fairness scheme was exploited and QoS fairness was evaluated according to the coefficient of variations measure. 

Specifically, in~\cite{9322381} the optimal $\alpha$-fair SA ILP algorithm was proposed, which was extended in~\cite{9347908} according to a near-optimal ILP-based algorithm for reducing problem complexity. The $\alpha$-fair SA algorithms were evaluated for several values of the inequality aversion parameter $\alpha$ that controls fairness.
It was shown that as parameter $\alpha$ increases towards max-min fairness, QoS fairness improves along with over- and under-provisioning, resource utilization, and connection blocking. The trade-off arising between fairness and network efficiency was subsequently examined in~\cite{9522187}.

\subsection{Adaptive Networks}\label{surv_adaptive_opti}
A final set of works deals with leveraging traffic predictions to mainly reduce connection blocking in adaptive networks~\cite{8501524,XIONG201999,Zhao:18,9203477,8436062,9042293}. In most of these works, optimization schemes developed for adaptive networks were compared with conventional optimization schemes in dynamic networks. Those works, indicated the potential of adaptive networks to outperform dynamic networks, especially when the predictions are of sufficient accuracy.     

Specifically, the authors in~\cite{8501524} proposed an on-line routing sapecturm and core allocation (RSCA) algorithm to dynamically provision connections in a DC-MC-EON, leveraging LSTM predictions, used to on-line update the mean residual life-time of connections.
In particular, the mean residual life-time information was used to adjust the link weights of the network towards minimizing network fragmentation and consequently connection blocking. Additionally, predictions were used to estimate future connection blocking events to a-priori re-allocate established connections towards avoiding connection blocking. The predictive RSCA approach was compared to a conventional on-line RSCA approach that was based on the first-fit SA scheme, most notably showing that the predictive approach outperforms the conventional dynamic approach, especially when the LSTM predictions are of sufficient accuracy; otherwise, predictive RSCA may lead to inappropriate resource allocation decisions.

In another work~\cite{9042293}, a predictive RSA algorithm was proposed to dynamically provision connections in EONs under tidal traffic. Tidal traffic was described through a multi-step trigonometric tidal traffic model (MSTM) to capture the traffic distributions for several types of areas (i.e., residential, business, etc.). The proposed predictive RSA heuristic accounted for the MSTM traffic by appropriately adjusting the network link weights according to their temporal and future spectrum utilization (i.e., to guide lightpath computations over the least congested links). This predictive approach was compared to a conventional RSA heuristic, showing that predictive RSA significantly outperforms the conventional dynamic approach in terms of connection blocking. 

Further, authors in~\cite{8436062} examined a predictive resource allocation scheme in multi-domain EONs. In this scheme, intra-domain DNN traffic predictions were utilized by an on-line RMSA heuristic to dynamically reconfigure intra-domain virtual links. Specifically, predictive RMSA jointly reduced the current and predicted utilization on all traversed inter-domain links to avoid the future generation of resource bottlenecks. The proposed scheme was again shown to outperform a conventional RMSA scheme in terms of connection blocking. 

However, in~\cite{9203477}, a GCN-GAN and an LSTM were applied to predict link loads to dynamically provision connection requests. Specifically, link load matrices were predicted hourly, guiding an RMSA heuristic to dynamically provision the arriving connections. It was shown that GCN-GAN-based RMSA outperforms the LSTM-based RMSA in terms of connection blocking. This is mainly due to GCN-GAN's higher prediction accuracy (as it captures spatio-temporal traffic dependencies), better guiding resource allocation decisions.  

Finally, in~\cite{XIONG201999} a predictive RSCA heuristic was proposed leveraging ENN traffic predictions to reduce spectrum fragmentation and crosstalk in MC-EONs. The RSCA heuristic was based on a 2D rectangular packing algorithm leveraging the predictions to adjust the network link weights towards avoiding fragmentation and crosstalk in future time points. This adaptive scheme was shown to outperform the conventional dynamic schemes in connection blocking, fragmentation, and spectrum utilization.           

\subsection{Main Outcomes and Research Challenges}~\label{chal2} 
The existing literature provides a stepping stone towards further research efforts, as all relevant works converged to clear conclusions regarding the advantages of predictive service provisioning in both proactive and adaptive networks. Overall, in proactive networks the existing works demonstrated significant improvements in energy consumption and spectrum utilization, as well as connection blocking and over-provisioning, as compared to static and reactive networks. A slight increase in under-provisioning has also been observed, but this was shown that it can be successfully handled with on-line operations. Furthermore, effective optimization approaches have been proposed for minimizing traffic loss arising by service disruptions incurred due to service reconfigurations. In adaptive network, existing works demonstrated the capabilities of predictive service provisioning on reducing connection blocking, as opposed to dynamic networks.

Of course, in both adaptive and proactive networks, to harvest the advantages of predictive service provisioning, the ML models must be, not only of sufficient accuracy (i.e., affected by the selection of an appropriate ML model), but their uncertainty over future observations must be appropriately represented, quantified, and considered during the optimization phase; an outcome that is evident in many existing works~\cite{8473978,8686111,9347908,8501524}. 

In most of the works, predictive provisioning builds mainly on top of efficient, baseline, rule-based optimization algorithms developed for conventional static, reactive, and dynamic networks; an approach that has shown to been effective. Related research challenges, however, include, amongst others the development of more advanced predictive approaches, considering for emerging service provisioning problems, such as the VNF-SC problem where arriving services  may be of diverse QoS requirements, while considering for the advances in the optical network domain (i.e., spectrally-spatially optical multiplexing technologies).            
    
\section{Prescriptive Provisioning: Survey on ML and Optimization}~\label{surv_prescr}
\begin{table*}[htbp]
\caption{Prescriptive optimization in proactive and adaptive networks (\scriptsize{*Blocking Probability (BP), Spectrum Utilization (SU), Wavelength Utilization (WU), Over-provisioning (OP), Under-provisioning (UP), Flaw Completion Time (FCT), Latency (L), Auxiliary Graph (AG)).}}
\label{prescr_survey}
\footnotesize
\begin{tabular} {p{0.5cm}|p{1.7cm}|p{1.5cm}|p{3.5cm}|p{2.5cm}|p{1.5cm}|p{0.8cm}|p{3cm}}
{\bf Ref. }& {\bf Network} &{ \bf Problem} & { \bf RL Formulation} & {\bf Comparisons} & {\bf Traffic Traces} &{\bf Eval. \newline Metric}  & {\bf Key \newline Findings} \\ \hline

\cite{8620207} &  EON &  SA & $\mathcal{S}$: Representing bandwidth demand \newline $\mathcal{A}$: Candidate bandwidth allocations for each s-d pair \newline $\mathcal{R}$ Obj.: Controls agent's bandwidth contention  & Proactive-Prescriptive, \newline Proactive-Predictive &  Synthetic & BP, UP, OP & Prescriptive outperforms predictive \newline Prescriptive controls OP and UP trade-off  \\ \hline

\cite{8738827} &  WDM &  WA & $\mathcal{S}$: Representing wavelength utilization and route \newline $\mathcal{A}$:  Set of network wavelengths \newline $\mathcal{R}$ Obj.: Min. BP & Adaptive, \newline Dynamic &  Synthetic & BP & Adaptive outperforms Dynamic  \\ \hline

\cite{8738827} &  EON &  RMSA & $\mathcal{S}$: Representing lightpaths' utilization at the link-level and conn. request \newline $\mathcal{A}$:  Set of candidate lightpaths for each s-d pair \newline $\mathcal{R}$ Obj. : Min. BP & Adaptive, \newline Dynamic &  Synthetic & BP & Adaptive outperforms Dynamic  \\ \hline

\cite{8738827} &  EON &  RMSA & $\mathcal{S}$: Representing lightpaths' utilization at the link-level and conn. request \newline $\mathcal{A}$: Candidate lightpaths for each s-d pair \newline $\mathcal{R}$ Obj.: Min. BP & Adaptive, \newline Dynamic &  Synthetic & BP & Adaptive outperforms Dynamic  \\ \hline

\cite{8696290} & Multi-Layer \newline OTN (generic) &  R & $\mathcal{S}$: Path-level lightpaths utilization and conn. request at $t$ \newline $\mathcal{A}$:  Candidate paths for each s-d pair \newline $\mathcal{R}$ Obj.: Max. SU & Adaptive, \newline Dynamic &  Synthetic & SU & Adaptive outperforms Dynamic \\ \hline

\cite{3229554} &  Multi-Layer DCN &  Optical switch link activation & $\mathcal{S}$: Representing active links \newline $\mathcal{A}$: Set of Links \newline $\mathcal{R}$ Obj.: Min. FCT  & Adaptive, \newline Dynamic (LP-based) &  Synthetic & FCT & Adaptive performs near-optimal  \\ \hline

\cite{9748476} &  Multi-Layer WDM &  RWA & $\mathcal{S}$: Residual Link Capacities and TD information \newline $\mathcal{A}$: Link weights of multi-layer AG \newline $\mathcal{R}$ Obj.: Min. WU  & Adaptive, \newline Dynamic &  Real &  WU, L & Adaptive outperforms Dynamic  \\ \hline

\cite{9083336} &  Multi-Domain EON &  RSA & $\mathcal{S}$: Representing adjacent domains utilization \newline $\mathcal{A}$: Set of RSA heuristics \newline $\mathcal{R}$ Obj.: Min. BP of inter-domain traffic & Adaptive with cooperative agents, \newline  Adaptive w/out cooperative agents, \newline  Dynamic &  Synthetic & BP & Adaptive with cooperative agents outperforms all  \\ \hline

\cite{9609608} &  DCN-EON &  VNF-SC & $\mathcal{S}$: VNF-SC request and network state \newline $\mathcal{A}$: A feasible provisioning \newline $\mathcal{R}$ Obj.: Min. BP  & Adaptive with Hierarchical Training, \newline Adaptive w/out Hierarchical Training \newline Dynamic &  Synthetic &  BP & Adaptive with hierarchical training outperforms all  \\ \hline

\end{tabular}

\end{table*}

This section surveys works dealing with prescriptive TE and/or service provisioning.
 Table~\ref{prescr_survey} provides information for some of these works. 
Briefly, Table~\ref{prescr_survey} provides, amongst others, information regarding the RL formulation in state and action spaces and what the reward function driving the optimization objective represents, i.e., the key ingredients of RL. The works chosen have diverse objectives, optical network technologies, and RL formulations, for both adaptive and proactive networks, so as to highlight the main problems the optical networks research community is currently addressing.

\subsection{Proactive Networks}~\label{presc_proact_survey}
Only a few works in the literature deal with prescriptive TE for proactive networks~\cite{8396130,8751369, 8620207,9748476}, advancing not only conventional ruled-based service provisioning schemes, but also outperforming predictive service provisioning approaches. Specifically, the authors in~\cite{8396130} examined the SA problem in EONs through a multi-agent distributed RL framework. Each agent, representing a pre-computed path between a source-destination ($s-d$) pair, independently learns a SA policy. Specifically, each RL agent learns its own policy by observing actual traffic fluctuations and appropriately penalizing or rewarding SA actions targeting on striking a balance between over- and under-provisioning. This work was subsequently extended in~\cite{8751369, 8620207} to centrally control the agents to cooperate and learn SA policies that account also for network congestion and agent's achievable QoS. The SA actions from the learned policies are subsequently used for off-line reoptimization. The proposed prescriptive provisioning approach was shown to outperform predictive provisioning approaches in terms of connection blocking, over- and under-provisioning. This is mainly due to the cooperation capabilities of prescriptive provisioning, based on RL, that is capable of appropriately controlling the SA decisions taken to better utilize spectrum resources and meet targeted QoS requirements; a capability missing from predictive provisioning approaches that directly utilize traffic predictions without considering how these predictions will eventually affect QoS requirements.  

It is worth mentioning that prescriptive TE for proactive networks has been encountered in the literature simply for estimating bit-rate needs for future planning intervals~\cite{9256655}. Even though this seems to be similar to the predictive approach (i.e., where supervised learning is applied), in essence the prescriptive policy has the capability to self-adapt without requiring a labeled dataset.           

\subsection{Adaptive Networks} \label{presc_adaptive_survey}
Most of works dealing with prescriptive service provisioning for adaptive networks opted for DRL frameworks~\cite{Zhao:18,8485853,8738827,8847548,3229554,9083336,9375634,9373585,SUN2021107891,Zhao:21,9279336,9507559,9609608,9779080,KHOSHKHOLGHI2022109451} with most of them considering actor-critic DRL implementations. This set of works examined different optimization objectives and optical network technologies. Furthermore, some works considered multi-layer and/or multi-domain networks, while the provisioning of NVF-SC in DC networks was also addressed. 

\subsubsection{Provisioning in Single-Domain, Single-Layer Networks}
The prescriptive RWA in WDM networks was examined in~\cite{Zhao:18,8644190}, aiming to minimize connection blocking of arriving requests. Indicatively, in~\cite{8644190} DRL actions concerned wavelength allocation decisions upon the arrival of a connection request, whereas the routing decisions were computed according to the shortest-path algorithm, subsequently passed to the agent's state information. In both~\cite{8644190,Zhao:18} DRL was shown to outperform the conventional rule-based RWA heuristic, requiring, however, several episodes to converge.   

The prescriptive RMSA problem, characterized by a higher complexity than the RWA problem, was examined in~\cite{8738827,9373585,Aibin2020MonteCT,9779080}, indicating the importance of appropriately addressing RL's curse of dimensionality as problem complexity increases. Specifically, in~\cite{8738827} the DRL framework (DeepRMSA) was defined over a state space representing the temporal network states (e.g., spectrum utilization, in-service lightpaths, etc.) upon the arrival of a connection request ($s-d$ pair, bit rate, service time), with actions representing possible RMSA options for each $s-d$ pair. DeepRMSA was trained according to several arriving requests (in the millions) with the results indicating that as the number of episodes increases, blocking probability improves, and eventually outperforms state-of-the-art on-line RMSA heuristics [i.e., shortest-path routing with first-fit SA (SP-FF), $k$-shortest-path routing with first-fit SA (KSP-FF)]. 

As a high convergence time (affected by the experience required due to the high problem complexity) may render the learned RMSA policy obsolete upon decision time, especially under non-stationary traffic, DeepRMSA was extended in~\cite{9373585} to incorporate TL techniques. It was shown that DeepRMSA with TL effectively reduced the experience (i.e., episodes) the agent required, from millions of requests to thousands, compared to DeepRMSA without TL. Importantly, DeepRMSA with TL was shown to be capable of additionally dealing with several DRL agents reinforced to master diverse tasks (i.e., RMSA for different network topologies), with their learned policies used for training a single multi-task learning (MTL) agent obtaining knowledge across the tasks (i.e., actor-mimic DRL). MTL agent knowledge can then be transferred to other networks (domains) for diverse tasks. As an example, one of the scenarios the authors investigated was leveraging the knowledge learned by RMSA agents (i.e., source tasks) to facilitate the training of DRL agents for anycast service provisioning tasks (i.e., target tasks).

Finally, the importance of appropriately selecting the NN models for prescriptive RSA was recently demonstrated in~\cite{9779080}, where GCN-RNN models were chosen in the actor-critic DRL setting with the purpose of capturing the spatio-temporal dependencies in the evolving EON (i.e., in the network states as these are transformed according to the arriving/departing lightpaths). Specifically, in this work, it was shown that the GCN-RNN based DRL outperforms in blocking probability the DeepRMSA algorithm~\cite{8738827} leveraging simple DNNs structures.           

\subsubsection{Provisioning in Multi-Layer Networks}
The more complex scenario of prescriptive provisioning in multi-layer networks was examined in~\cite{8847548,3229554,9748476}. In general, in multi-layer networks, the RL problem deals with finding an optimal service provisioning policy by jointly optimizing the resources utilized in both logical and physical layers. Hence, related work is mostly concerned with how to appropriately represent the multi-layer environment in an abstracted form to reduce RL problem dimensionality,  while finding a policy that outperforms rule-based provisioning heuristics. 

Specifically, to achieve this, the authors in~\cite{8847548} proposed a DRL framework for IP-over-Optical Transport Network (OTN), where the agent operated only over the logical topology. Hence, a number of lightpaths were pre-computed and the agent actions concerned selecting, for each arriving connection request, a route (path) that maximizes bandwidth utilization over the fiber links. Hence, upon decision time, the agent observed an abstracted OTN state consisting of path-level utilization information. The proposed DRL framework was compared against other DRL frameworks (i.e., similar to DeepRMSA) operating according to (less abstracted) link-level information and to the shortest path routing algorithm. It was shown that the proposed DRL converges to a routing policy that significantly outperforms benchmark approaches. It should be noted, however, that the proposed DRL framework still required thousands of episodes to converge (i.e., requiring possibly an experience of millions of connection requests), leaving room for further improvements. The work in~\cite{8847548} was further extended in~\cite{almasan2020deep} by considering instead of a DNN-based DRL model, a GNN-based DRL model, with the latter, capturing the spatial lightpath dependencies, outperforming the former in bandwidth utilization.   
  
In the same vein, the authors in~\cite{3229554} proposed a DRL-based framework (DeepConfig) for DCNs, where the agent operated only over the logical topology by observing IP flows arriving at each optical switch. The objective in that work was to minimize the flow completion time of large flows (i.e., duration between the first and last packet of a flow) by finding a policy that appropriately activates the optical switch links at each flow event. It was shown that the DRL model performed near optimal, as it achieved flow completion times that are close to a linear programming algorithm that is a-priori aware of several future traffic flows (i.e., it is an LP-based dynamic scheme). The proposed DRL model converged to a near-optimal model only within a few training episodes (i.e., below $50$). However, only small network topologies were considered (i.e., consisting of $10$ optical nodes/switches) for a relatively simple optimization problem. 

Finally, the authors in~\cite{9748476}, recently developed a DRL framework (ADMIRE) for multi-layer service provisioning in X-Haul networks. The purpose of the DRL was to learn a link weight allocation policy, subsequently driving an on-line RWA heuristic, to minimize wavelength utilization. The multi-layer X-Haul network was represented by an auxiliary graph (AG), over which ADMIRE learns the link weight allocation policy by utilizing real traffic traces (i.e., capturing traffic demand variations over time). Hence, in that work, unlike~\cite{8847548,3229554}, where the agent operated only over the logical layer, the agent operates over the entire multi-layer network (i.e., cross-layer optimization) by observing the AG state. It was shown that ADMIRE outperforms a conventional rule-based RWA heuristic that operates over an AG with unchanged (static) weights in terms of wavelength utilization, as traffic varied over time. Additionally, it was shown, through a real X-Haul testbed, that ADMIRE achieves lower end-to-end latency than the rule-based RWA as it creates longer lightpaths (longer propagation latency) than ADMIRE. Regarding the number of episodes required, these were less than $100$, but this number is not representative due to the small testbed size ($9$ nodes, $12$ fiber links, with $3$ wavelengths each). 
   
\subsubsection{Provisioning in Multi-Domain Networks}
The DeepCoop framework was proposed in~\cite{9083336} to realize service provisioning in multi-domain EONs with cooperative deep reinforcement learning (DRL) agents. DeepCoop aims at minimizing inter-domain blocking probability by finding a policy that appropriately selects a RSA algorithm for each arriving connection request. Each domain is controlled by its own DRL agent, trained in a cooperative fashion to share domain utilization with other agents at each time point of interest. Hence, the DRL agent action space is in essence a set of well-known dynamic RSA heuristics [SP-FF, KSP-FF, and k-shortest path and load balancing (KSP-LB)], that, when  considered as the action space, effectively reduce the  dimensionality of the RL problem (i.e., instead of letting the RL to learn RSA heuristics from scratch, effective heuristics are considered as actions to solve the RSA problem).

DeepCoop was compared against these on-line heuristics and against a DRL framework in which multi-domain agents did not share information (non-cooperative). It was shown that DeepCoop outperforms all benchmarks after a relatively low number of training episodes for the size of EONs considered. Interestingly, non-cooperative DRL fails to converge, indicating the importance of cooperative learning (i.e., sharing domain information during training). In~\cite{9507559}, the performance of DeepCoop was further improved by considering instead of the A2C algorithm, the SAC algorithm (i.e., capable of achieving a better trade-off between exploration and exploitation than A2C), indicating the importance of appropriately selecting a DRL optimization algorithm, especially in complex environments such as multi-domain networks. 

It is worth mentioning, that recently SL was also applied to train a DNN that selects the best routing heuristic (i.e., the one that results in a lower blocking probability) upon the arrival of a connection request~\cite{HONG2022100677}. Specifically, the DNN was off-line trained according to various DC network states and arriving requests, with the trained model utilized for on-line provisioning. Even through both SL in~\cite{HONG2022100677} and DRL in~\cite{9083336,9507559} were used towards achieving similar targets (i.e., the selection of a routing strategy), it is important to note that DRL is considered more appropriate towards this objective, since, unlike SL, it is capable of capturing the long-term dependencies of the on-line provisioning decisions over time, consequently leading to a lower blocking probability. Of course, the application of SL may be a good approach when the dimensionality of the problem is prohibitive for the DRL to converge to a good provisioning policy. In general, however, such comparisons, merit further investigation.     

\subsubsection{Provisioning VNF-SC in Data Center Networks}
Recently, the prescriptive VNF-SC provisioning problem was examined in~\cite{9609608,KHOSHKHOLGHI2022109451}. In general, the VNF provisioning problem is becoming more and more popular as it enables agile resource allocation and fast deployment of new services. While several effective rule-based heuristics~\cite{7490359,7293303} have been developed to address this problem, policy-driven schemes, enhanced by RL, are shown to outperform conventional approaches in blocking probability, utilization of IT resources in DCs, and utilization of spectrum resources on fiber links.  

Indicatively, the authors in~\cite{9609608} addressed the VNF-SC provisioning problem in inter-DC EONs with the purpose of minimizing the blocking probability of arriving VNF-SC requests. A hierarchical GNN-based DRL framework was developed consisting of two hierarchical but collaborative training processes (i.e., upper- and lower-level training processes). The lower-level training process aimed to minimize the resource utilization (i.e., IT and spectrum resources) of VNF-SC requests, subsequently assisting the upper-level training process in minimizing the VNF-SC blocking probability (i.e., by further fine tuning GNN model parameters). Hence, by doing so, the authors were able to reduce the problem complexity arising when resource utilization and blocking probability are jointly minimized. It was shown that a lower blocking probability is achieved, as opposed to a non-hierarchical DRL framework that is based on a DNN structure, and as opposed to rule-based on-line VNF-SC heuristics. Furthermore, the hierarchical approach demonstrated better convergence performance (i.e., fewer episodes) than the non-hierarchical one. Overall, the outcomes of this work are indicators of the importance of appropriately selecting the NN structure of DRL frameworks (e.g., to capture spatio-temporal information in EONs), of the importance of effectively reducing DRL problem complexity, and of the capabilities of prescriptive schemes to advance rule-based heuristics.

\subsection{Main Outcomes and Research Challenges}\label{chal3}
Prescriptive provisioning, based on policy-driven optimization, was shown to be a promising approach not only for advancing rule-based optimization baselines but also for outperforming predictive provisioning based on traffic-driven optimization~\cite{8751369, 8620207}. This is mainly due to the capabilities of RL that, unlike SL, that captures only the time-varying traffic trend, it also controls the learned policies towards meeting the environment's targeted QoS requirements. 

Even though the related literature has demonstrated promising results, it has also demonstrated that challenges exist on effectively training the targeted RL policies especially as it concerns hyperparameter tuning, selecting an efficient RL optimization algorithm, selecting an appropriate NN structure (i.e., when DRL is opted), and most importantly  appropriately representing action-state-reward. Regarding the latter, the importance of reducing the action-state space as much as possible (e.g., through AGs representing the multi-layer network~\cite{9748476}) is evident in the literature, especially as the agent's environment becomes more complex (e.g., multi-domain, multi-layer networks)~\cite{8847548,3229554}. Specifically, reducing problem dimensionality allows RL to converge to an optimal policy faster (i.e., reduces the experience required), which is an important consideration under non-stationary traffic where the model may become obsolete after convergence and upon decision-making. Additionally, the consideration of TL techniques was shown to be a promising approach towards accelerating policy convergence~\cite{9373585}.

\subsubsection{Improving RL Convergence}
Further advances targeting the improvement of RL convergence are critical to accelerate the adoption of prescriptive provisioning frameworks in the field, especially when more complex problems are considered. Indicatively, to further empower RL, SL can be applied to allow the agent to learn, by imitation, baseline algorithms, and heuristics. The initial model can then be used to seed the agent's RL model (i.e., by means of transfer learning) to converge faster to a policy that eventually outperforms the baseline algorithms. This approach was recently proposed by DeepMind in their AlphaStar implementation, with the multi-agent system defeating word champions in the challenging StarCraft game~\cite{s41586-019-1724-z}. 

\subsubsection{Cooperative and Fair Learning}
Multi-agent DRL (MARL), in the sense that the agents are cooperative during learning, is a challenging future direction, especially due to its higher complexity, that it is also imposed to the control and management operation. Even though multi-agent network environments have recently been considered~\cite{8620207,9083336}, the underlying DRL framework is not truly cooperative as the agents are trained interdependently and controlled externally (i.e., by manually controlling their behavior), or cooperation is achieved by exchanging information through their observable states. Multi-agent RL frameworks where the agents are controlled and learn through a common reward function, have not been examined in as of yet. Towards this direction, promising multi-agent approaches constitute frameworks that are based on game-theoretic principles~\cite{s41586-019-1724-z,pmlr-v80-yang18d, mguni2018decentralised}, potentially also addressing multi-agent contention for network resources (i.e., learning fair policies in multi-agent RL~\cite{https://doi.org/10.48550/arxiv.1611.03071}). A comprehensive survey on applications of MARL in various other types of networks (e.g., 5G wireless networks, vehicular networks) can be found in~\cite{9738819}.

\subsubsection{Human-Centric RL}
Even though RL is capable of learning model-free polices (i.e., learned in the field by trial and error), there are particular use-cases where this may lead to disastrous outcomes in terms of network operations. Network control and management use cases belong to this category, as erroneous actions, required to learn a policy, may lead to inappropriate allocation of network resources and violations of QoS requirements. Hence, RL policies need to be learned off-line, before their actually deployment in the field. To this end, to mimic the environment off-line and to evaluate efficiency of the policy learned in the simulated environment, the development of models is required (e.g., traffic demand models, resource optimization algorithms, etc.).

Given this, it is clear that network control and management can never be really autonomous. Human intervention is required to design the aforementioned models and to redefine the agent's performance targets when necessary (e.g., when QoS requirements change). Importantly, human innervation (intelligence) is required to handle situations (e.g., unexpected events) that RL intelligence cannot effectively deal with. Towards this direction, HITL~\cite{Monarch21} can be considered during the RL formulation. As an example, with HITL, an agent can be trained with real-time feedback from a human observer who gives rewards for some actions~\cite{https://doi.org/10.48550/arxiv.1810.11748}, or alternatively RL can be formulated to include actions that switch the decision-making to the human [e.g., when the agent observes a state that is far (i.e., unknown) from what it has observed so far]. 

\subsubsection{RL for Realistic Networks at Scale}
Currently, the vast majority of existing works examined prescriptive provisioning on simulated environments, considering stationary, synthetic traffic demand traces. Only in~\cite{9748476} real traffic traces are considered over a real, small-scale, testbed. Hence, the biggest challenge is in the application of RL for real network environments at scale to validate the benefits of prescriptive provisioning over rule-based provisioning schemes under realistic conditions [i.e., non-stationary traffic, unexpected events (e.g., disasters), uncertainty, etc.] and under cooperative decentralized training paradigms enabled by the recent advances in the edge-to-cloud continuum. Furthermore, this will allow revealing new challenges related with the practical applicability of prescriptive provisioning mainly as it concerns the network telemetry requirements (i.e., for real-time monitoring and learning), requirements in computing resources, operational complexity and communication overhead, and privacy preservation during the exchange of information between several agents (e.g., in multi-domain networks).

\section{Summary Remarks and Future Directions}\label{conclusion}
The existing literature has demonstrated the ability of both predictive and prescriptive provisioning to outperform state-of-the art, rule-based, approaches in both proactive and adaptive networks. This is mainly due to the capabilities of ML to more accurately capture/model time-varying traffic behavior towards long-term efficient decision making, as opposed to statistical models and/or myopic rule-based service provisioning approaches. 
Specifically, ML-aided service provisioning for proactive and adaptive networks has shown to bring significant advantages in terms of resource utilization, energy consumption, network throughput, and latency, as opposed to traditional static, reactive, and dynamic networks provisioned according to rule-based optimization algorithms. 

While it is true that both predictive and prescriptive provisioning may lead to several undesired effects (e.g., service disruptions, inappropriate decision making due to ML-model uncertainty, etc.), these can be successfully controlled at acceptable levels either as part of a better ML treatment (e.g., considering of uncertainty) or as part of an appropriate optimization procedure. The reader should note, that the main outcomes, limitations, and future research directions for each ML-aided provisioning approach are extensively discussed in Sections~\ref{chal1},~\ref{chal2}, and~\ref{chal3}. Briefly, the most notable research challenges are related with the ML treatment, especially as it concerns the disaggregation of traffic, the consideration of FL (i.e., training at the edge), fairness considerations in both resource allocation decisions and accuracies achieved in decentralized training (i.e., in FL and multi-agent RL), the appropriate representation and quantification of ML-model uncertainty, the identification of ML life-cycle under non-stationary traffic, and the development of more human-centric schemes (e.g., HITL, explainable ML).   

Furthermore, field-trials at scale, demonstrating the feasibility of both prescriptive and predictive provisioning in proactive and adaptive networks are currently missing but are necessary to validate the feasibility of such approaches under realistic conditions and under the technological readiness level of SDN/NFV frameworks and protocols, optical hardware flexibility, network telemetry equipment and capabilities, availability of computing resources in the control and management units, as well as under the limitations of ML algorithms in terms of what they can really learn from the data towards real-time and confident decision making. 
    
Another important aspect that has not yet been considered in the literature is that ML systems present a new type of attack surface (i.e., adversarial ML attacks), increasing security risks through the possibility of data (e.g., traffic- or network state-related) manipulation and exploitation by attackers. In general, the ML attacks aim to cause malfunctions to the ML models, eventually leading to inappropriate resource allocation decisions (e.g., leading to outages, violation of targeted QoS requirements, congestion, packet loss, or increased latency). Evidently, in the presence of critical services and applications (e.g., virtual reality for healthcare, self-driving cars) supported by end-to-end 5G (or future 6G) networks, violation of sensitive QoS requirements (e.g., latency) may cause severe or even catastrophic effects. Therefore, the consideration of appropriate defence mechanisms (e.g., adversarial training) towards trustworthy ML systems must be considered as an inextricable part of the ML system. 

Finally, it must be noted that the abundance of challenges raised in this survey provide similarly a number of opportunities for research and development at all levels. Attempting to solve these problems requires, however, multidisciplinary approaches that cover all layers, from software to algorithms, to architectures and micro-architectures, and to technologies.

These efforts require pushing the limits towards improving the overall design stack, by modifying the training approaches to consider the constraints of the targeted architectures; by utilizing better and customized datasets; by identifying innovative solutions such as applying bio-inspired optimizations in an effort to minimize unnecessary computations; and by innovative optimization techniques. Alternative NN models must also be considered, that are customized based on the application domain (e.g., spiking NNs~\cite{8747378}), especially when considering much simpler problems where DNNs have been proven to be an overkill. The consideration of learning at the edge along with the recent advances in tiny ML are also expected to play a pivotal role, importantly to reduce latency time between model inference and action taken in the network environment, energy consumption, and computing requirements for model training.  

Clearly, the road towards the practical realization of proactive or even adaptive networks is still long, with the various related research advantages as well as gaps identified throughout this survey. 

\section*{Acknowledgment}
This work was supported by the European Union's Horizon 2020 research and innovation programme under grant agreement No 739551 (KIOS CoE - TEAMING) and from the Republic of Cyprus through the Deputy Ministry of Research, Innovation and Digital Policy.

\ifCLASSOPTIONcaptionsoff
  \newpage
\fi

\bibliographystyle{IEEEtran}
\bibliography{IEEEabrv,mybib_Submitted}

\begin{thebibliography}{100}
\providecommand{\url}[1]{#1}
\csname url@samestyle\endcsname
\providecommand{\newblock}{\relax}
\providecommand{\bibinfo}[2]{#2}
\providecommand{\BIBentrySTDinterwordspacing}{\spaceskip=0pt\relax}
\providecommand{\BIBentryALTinterwordstretchfactor}{4}
\providecommand{\BIBentryALTinterwordspacing}{\spaceskip=\fontdimen2\font plus
\BIBentryALTinterwordstretchfactor\fontdimen3\font minus
  \fontdimen4\font\relax}
\providecommand{\BIBforeignlanguage}[2]{{%
\expandafter\ifx\csname l@#1\endcsname\relax
\typeout{** WARNING: IEEEtran.bst: No hyphenation pattern has been}%
\typeout{** loaded for the language `#1'. Using the pattern for}%
\typeout{** the default language instead.}%
\else
\language=\csname l@#1\endcsname
\fi
#2}}
\providecommand{\BIBdecl}{\relax}
\BIBdecl

\bibitem{cisco20}
\BIBentryALTinterwordspacing
``Cisco visual networking index: {F}orecast and trends, 2017–2022,'' Cisco
  White Paper, 2019. [Online]. Available:
  \url{https://twiki.cern.ch/twiki/pub/HEPIX/TechwatchNetwork/HtwNetworkDocuments/white-paper-c11-741490.pdf}
\BIBentrySTDinterwordspacing

\bibitem{xiao2020selflearning}
Y.~Xiao, G.~Shi, Y.~Li, W.~Saad, and H.~V. Poor, ``Towards self-learning edge
  intelligence in {6G},'' \emph{arXiv:2010.00176 [cs.NI]}, 2020.

\bibitem{5G-PPP}
\BIBentryALTinterwordspacing
``European vision for the {6G} network ecosystem,'' 5G-PPP Tech. Report, 2021.
  [Online]. Available:
  \url{https://5g-ppp.eu/european-vision-for-the-6g-network-ecosystem/}
\BIBentrySTDinterwordspacing

\bibitem{7444562}
Z.~{Zhong}, N.~{Hua}, M.~{Tornatore}, Y.~{Li}, H.~{Liu}, C.~{Ma}, Y.~{Li},
  X.~{Zheng}, and B.~{Mukherjee}, ``Energy efficiency and blocking reduction
  for tidal traffic via stateful grooming in {IP}-over-optical networks,''
  \emph{IEEE/OSA J. Opt. Commun. Netw.}, vol.~8, no.~3, pp. 175--189, 2016.

\bibitem{Furdek:21}
M.~Furdek, C.~Natalino, A.~D. Giglio, and M.~Schiano, ``Optical network
  security management: {R}equirements, architecture, and efficient machine
  learning models for detection of evolving threats,'' \emph{IEEE/OSA J. Opt.
  Commun. Netw.}, vol.~13, no.~2, pp. A144--A155, 2021.

\bibitem{Musumeci:19}
F.~Musumeci, C.~Rottondi, G.~Corani, S.~Shahkarami, F.~Cugini, and
  M.~Tornatore, ``A tutorial on machine learning for failure management in
  optical networks,'' \emph{IEEE/OSA J. Light. Technol.}, vol.~37, no.~16, pp.
  4125--4139, 2019.

\bibitem{Pointurier:21}
Y.~Pointurier, ``Machine learning techniques for quality of transmission
  estimation in optical networks,'' \emph{IEEE/OSA J. Opt. Commun. Netw.},
  vol.~13, no.~4, pp. B60--B71, 2021.

\bibitem{9799746}
R.~Ayassi, A.~Triki, N.~Crespi, R.~Minerva, and M.~Laye, ``Survey on the use of
  machine learning for quality of transmission estimation in optical transport
  networks,'' \emph{IEEE/OSA J. Light. Technol.}, pp. 1--13, 2022.

\bibitem{ZHANG2022102804}
L.~Zhang, X.~Li, Y.~Tang, J.~Xin, and S.~Huang, ``A survey on {Q}o{T}
  prediction using machine learning in optical networks,'' \emph{Opt. Fiber
  Technol.}, vol.~68, no. 102804, 2022.

\bibitem{MATA201843}
J.~Mata, I.~{de Miguel}, R.~J. Durán, N.~Merayo, S.~K. Singh, A.~Jukan, and
  M.~Chamania, ``Artificial intelligence ({AI}) methods in optical networks:
  {A} comprehensive survey,'' \emph{Opt. Switch. Netw.}, vol.~28, pp. 43 -- 57,
  2018.

\bibitem{8527529}
F.~Musumeci, C.~Rottondi, A.~Nag, I.~Macaluso, D.~Zibar, M.~Ruffini, and
  M.~Tornatore, ``An overview on application of machine learning techniques in
  optical networks,'' \emph{IEEE Commun. Surv. Tutor.}, vol.~21, no.~2, pp.
  1383--1408, 2019.

\bibitem{rafique}
D.~Rafique and L.~Velasco, ``Machine learning for network automation:
  {O}verview, architecture, and applications,'' \emph{IEEE/OSA J. Opt. Commun.
  Netw.}, vol.~10, no.~10, pp. D126--D143, 2018.

\bibitem{920841}
I.~{Baldine} and G.~N. {Rouskas}, ``Traffic adaptive {WDM} networks: {A} study
  of reconfiguration issues,'' \emph{IEEE/OSA J. Light. Technol.}, vol.~19,
  no.~4, pp. 433--455, 2001.

\bibitem{7273738}
Z.~Dong, F.~N. Khan, Q.~Sui, K.~Zhong, C.~Lu, and A.~P.~T. Lau, ``Optical
  performance monitoring: {A} review of current and future technologies,''
  \emph{IEEE/OSA J. Light. Technol.}, vol.~34, no.~2, pp. 525--543, 2016.

\bibitem{8770528}
L.~Velasco, A.~C. Piat, O.~Gonzlez, A.~Lord, A.~Napoli, P.~Layec, D.~Rafique,
  A.~D'Errico, D.~King, M.~Ruiz, F.~Cugini, and R.~Casellas, ``Monitoring and
  data analytics for optical networking: {B}enefits, architectures, and use
  cases,'' \emph{IEEE Netw.}, vol.~33, no.~6, pp. 100--108, 2019.

\bibitem{6083231}
N.~{Charbonneau} and V.~M. {Vokkarane}, ``A survey of advance reservation
  routing and wavelength assignment in wavelength-routed {WDM} networks,''
  \emph{IEEE Commun. Surv. Tutor.}, vol.~14, no.~4, pp. 1037--1064, 2012.

\bibitem{Zang00areview}
H.~Zang and J.~P. Jue, ``A review of routing and wavelength assignment
  approaches for wavelength-routed optical {WDM} networks,'' \emph{Opt. Netw.
  Mag.}, vol.~1, pp. 47--60, 2000.

\bibitem{7105364}
B.~C. {Chatterjee}, N.~{Sarma}, and E.~{Oki}, ``Routing and spectrum allocation
  in elastic optical networks: {A} tutorial,'' \emph{IEEE Commun. Surv.
  Tutor.}, vol.~17, no.~3, pp. 1776--1800, 2015.

\bibitem{9713679}
N.~D. Cicco, E.~F. Mercan, O.~Karandin, O.~Ayoub, S.~Troia, F.~Musumeci, and
  M.~Tornatore, ``On deep reinforcement learning for static routing and
  wavelength assignment,'' \emph{IEEE J. Sel. Top. Quantum Electron.}, vol.~28,
  no.~4, pp. 1--12, 2022.

\bibitem{9766412}
M.~Zhu, Q.~Chen, J.~Gu, and P.~Gu, ``Deep reinforcement learning for
  provisioning virtualized network function in inter-datacenter elastic optical
  networks,'' \emph{IEEE Trans. Netw. Service Manag.}, pp. 1--1, 2022.

\bibitem{9895124}
R.~Wang, J.~Zhang, Z.~Gu, S.~Yan, Y.~Xiao, and Y.~Ji, ``Edge-enhanced graph
  neural network for {DU}-{CU} placement and lightpath provision in {X}-{H}aul
  networks,'' \emph{IEEE/OSA Journal of Optical Communications and Networking},
  vol.~14, no.~10, pp. 828--839, 2022.

\bibitem{ciena18adaptive}
\BIBentryALTinterwordspacing
G.~{Yigit} and D.~{Cooperson}, ``From autonomous to adaptive: {T}he next
  evolution in networking,'' Ciena White Paper, 2018. [Online]. Available:
  \url{https://www.analysysmason.com/research/content/reports/autonomous-to-adaptive-networks-white-paper/}
\BIBentrySTDinterwordspacing

\bibitem{7503119}
A.~S. Thyagaturu, A.~Mercian, M.~P. McGarry, M.~Reisslein, and W.~Kellerer,
  ``Software defined optical networks ({SDON}s): {A} comprehensive survey,''
  \emph{IEEE Commun. Surv. Tutor.}, vol.~18, no.~4, pp. 2738--2786, 2016.

\bibitem{Bishop06}
C.~Bishop, \emph{Pattern Recognition and Machine Learning}.\hskip 1em plus
  0.5em minus 0.4em\relax Springer, 2006.

\bibitem{7917576}
S.~{Troia}, {Gao Sheng}, R.~{Alvizu}, G.~A. {Maier}, and A.~{Pattavina},
  ``Identification of tidal-traffic patterns in metro-area mobile networks via
  matrix factorization based model,'' in \emph{Proc. IEEE Int. Conf. Pervasive
  Comput. Commun. Wkshps (PerCom Workshops)}, 2017, pp. 297--301.

\bibitem{KLINKOWSKI201858}
M.~Klinkowski, P.~Lechowicz, and K.~Walkowiak, ``Survey of resource allocation
  schemes and algorithms in spectrally-spatially flexible optical networking,''
  \emph{Opt. Switch. Netw.}, vol.~27, pp. 58--78, 2018.

\bibitem{328952}
J.~{Labourdette}, G.~{Hart}, and A.~{Acampora}, ``Branch-exchange sequences for
  reconfiguration of lightwave networks,'' \emph{IEEE Trans. Commun.}, vol.~42,
  no.~10, pp. 2822--2832, 1994.

\bibitem{594381}
K.~Bala, G.~Ellinas, M.~Post, C.-C. Shen, J.~Wei, and N.~Antoniades, ``Towards
  hitless reconfiguration in {WDM} optical networks for {ATM} transport,'' in
  \emph{Proc. IEEE Global Commun. Conf. (GLOBECOM)}, vol.~1, 1996, pp.
  316--320.

\bibitem{1194820}
A.~{Gencata} and B.~{Mukherjee}, ``Virtual-topology adaptation for {WDM} mesh
  networks under dynamic traffic,'' \emph{IEEE/ACM Trans. Netw.}, vol.~11,
  no.~2, pp. 236--247, 2003.

\bibitem{891340}
B.~{Ramamurthy} and A.~{Ramakrishnan}, ``Virtual topology reconfiguration of
  wavelength-routed optical {WDM} networks,'' in \emph{Proc. IEEE Global
  Commun. Conf. (GLOBECOM)}, vol.~2, 2000, pp. 1269--1275.

\bibitem{7308092}
M.~{Zhang}, C.~{You}, and Z.~{Zhu}, ``On the parallelization of spectrum
  defragmentation reconfigurations in elastic optical networks,''
  \emph{IEEE/ACM Trans. Netw.}, vol.~24, no.~5, pp. 2819--2833, 2016.

\bibitem{5272256}
Y.~{Ohsita}, T.~{Miyamura}, S.~{Arakawa}, S.~{Ata}, E.~{Oki}, K.~{Shiomoto},
  and M.~{Murata}, ``Gradually reconfiguring virtual network topologies based
  on estimated traffic matrices,'' \emph{IEEE/ACM Trans. Netw.}, vol.~18,
  no.~1, pp. 177--189, 2010.

\bibitem{Yin:12}
Y.~Yin, K.~Wen, D.~J. Geisler, R.~Liu, and S.~J.~B. Yoo, ``Dynamic on-demand
  defragmentation in flexible bandwidth elastic optical networks,'' \emph{OSA
  Opt. Express}, vol.~20, no.~2, pp. 1798--1804, 2012.

\bibitem{CASTRO20122869}
A.~Castro, L.~Velasco, M.~Ruiz, M.~Klinkowski, J.~P. Fernández-Palacios, and
  D.~Careglio, ``Dynamic routing and spectrum (re)allocation in future flexgrid
  optical networks,'' \emph{Comput. Netw.}, vol.~56, no.~12, pp. 2869 -- 2883,
  2012.

\bibitem{6524878}
R.~{Wang} and B.~{Mukherjee}, ``Provisioning in elastic optical networks with
  non-disruptive defragmentation,'' in \emph{Proc. IEEE Int. Conf. Opt. Netw.
  Des. Model. (ONDM)}, 2013, pp. 223--228.

\bibitem{6532668}
S.~{Shakya} and X.~{Cao}, ``Spectral defragmentation in elastic optical path
  networks using independent sets,'' in \emph{Proc. IEEE/OSA Opt. Fiber Commun.
  Conf. (OFC)}, 2013, pp. 1--3.

\bibitem{8501527}
G.~{Choudhury}, D.~{Lynch}, G.~{Thakur}, and S.~{Tse}, ``Two use cases of
  machine learning for {SDN}-enabled {IP}/optical networks: {T}raffic matrix
  prediction and optical path performance prediction,'' \emph{IEEE/OSA J. Opt.
  Commun. Netw.}, vol.~10, no.~10, pp. D52--D62, 2018.

\bibitem{9204965}
A.~{Bayati}, K.~K. {Nguyen}, and M.~{Cheriet}, ``Gaussian process regression
  ensemble model for network traffic prediction,'' \emph{IEEE Access}, vol.~8,
  pp. 176\,540--176\,554, 2020.

\bibitem{8047676}
R.~{Alvizu}, S.~{Troia}, G.~{Maier}, and A.~{Pattavina}, ``Matheuristic with
  machine-learning-based prediction for software-defined mobile metro-core
  networks,'' \emph{IEEE/OSA J. Opt. Commun. Netw.}, vol.~9, no.~9, pp.
  D19--D30, 2017.

\bibitem{7830260}
F.~{Morales}, M.~{Ruiz}, L.~{Gifre}, L.~M. {Contreras}, V.~{Lopez}, and
  L.~{Velasco}, ``Virtual network topology adaptability based on data analytics
  for traffic prediction,'' \emph{IEEE/OSA J. Opt. Commun. Netw.}, vol.~9,
  no.~1, pp. A35--A45, 2017.

\bibitem{8473978}
S.~{Troia}, R.~{Alvizu}, Y.~{Zhou}, G.~{Maier}, and A.~{Pattavina}, ``Deep
  learning-based traffic prediction for network optimization,'' in \emph{Proc.
  Int. Conf. Transp. Opt. Netw. (ICTON)}, 2018.

\bibitem{8430520}
J.~{Guo} and Z.~{Zhu}, ``When deep learning meets inter-datacenter optical
  network management: Advantages and vulnerabilities,'' \emph{IEEE/OSA J.
  Light. Technol.}, vol.~36, no.~20, pp. 4761--4773, 2018.

\bibitem{8386186}
A.~{Yu}, H.~{Yang}, W.~{Bai}, L.~{He}, H.~{Xiao}, and J.~{Zhang}, ``Leveraging
  deep learning to achieve efficient resource allocation with traffic
  evaluation in datacenter optical networks,'' in \emph{Proc. IEEE/OSA Opt.
  Fiber Commun. Conf. (OFC)}, 2018, pp. 1--3.

\bibitem{8696381}
M.~{Balanici} and S.~{Pachnicke}, ``Machine learning-based traffic prediction
  for optical switching resource allocation in hybrid intra-data center
  networks,'' in \emph{Proc. IEEE/OSA Opt. Fiber Commun. Conf. (OFC)}, 2019,
  pp. 1--3.

\bibitem{Balanici:21}
M.~Balanici and S.~Pachnicke, ``Classification and forecasting of real-time
  server traffic flows employing long short-term memory for hybrid {{E/O}} data
  center networks,'' \emph{IEEE/OSA J. Opt. Commun. Netw.}, vol.~13, no.~5, pp.
  85--93, 2021.

\bibitem{doi:10.1002/dac.4516}
C.~Kyriakopoulos, P.~Nicopolitidis, G.~Papadimitriou, and E.~Varvarigos,
  ``Exploiting {IP}-layer traffic prediction analytics to allocate spectrum
  resources using swarm intelligence,'' \emph{Int. J. Commun. Syst.}, vol.~33,
  pp. 1--14, 2020.

\bibitem{9748600}
T.~Panayiotou and G.~Ellinas, ``Addressing traffic prediction uncertainty in
  multi-period planning optical networks,'' in \emph{Proc. IEEE/OSA Opt. Fiber
  Commun. Conf. (OFC)}, 2022, pp. 1--3.

\bibitem{8845132}
D.~{Andreoletti}, S.~{Troia}, F.~{Musumeci}, S.~{Giordano}, G.~{Maier}, and
  M.~{Tornatore}, ``Network traffic prediction based on diffusion convolutional
  recurrent neural networks,'' in \emph{Proc. IEEE Conf. Comput. Commun. Wkshps
  (INFOCOM WKSHPS)}, 2019, pp. 246--251.

\bibitem{9782838}
S.~K. Singh, C.-Y. Liu, S.~J. Ben~Yoo, and R.~Proietti,
  ``Machine-learning-aided dynamic reconfiguration in optical {DC}/{HPC}
  networks,'' in \emph{Proc. International Conference on Optical Network Design
  and Modeling (ONDM)}, 2022, pp. 1--6.

\bibitem{MARYAM202313}
H.~Maryam, T.~Panayiotou, and G.~Ellinas, ``Uncertainty quantification and
  consideration in {ML}-aided traffic-driven service provisioning,''
  \emph{Computer Communications}, vol. 202, pp. 13--22, 2023.

\bibitem{ZHANG2011171}
Y.~Zhang, M.~Tornatore, P.~Chowdhury, and B.~Mukherjee, ``Energy optimization
  in {IP-over-WDM} networks,'' \emph{Opt. Switch. Netw.}, vol.~8, no.~3, pp.
  171 -- 180, 2011.

\bibitem{5540249}
A.~{Coiro}, M.~{Listanti}, A.~{Valenti}, and F.~{Matera}, ``Reducing power
  consumption in wavelength routed networks by selective switch off of optical
  links,'' \emph{IEEE J. Sel. Top. Quantum Electron.}, vol.~17, no.~2, pp.
  428--436, 2011.

\bibitem{6515884}
E.~{Bonetto}, L.~{Chiaraviglio}, F.~{Idzikowski}, and E.~{Le Rouzic},
  ``Algorithms for the multi-period power-aware logical topology design with
  reconfiguration costs,'' \emph{IEEE/OSA J. Opt. Commun. Netw.}, vol.~5,
  no.~5, pp. 394--410, 2013.

\bibitem{6384633}
A.~{Morea}, O.~{Rival}, N.~{Brochier}, and E.~{Le Rouzic}, ``Datarate
  adaptation for night-time energy savings in core networks,'' \emph{IEEE/OSA
  J. Light. Technol.}, vol.~31, no.~5, pp. 779--785, 2013.

\bibitem{Alvizu2017EnergyED}
R.~Alvizu, X.~Zhao, G.~Maier, Y.~Xu, and A.~Pattavina, ``Energy efficient
  dynamic optical routing for mobile metro-core networks under tidal traffic
  patterns,'' \emph{IEEE/OSA J. Light. Technol.}, vol.~35, no.~2, pp. 325--333,
  2017.

\bibitem{s11227-020-03493-7}
{L.A.J. Mesquita, K.D.R. Assis, and R.C. Almeida}, ``Multi-period traffic on
  elastic optical networks planning: {A}lleviating the capacity crunch,''
  \emph{J. Supercomput.}, vol.~77, pp. 5468--5491, 2021.

\bibitem{6381741}
M.~{Klinkowski}, M.~{Ruiz}, L.~{Velasco}, D.~{Careglio}, V.~{Lopez}, and
  J.~{Comellas}, ``Elastic spectrum allocation for time-varying traffic in
  flexgrid optical networks,'' \emph{IEEE J. Sel. Areas Commun.}, vol.~31,
  no.~1, pp. 26--38, 2013.

\bibitem{6831425}
S.~{Shakya}, Y.~{Wang}, X.~{Cao}, Z.~{Ye}, and C.~{Qiao}, ``Minimize
  sub-carrier reallocation in elastic optical path networks using traffic
  prediction,'' in \emph{Proc. IEEE Global Commun. Conf. (GLOBECOM)}, 2013, pp.
  2352--2357.

\bibitem{8686111}
P.~{Kokkinos}, P.~{Soumplis}, and E.~A. {Varvarigos}, ``Pattern-driven resource
  allocation in optical networks,'' \emph{IEEE Trans. Netw. Service Manag.},
  vol.~16, no.~2, pp. 489--504, 2019.

\bibitem{9322381}
T.~{Panayiotou} and G.~{Ellinas}, ``Fair resource allocation in optical
  networks under tidal traffic,'' in \emph{Proc. IEEE Global Commun. Conf.
  (GLOBECOM)}, 2020.

\bibitem{9347908}
------, ``Optimal and near-optimal alpha-fair resource allocation algorithms
  based on traffic demand predictions for optical network planning,''
  \emph{IEEE/OSA J. Opt. Commun. Netw.}, vol.~13, no.~3, pp. 53--68, 2021.

\bibitem{9522187}
T.~Panayiotou and G.~Ellinas, ``A reward-based fair resource allocation in
  {EONs} considering traffic demand behavior,'' in \emph{Proc. IEEE Int. Conf.
  Comput. Commun. Netw. (ICCCN)}, 2021.

\bibitem{8396130}
T.~{Panayiotou}, K.~{Manousakis}, S.~P. {Chatzis}, and G.~{Ellinas}, ``On
  learning bandwidth allocation models for time-varying traffic in flexible
  optical networks,'' in \emph{Proc. Int. Conf. Opt. Netw. Des. Model. (ONDM)},
  2018, pp. 194--199.

\bibitem{8751369}
T.~{Panayiotou} and G.~{Ellinas}, ``Data-driven bandwidth allocation in
  {EONs},'' in \emph{Proc. Photon. Switch. Comput. (PSC)}, 2018, pp. 1--3.

\bibitem{8620207}
T.~{Panayiotou}, K.~{Manousakis}, S.~P. {Chatzis}, and G.~{Ellinas}, ``A
  data-driven bandwidth allocation framework with {QoS} considerations for
  {EONs},'' \emph{IEEE/OSA J. Light. Technol.}, vol.~37, no.~9, pp. 1853--1864,
  2019.

\bibitem{9256655}
A.~{Valkanis}, G.~A. {Beletsioti}, P.~{Nicopolitidis}, G.~{Papadimitriou}, and
  E.~{Varvarigos}, ``Reinforcement learning in traffic prediction of core
  optical networks using learning automata,'' in \emph{Proc. Int. Conf.
  Commun., Comput., Cybersecur., Inform. (CCCI)}, 2020, pp. 1--6.

\bibitem{7490359}
W.~{Fang}, M.~{Zeng}, X.~{Liu}, W.~{Lu}, and Z.~{Zhu}, ``Joint spectrum and
  {IT} resource allocation for efficient {VNF} service chaining in
  inter-datacenter elastic optical networks,'' \emph{IEEE Commun. Lett.},
  vol.~20, no.~8, pp. 1539--1542, 2016.

\bibitem{7293303}
P.~Lu, L.~Zhang, X.~Liu, J.~Yao, and Z.~Zhu, ``Highly efficient data migration
  and backup for big data applications in elastic optical inter-data-center
  networks,'' \emph{IEEE Netw.}, vol.~29, no.~5, pp. 36--42, 2015.

\bibitem{8501524}
S.~K. {Singh} and A.~{Jukan}, ``Machine-learning-based prediction for resource
  (re)allocation in optical data center networks,'' \emph{IEEE/OSA J. Opt.
  Commun. Netw.}, vol.~10, no.~10, pp. D12--D28, 2018.

\bibitem{XIONG201999}
Y.~Xiong, Y.~Yang, Y.~Ye, and G.~N. Rouskas, ``A machine learning approach to
  mitigating fragmentation and crosstalk in space division multiplexing elastic
  optical networks,'' \emph{Opt. Fiber Technol.}, vol.~50, pp. 99--107, 2019.

\bibitem{8737631}
K.~{Lei}, M.~{Qin}, B.~{Bai}, G.~{Zhang}, and M.~{Yang}, ``{GCN-GAN: A}
  non-linear temporal link prediction model for weighted dynamic networks,'' in
  \emph{Proc. IEEE Conf. Comput. Commun. (INFOCOM)}, 2019, pp. 388--396.

\bibitem{Zhao:18}
Y.~Zhao, B.~Yan, D.~Liu, Y.~He, D.~Wang, and J.~Zhang, ``{SOON:
  S}elf-optimizing optical networks with machine learning,'' \emph{OSA Opt.
  Express}, vol.~26, no.~22, pp. 28\,713--28\,726, 2018.

\bibitem{9203477}
C.~{Vinchoff}, N.~{Chung}, T.~{Gordon}, L.~{Lyford}, and M.~{Aibin}, ``Traffic
  prediction in optical networks using graph convolutional generative
  adversarial networks,'' in \emph{Proc. Int. Conf. Transp. Opt. Netw.
  (ICTON)}, 2020.

\bibitem{8436062}
X.~{Chen}, R.~{Proietti}, H.~{Lu}, A.~{Castro}, and S.~J.~B. {Yoo},
  ``Knowledge-based autonomous service provisioning in multi-domain elastic
  optical networks,'' \emph{IEEE Commun. Mag.}, vol.~56, no.~8, pp. 152--158,
  2018.

\bibitem{9042293}
B.~{Yan}, Y.~{Zhao}, W.~{Chen}, and J.~{Zhang}, ``Area-aware routing and
  spectrum allocation for the tidal traffic pattern in metro optical
  networks,'' \emph{IEEE Access}, vol.~8, pp. 56\,501--56\,509, 2020.

\bibitem{8738827}
X.~{Chen}, B.~{Li}, R.~{Proietti}, H.~{Lu}, Z.~{Zhu}, and S.~J.~B. {Yoo},
  ``{DeepRMSA: A} deep reinforcement learning framework for routing, modulation
  and spectrum assignment in elastic optical networks,'' \emph{IEEE/OSA J.
  Light. Technol.}, vol.~37, no.~16, pp. 4155--4163, 2019.

\bibitem{8847548}
J.~{Suarez-Varela}, A.~{Mestres}, J.~{Yu}, L.~{Kuang}, H.~{Feng},
  A.~{Cabellos-Aparicio}, and P.~{Barlet-Ros}, ``Routing in optical transport
  networks with deep reinforcement learning,'' \emph{IEEE/OSA J. Opt. Commun.
  Netw.}, vol.~11, no.~11, pp. 547--558, 2019.

\bibitem{3229554}
S.~Salman, C.~Streiffer, H.~Chen, T.~Benson, and A.~Kadav, ``{DeepConf:
  A}utomating data center network topologies management with machine
  learning,'' in \emph{Proc. ACM Wkshp on Network Meets {AI} \& {ML}}, 2018.

\bibitem{8485853}
Z.~{Xu}, J.~{Tang}, J.~{Meng}, W.~{Zhang}, Y.~{Wang}, C.~H. {Liu}, and
  D.~{Yang}, ``Experience-driven networking: {A} deep reinforcement learning
  based approach,'' in \emph{Proc. IEEE Conf. Comput. Commun. (INFOCOM)}, 2018,
  pp. 1871--1879.

\bibitem{9083336}
B.~{Li} and Z.~{Zhu}, ``{DeepCoop}: Leveraging cooperative {DRL} agents to
  achieve scalable network automation for multi-domain {SD-EONs},'' in
  \emph{Proc. IEEE/OSA Opt. Fiber Commun. Conf. (OFC)}, 2020, pp. 1--3.

\bibitem{9375634}
X.~{Guo}, F.~{Yan}, X.~{Xue}, B.~{Pan}, G.~{Exarchakos}, and N.~{Calabretta},
  ``{QoS}-aware data center network reconfiguration method based on deep
  reinforcement learning,'' \emph{IEEE/OSA J. Opt. Commun. Netw.}, vol.~13,
  no.~5, pp. 94--107, 2021.

\bibitem{9373585}
X.~{Chen}, R.~{Proietti}, C.~Y. {Liu}, and S.~J.~B. {Yoo}, ``A
  multi-task-learning-based transfer deep reinforcement learning design for
  autonomic optical networks,'' \emph{IEEE J. Sel. Areas Commun.}, vol.~39,
  no.~9, pp. 2878--2889, 2021.

\bibitem{SUN2021107891}
P.~Sun, Z.~Guo, J.~Lan, J.~Li, Y.~Hu, and T.~Baker, ``{ScaleDRL: A} scalable
  deep reinforcement learning approach for traffic engineering in {SDN} with
  pinning control,'' \emph{Comput. Netw.}, vol. 190, no. 107891, 2021.

\bibitem{Zhao:21}
Z.~Zhao, Y.~Zhao, Y.~Li, F.~Wang, X.~Li, D.~Han, and J.~Zhang, ``Service
  restoration in multi-modal optical transport networks with reinforcement
  learning,'' \emph{OSA Opt. Express}, vol.~29, no.~3, pp. 3825--3840, 2021.

\bibitem{Aibin2020MonteCT}
M.~Aibin and K.~Walkowiak, ``Monte {C}arlo tree search with last-good-reply
  policy for cognitive optimization of cloud-ready optical networks,'' \emph{J.
  Netw. Syst. Manag.}, vol.~28, pp. 1722--1744, 2020.

\bibitem{9507559}
B.~Li, R.~Zhang, X.~Tian, and Z.~Zhu, ``Multi-agent and cooperative deep
  reinforcement learning for scalable network automation in mult-domain
  {SD-EONs},'' \emph{IEEE Trans. Netw. Service Manag.}, vol.~18, no.~4, pp.
  4801--4813, 2021.

\bibitem{9609608}
B.~Li and Z.~Zhu, ``{GNN}-based hierarchical deep reinforcement learning for
  {NFV}-oriented online resource orchestration in elastic optical {DCIs},''
  \emph{IEEE/OSA J. Light. Technol.}, vol.~40, no.~4, pp. 935--946, 2022.

\bibitem{9748476}
Z.~Chen, J.~Zhang, B.~Zhang, R.~Wang, H.~Ma, and Y.~Ji, ``{ADMIRE:
  D}emonstration of collaborative data-driven and model-driven intelligent
  routing engine for {IP}/optical cross-layer optimization in {X}-haul
  networks,'' in \emph{Proc. IEEE/OSA Opt. Fiber Commun. Conf. (OFC)}, 2022,
  pp. 1--3.

\bibitem{9279336}
S.~Troia, F.~Sapienza, L.~Varé, and G.~Maier, ``On deep reinforcement learning
  for traffic engineering in {SD-WAN},'' \emph{IEEE J. Sel. Areas Commun.},
  vol.~39, no.~7, pp. 2198--2212, 2021.

\bibitem{9779080}
L.~Xu, Y.-C. Huang, Y.~Xue, and X.~Hu, ``Deep reinforcement learning-based
  routing and spectrum assignment of {EON}s by exploiting {GCN} and {RNN} for
  feature extraction,'' \emph{Journal of Lightwave Technology}, vol.~40,
  no.~15, pp. 4945--4955, 2022.

\bibitem{KHOSHKHOLGHI2022109451}
M.~A. Khoshkholghi and T.~Mahmoodi, ``Edge intelligence for service function
  chain deployment in {NFV}-enabled networks,'' \emph{Computer Networks}, vol.
  219, no. 109451, 2022.

\bibitem{5948951}
Y.~Wang and X.~Cao, ``Multi-granular optical switching: {A} classified overview
  for the past and future,'' \emph{IEEE Commun. Surv. Tutor.}, vol.~14, no.~3,
  pp. 698--713, 2012.

\bibitem{SANG2002329}
A.~{Sang} and S.~{ Li}, ``A predictability analysis of network traffic,''
  \emph{Comput. Netw.}, vol.~39, no.~4, pp. 329--345, 2002.

\bibitem{6381046}
X.~{Zhou}, Z.~{Zhao}, R.~{Li}, {Yifan Zhou}, and H.~{Zhang}, ``The
  predictability of cellular networks traffic,'' in \emph{Proc. Int. Symp.
  Commun. Inform. Technol. (ISCIT)}, 2012, pp. 973--978.

\bibitem{DBLP:journals/corr/abs-1302-6613}
R.~Adhikari and R.~K. Agrawal, ``An introductory study on time series modeling
  and forecasting,'' \emph{arXiv:1302.6613 [cs.LG]}, 2013.

\bibitem{1510713}
K.~{Papagiannaki}, N.~{Taft}, {Zhi-Li Zhang}, and C.~{Diot}, ``Long-term
  forecasting of {I}nternet backbone traffic,'' \emph{IEEE Trans. Neural
  Netw.}, vol.~16, no.~5, pp. 1110--1124, 2005.

\bibitem{10.1007/978-3-540-31956-6_58}
G.~{Mao}, ``Real-time network traffic prediction based on a multiscale
  decomposition,'' in \emph{Proc. Int. Conf. Netw. (ICN)}, 2005, pp. 492--499.

\bibitem{10.1007/s12544-015-0170-8}
S.~{Kumar} and L.~{Vanajakshi}, ``Short-term traffic flow prediction using
  seasonal {ARIMA} model with limited input data,'' \emph{Eur. Transp. Res.
  Rev.}, vol.~7, no.~21, pp. 1--9, 2015.

\bibitem{6887148}
D.~{Siracusa}, A.~{Francescon}, N.~{Fernández}, I.~{de Miguel}, R.~J.
  {Durán}, J.~C. {Aguado}, and E.~{Salvadori}, ``Experimental evaluation of
  virtual topology design and reconfiguration in optical networks by means of
  cognition,'' in \emph{Proc. IEEE/OSA Opt. Fiber Commun. Conf. (OFC)}, 2014,
  pp. 1--3.

\bibitem{7023486}
N.~{Fernández}, R.~J.~D. {Barroso}, D.~{Siracusa}, A.~{Francescon}, I.~{de
  Miguel}, E.~{Salvadori}, J.~C. {Aguado}, and R.~M. {Lorenzo}, ``Virtual
  topology reconfiguration in optical networks by means of cognition:
  {E}valuation and experimental validation,'' \emph{IEEE/OSA J. Opt. Commun.
  Netw.}, vol.~7, no.~1, pp. A162--A173, 2015.

\bibitem{DEGOOIJER2006443}
J.~{De Gooijer} and R.~Hyndman, ``25 years of time series forecasting,''
  \emph{Int. J. Forecast.}, vol.~22, no.~3, pp. 443--473, 2006.

\bibitem{ZHANG2003159}
G.~Zhang, ``Time series forecasting using a hybrid {ARIMA} and neural network
  model,'' \emph{Neurocomputing}, vol.~50, pp. 159--175, 2003.

\bibitem{6047849}
M.~{Barabas}, G.~{Boanea}, A.~B. {Rus}, V.~{Dobrota}, and J.~{Domingo-Pascual},
  ``Evaluation of network traffic prediction based on neural networks with
  multi-task learning and multiresolution decomposition,'' in \emph{Proc. IEEE
  Int. Conf. Intell. Comput. Commun. Process.}, 2011, pp. 95--102.

\bibitem{1544219}
{H. Feng} and {Y. Shu}, ``Study on network traffic prediction techniques,'' in
  \emph{Proc. Int. Conf. Wireless Commun., Netw. Mob. Comput.}, vol.~2, 2005,
  pp. 1041--1044.

\bibitem{8530608}
R.~{Madan} and P.~S. {Mangipudi}, ``Predicting computer network traffic: {A}
  time series forecasting approach using {DWT, ARIMA and RNN},'' in \emph{Proc.
  Int. Conf. Contemporary Comput. (IC3)}, 2018, pp. 1--5.

\bibitem{9012669}
A.~{Lazaris} and V.~K. {Prasanna}, ``Deep learning models for aggregated
  network traffic prediction,'' in \emph{Proc. Int. Conf. Netw. Serv. Manag.
  (CNSM)}, 2019, pp. 1--5.

\bibitem{kingma2017adam}
D.~P. Kingma and J.~Ba, ``Adam: A method for stochastic optimization,''
  \emph{arXiv:1412.6980 [cs.LG]}, 2017.

\bibitem{9782850}
A.~Knapi\'{n}ska, K.~P\'{o}{\l}torak, D.~Por\k{e}ba, J.~Miszczyk, M.~Daniluk,
  and K.~Walkowiak, ``On feature selection in short-term prediction of backbone
  optical network traffic,'' in \emph{Proc. International Conference on Optical
  Network Design and Modeling (ONDM)}, 2022, pp. 1--6.

\bibitem{804564}
A.~{Varga}, ``{Using the OMNeT++ discrete event simulation system in
  education},'' \emph{IEEE Trans. Educ.}, vol.~42, no.~4, 1999.

\bibitem{Abilene_dataset}
\BIBentryALTinterwordspacing
Abilene dataset. [Online]. Available: \url{http://sndlib.zib.de/home.action}
\BIBentrySTDinterwordspacing

\bibitem{totem_project}
\BIBentryALTinterwordspacing
``The {Totem} project. {GEANT} traffic matrices,'' Tech. Rep., 2006. [Online].
  Available: \url{https://totem.info.ucl.ac.be/dataset.html}
\BIBentrySTDinterwordspacing

\bibitem{CAIDA}
\BIBentryALTinterwordspacing
{CAIDA}. [Online]. Available: \url{http://www.caida.org/home}
\BIBentrySTDinterwordspacing

\bibitem{telus_fibre}
\BIBentryALTinterwordspacing
``{TELUS PureFibre} areas,'' Tech. Rep. [Online]. Available:
  \url{https://www.telus.com/en/internet/fibre/new/areas}
\BIBentrySTDinterwordspacing

\bibitem{datamarket}
\BIBentryALTinterwordspacing
``{DataMarket},'' Tech. Rep., 2020. [Online]. Available:
  \url{https://data.is/TSDLdemo/}
\BIBentrySTDinterwordspacing

\bibitem{Waikato}
\BIBentryALTinterwordspacing
{Waikato VIII}. [Online]. Available: \url{http://wand.net.nz/wits/ waikato/8/}
\BIBentrySTDinterwordspacing

\bibitem{Titalia}
\BIBentryALTinterwordspacing
{Telecom Italia}. A multi-source dataset of urban life in the city of {M}ilan
  and the province of {T}rentino dataverse. [Online]. Available:
  \url{https://dataverse.harvard.edu/dataset.xhtml?persistentId=doi:10.7910/DVN/EGZHFV}
\BIBentrySTDinterwordspacing

\bibitem{ENAiKOON16}
\BIBentryALTinterwordspacing
ENAiKOON. (2016) {Open Cell ID}. [Online]. Available:
  \url{http://opencellid.org/}
\BIBentrySTDinterwordspacing

\bibitem{crawdaddatasets}
\BIBentryALTinterwordspacing
{CRAWDAD} datasets. [Online]. Available: \url{https://crawdad.org}
\BIBentrySTDinterwordspacing

\bibitem{OrlowskiPioroTomaszewskiWessaely2010}
S.~Orlowski, M.~Pi{\'o}ro, A.~Tomaszewski, and R.~Wess{\"a}ly,
  ``\BIBforeignlanguage{English}{{SNDlib} 1.0--{S}urvivable {N}etwork {D}esign
  {L}ibrary},'' \emph{\BIBforeignlanguage{English}{Networks}}, vol.~55, no.~3,
  pp. 276--286, 2010.

\bibitem{8352941}
J.~{Zhao}, H.~{Qu}, J.~{Zhao}, and D.~{Jiang}, ``Towards traffic matrix
  prediction with {LSTM} recurrent neural networks,'' \emph{Electron. Lett.},
  vol.~54, no.~9, pp. 566--568, 2018.

\bibitem{LI2020102258}
M.~Li, Y.~Wang, Z.~Wang, and H.~Zheng, ``A deep learning method based on an
  attention mechanism for wireless network traffic prediction,'' \emph{Ad Hoc
  Netw.}, vol. 107, no. 102258, 2020.

\bibitem{vaswani2017attention}
A.~Vaswani, N.~Shazeer, N.~Parmar, J.~Uszkoreit, L.~Jones, A.~N. Gomez,
  L.~Kaiser, and I.~Polosukhin, ``Attention is all you need,''
  \emph{arXiv:1706.03762 [cs.CL]}, 2017.

\bibitem{bahdanau2016neural}
D.~Bahdanau, K.~Cho, and Y.~Bengio, ``Neural machine translation by jointly
  learning to align and translate,'' \emph{arXiv:1409.0473 [cs.CL]}, 2016.

\bibitem{8126198}
R.~{Vinayakumar}, K.~P. {Soman}, and P.~{Poornachandran}, ``Applying deep
  learning approaches for network traffic prediction,'' in \emph{Proc. Int.
  Conf. Adv. Comput. Commun. Inform. (ICACCI)}, 2017, pp. 2353--2358.

\bibitem{7785324}
P.~{Poupart}, Z.~{Chen}, P.~{Jaini}, F.~{Fung}, H.~{Susanto}, Y.~{Geng},
  L.~{Chen}, K.~{Chen}, and H.~{Jin}, ``Online flow size prediction for
  improved network routing,'' in \emph{Proc. IEEE Int. Conf. Netw. Protocols
  (ICNP)}, 2016, pp. 1--6.

\bibitem{2016220}
H.~Zhou, L.~Tan, Q.~Zeng, and C.~Wu, ``Traffic matrix estimation: {A} neural
  network approach with extended input and expectation maximization
  iteration,'' \emph{J. Netw. Comput. Appl.}, vol.~60, pp. 220--232, 2016.

\bibitem{JIANG201175}
D.~Jiang, X.~Wang, L.~Guo, H.~Ni, and Z.~Chen, ``Accurate estimation of
  large-scale {IP} traffic matrix,'' \emph{AEU - Int. J. Electron. Commun.},
  vol.~65, no.~1, pp. 75--86, 2011.

\bibitem{201902004}
S.~Wang, Q.~Zhuo, H.~Yan, Q.~L. Qianmu, and Q.~Yong, ``A network traffic
  prediction method based on {LSTM},'' \emph{ZTE Commun.}, vol.~17, no.~2, pp.
  19--25, 2019.

\bibitem{8553653}
I.~{Alawe}, A.~{Ksentini}, Y.~{Hadjadj-Aoul}, and P.~{Bertin}, ``Improving
  traffic forecasting for {5G} core network scalability: {A} machine learning
  approach,'' \emph{IEEE Netw.}, vol.~32, no.~6, pp. 42--49, 2018.

\bibitem{0191939}
A.~Mozo, B.~Ordozgoiti, and S.~Gómez-Canaval, ``Forecasting short-term data
  center network traffic load with convolutional neural networks,'' \emph{PLOS
  ONE}, vol.~13, no.~2, pp. 1--31, 2018.

\bibitem{Huo2019}
L.~Huo, D.~Jiang, S.~Qi, H.~Song, and L.~Miao, ``An {AI}-based adaptive
  cognitive modeling and measurement method of network traffic for {EIS},''
  \emph{Mob. Netw. Appl.}, vol.~26, pp. 575--585, 2021.

\bibitem{13006}
O.~Barut, Y.~Luo, T.~Zhang, W.~Li, and P.~Li, ``{NetML}: {A} challenge for
  network traffic analytics,'' \emph{arXiv:2004.13006 [cs.CR]}, 2020.

\bibitem{8386209}
W.~{Mo}, C.~L. {Gutterman}, Y.~{Li}, G.~{Zussman}, and D.~C. {Kilper}, ``Deep
  neural network based dynamic resource reallocation of {BBU} pools in {5G
  C-RAN ROADM} networks,'' in \emph{Proc. IEEE/OSA Opt. Fiber Commun. Conf.
  (OFC)}, 2018, pp. 1--3.

\bibitem{8406199}
A.~{Azzouni} and G.~{Pujolle}, ``Neu{TM}: A neural network-based framework for
  traffic matrix prediction in {SDN},'' in \emph{Proc. IEEE/IFIP Netw. Oper.
  Manag. Symp. (NOMS)}, 2018, pp. 1--5.

\bibitem{azzouni2017long}
A.~Azzouni and G.~Pujolle, ``A long short-term memory recurrent neural network
  framework for network traffic matrix prediction,'' \emph{arXiv:1705.05690
  [cs.NI]}, 2017.

\bibitem{9163001}
K.~{Gao}, D.~{Li}, L.~{Chen}, J.~{Geng}, F.~{Gui}, Y.~{Cheng}, and Y.~{Gu},
  ``Predicting traffic demand matrix by considering inter-flow correlations,''
  in \emph{Proc. IEEE Conf. Comput. Commun. Wkshps (INFOCOM WKSHPS)}, 2020, pp.
  165--170.

\bibitem{NIE201616}
L.~Nie, D.~Jiang, L.~Guo, and S.~Yu, ``Traffic matrix prediction and estimation
  based on deep learning in large-scale {IP} backbone networks,'' \emph{J.
  Netw. Comput. Appl.}, vol.~76, pp. 16--22, 2016.

\bibitem{7069393}
W.~{Yoo} and A.~{Sim}, ``Network bandwidth utilization forecast model on high
  bandwidth networks,'' in \emph{Proc. Int. Conf. Comput., Netw. Commun.
  (ICNC)}, 2015, pp. 494--498.

\bibitem{5693766}
Y.~{Wei}, J.~{Wang}, and C.~{Wang}, ``A traffic prediction based bandwidth
  management algorithm of a future {I}nternet architecture,'' in \emph{Proc.
  Int. Conf. Intell. Netw. Intell. Systems}, 2010, pp. 560--563.

\bibitem{8757058}
H.~{Xia}, X.~{Wei}, Y.~{Gao}, and H.~{Lv}, ``Traffic prediction based on
  ensemble machine learning strategies with bagging and light{GBM},'' in
  \emph{Proc. IEEE Int. Conf. Commun. Wkshps (ICC Wkshps)}, 2019, pp. 1--6.

\bibitem{20030603}
X.~{Zheng}, W.~{Lai}, H.~{Chen}, and S.~{Fang}, ``Data prediction of mobile
  network traffic in public scenes by {SOS}-v{SVR} method,'' \emph{Sensors},
  vol.~20, no.~3, 2020.

\bibitem{8254808}
Y.~{Xu}, W.~{Xu}, F.~{Yin}, J.~{Lin}, and S.~{Cui}, ``High-accuracy wireless
  traffic prediction: {A GP}-based machine learning approach,'' in \emph{Proc.
  IEEE Global Commun. Conf. (GLOBECOM)}, 2017, pp. 1--6.

\bibitem{PhysRevLett.59.2229}
F.~J. Pineda, ``Generalization of back-propagation to recurrent neural
  networks,'' \emph{Phys. Rev. Lett.}, vol.~59, no.~19, pp. 2229--2232, 1987.

\bibitem{10.1142/S0218488598000094}
S.~Hochreiter, ``The vanishing gradient problem during learning recurrent
  neural nets and problem solutions,'' \emph{Int. J. Uncertain. Fuzziness
  Knowlege-Based Syst.}, vol.~6, no.~2, pp. 107--116, 1998.

\bibitem{10.1162/neco.1997.9.8.1735}
S.~Hochreiter and J.~Schmidhuber, ``Long short-term memory,'' \emph{Neural
  Comput.}, vol.~9, no.~8, pp. 1735--1780, 1997.

\bibitem{https://doi.org/10.48550/arxiv.1412.3555}
J.~Chung, C.~Gulcehre, K.~Cho, and Y.~Bengio, ``Empirical evaluation of gated
  recurrent neural networks on sequence modeling,'' \emph{arXiv:1412.3555
  [cs.NE]}, 2014.

\bibitem{cho2014learning}
K.~Cho, B.~van Merrienboer, C.~Gulcehre, D.~Bahdanau, F.~Bougares, H.~Schwenk,
  and Y.~Bengio, ``Learning phrase representations using {RNN} encoder-decoder
  for statistical machine translation,'' \emph{arXiv:1406.1078 [cs.CL]}, 2014.

\bibitem{sutskever2014sequence}
I.~Sutskever, O.~Vinyals, and Q.~V. Le, ``Sequence to sequence learning with
  neural networks,'' \emph{arXiv:1409.3215 [cs.CL]}, 2014.

\bibitem{10.1145/3065386}
A.~Krizhevsky, I.~Sutskever, and G.~E. Hinton, ``{ImageNet} classification with
  deep convolutional neural networks,'' \emph{Commun. ACM}, vol.~60, no.~6, pp.
  84--90, 2017.

\bibitem{7178838}
T.~N. {Sainath}, O.~{Vinyals}, A.~{Senior}, and H.~{Sak}, ``Convolutional, long
  short-term memory, fully connected deep neural networks,'' in \emph{Proc.
  IEEE Int. Conf. Acoust. Speech Signal Process. (ICASSP)}, 2015.

\bibitem{vel2018graph}
P.~Veličković, G.~Cucurull, A.~Casanova, A.~Romero, P.~Liò, and Y.~Bengio,
  ``Graph attention networks,'' \emph{arXiv:1710.10903 [stat.ML]}, 2018.

\bibitem{10.1007/978-3-319-93417-4_38}
M.~Schlichtkrull, T.~N. Kipf, P.~Bloem, R.~van~den Berg, I.~Titov, and
  M.~Welling, ``Modeling relational data with graph convolutional networks,''
  in \emph{Proc. Eur. Semantic Web Conf. (ESWC)}, 2018, pp. 593--607.

\bibitem{4700287}
F.~{Scarselli}, M.~{Gori}, A.~C. {Tsoi}, M.~{Hagenbuchner}, and
  G.~{Monfardini}, ``The graph neural network model,'' \emph{IEEE Trans. Neural
  Netw.}, vol.~20, no.~1, pp. 61--80, 2009.

\bibitem{s40649-019-0069-y}
S.~Zhang, H.~Tong, J.~Xu, and R.~Maciejewski, ``Graph convolutional networks:
  {A} comprehensive review,'' \emph{Comput. Soc. Netw.}, vol.~6, no.~11, pp.
  1--23, 2019.

\bibitem{theodoridis_book}
S.~Theodoridis, \emph{Machine Learning: {A} {B}ayesian and Optimization
  Perspective}.\hskip 1em plus 0.5em minus 0.4em\relax Academic Press, 2020.

\bibitem{3569}
C.~Rasmussen and C.~Williams, \emph{Gaussian Processes for Machine
  Learning}.\hskip 1em plus 0.5em minus 0.4em\relax MIT Press, 2006.

\bibitem{5966349}
S.~P. {Chatzis} and Y.~{Demiris}, ``Echo state {G}aussian process,'' \emph{IEEE
  Trans. Neural Netw.}, vol.~22, no.~9, pp. 1435--1445, 2011.

\bibitem{10.1145/3409383}
H.~Wang and D.-Y. Yeung, ``A survey on {B}ayesian deep learning,'' \emph{ACM
  Comput. Surv.}, vol.~53, no.~5, pp. 1--37, 2020.

\bibitem{MaryamMed22}
H.~Maryam, T.~Panayiotou, and G.~Ellinas, ``Representing uncertainty in deep
  {Q}o{T} models,'' in \emph{Proc. IEEE Mediterranean Commun. Comput. Netw.
  Conf. (MedComNet)}, 2022, pp. 113--121.

\bibitem{fortunato2019bayesian}
M.~Fortunato, C.~Blundell, and O.~Vinyals, ``Bayesian recurrent neural
  networks,'' \emph{arXiv:1704.02798 [cs.LG]}, 2019.

\bibitem{damianou2013deep}
A.~C. Damianou and N.~D. Lawrence, ``Deep {G}aussian processes,''
  \emph{arXiv:1211.0358 [stat.ML]}, 2013.

\bibitem{salimbeni2017doubly}
H.~Salimbeni and M.~Deisenroth, ``Doubly stochastic variational inference for
  deep {G}aussian processes,'' \emph{arXiv:1705.08933 [stat.ML]}, 2017.

\bibitem{lee2018deep}
J.~Lee, Y.~Bahri, R.~Novak, S.~S. Schoenholz, J.~Pennington, and
  J.~Sohl-Dickstein, ``Deep neural networks as {G}aussian processes,''
  \emph{arXiv:1711.00165 [stat.ML]}, 2018.

\bibitem{gal2016dropout}
Y.~Gal and Z.~Ghahramani, ``Dropout as a {B}ayesian approximation:
  {R}epresenting model uncertainty in deep learning,'' \emph{arXiv:1506.02142
  [stat.ML]}, 2016.

\bibitem{8714026}
N.~C. {Luong}, D.~T. {Hoang}, S.~{Gong}, D.~{Niyato}, P.~{Wang}, Y.~{Liang},
  and D.~I. {Kim}, ``Applications of deep reinforcement learning in
  communications and networking: {A} survey,'' \emph{IEEE Commun. Surv.
  Tutor.}, vol.~21, no.~4, pp. 3133--3174, 2019.

\bibitem{9403369}
W.~Chen, X.~Qiu, T.~Cai, H.-N. Dai, Z.~Zheng, and Y.~Zhang, ``Deep
  reinforcement learning for internet of things: {A} comprehensive survey,''
  \emph{IEEE Commun. Surv. Tutor.}, vol.~23, no.~3, pp. 1659--1692, 2021.

\bibitem{nature14539}
Y.~{LeCun}, Y.~{Bengio}, and G.~{Hinton}, ``Deep learning,'' \emph{Nature},
  vol. 521, pp. 436--444, 2015.

\bibitem{nature24270}
D.~{Silver}, J.~{Schrittwieser}, K.~{Simonyan}, I.~{Antonoglou}, A.~{Huang},
  A.~{Guez}, T.~{Hubert}, L.~{Baker}, M.~{Lai}, A.~{Bolton}, Y.~{Chen},
  T.~{Lillicrap}, F.~{Hui}, L.~{Sifre}, G.~{van den Driessche}, T.~{Graepel},
  and D.~{Hassabis}, ``Mastering the game of {GO} without human knowledge,''
  \emph{Nature}, vol. 550, pp. 354--359, 2017.

\bibitem{s41586-019-1724-z}
O.~{Vinyals}, I.~{Babuschkin}, W.~{Czarnecki} \emph{et~al.}, ``Grandmaster
  level in {StarCraft II} using multi-agent reinforcement learning,''
  \emph{Nature}, vol. 575, pp. 350--354, 2019.

\bibitem{8644190}
B.~{Yan}, Y.~{Zhao}, Y.~{Li}, X.~{Yu}, J.~{Zhang}, Y.~{Wang}, L.~{Yan}, and
  S.~{Rahman}, ``Actor-critic-based resource allocation for multi-modal optical
  networks,'' in \emph{Proc. IEEE Global Commun. Conf. Wkshps (GC Wkshps)},
  2018, pp. 1--6.

\bibitem{9308056}
J.~{Li}, L.~{Chen}, and J.~{Chen}, ``Enabling technologies for low-latency
  service migration in {5G} transport networks,'' \emph{IEEE/OSA J. Opt.
  Commun. Netw.}, vol.~13, no.~2, pp. A200--A210, 2021.

\bibitem{almasan2020deep}
P.~Almasan, J.~Suárez-Varela, A.~Badia-Sampera, K.~Rusek, P.~Barlet-Ros, and
  A.~Cabellos-Aparicio, ``Deep reinforcement learning meets graph neural
  networks: {E}xploring a routing optimization use case,''
  \emph{arXiv:1910.07421 [cs.NI]}, 2020.

\bibitem{BF00992698}
C.~J. C.~H. Watkins and P.~Dayan, ``Q-learning,'' \emph{Mach. Learning},
  vol.~8, no. 3-4, pp. 279--29, 1992.

\bibitem{nature14236}
V.~{Mnih}, K.~{Kavukcuoglu}, D.~{Silver}, A.~{Rusu}, J.~{Veness},
  M.~{Bellemare}, A.~{Graves}, M.~{Riedmiller}, A.~{Fidjeland}, G.~{Ostrovski},
  S.~{Petersen}, C.~{Beattie}, A.~{Sadik}, I.~{Antonoglou}, H.~{King},
  D.~{Kumaran}, D.~{Wierstra}, S.~{Legg}, and D.~{Hassabis}, ``Human-level
  control through deep reinforcement learning,'' \emph{Nature}, vol. 518, pp.
  529--533, 2015.

\bibitem{Williams92}
R.~Williams, ``Simple statistical gradient-following algorithms for
  connectionist reinforcement learning,'' \emph{Mach. Learn.}, vol.~8, pp.
  229--256, 1992.

\bibitem{8186816}
M.~P. Deisenroth, G.~Neumann, and J.~Peters, \emph{A Survey on Policy Search
  for Robotics}.\hskip 1em plus 0.5em minus 0.4em\relax Now Publishers, 2013.

\bibitem{Sutton18}
R.~S. Sutton and A.~G. Barto, \emph{Reinforcement Learning: {A}n
  Introduction}.\hskip 1em plus 0.5em minus 0.4em\relax MIT Press, 2018.

\bibitem{8103164}
K.~{Arulkumaran}, M.~P. {Deisenroth}, M.~{Brundage}, and A.~A. {Bharath},
  ``Deep reinforcement learning: {A} brief survey,'' \emph{IEEE Signal Process.
  Mag.}, vol.~34, no.~6, pp. 26--38, 2017.

\bibitem{Henderson_2018}
P.~Henderson, R.~Islam, P.~Bachman, J.~Pineau, D.~Precup, and D.~Meger, ``Deep
  reinforcement learning that matters,'' in \emph{Proc. AAAI Conf. Artif.
  Intell.}, 2018.

\bibitem{pmlr-v80-yang18d}
Y.~Yang, R.~Luo, M.~Li, M.~Zhou, W.~Zhang, and J.~Wang, ``Mean field
  multi-agent reinforcement learning,'' in \emph{Proc. Int. Conf. Mach. Learn.
  (ICML)}, vol.~80, 2018, pp. 5571--5580.

\bibitem{MARYAM2022108992}
H.~Maryam, T.~Panayiotou, and G.~Ellinas, ``Learning quantile {Q}o{T} models to
  address uncertainty over unseen lightpaths,'' \emph{Comput. Netw.}, vol. 212,
  no. 108992, 2022.

\bibitem{citeulike:105953}
J.~D. Hamilton, \emph{{Time Series Analysis}}.\hskip 1em plus 0.5em minus
  0.4em\relax Princeton University Press, 1994.

\bibitem{li2018diffusion}
Y.~Li, R.~Yu, C.~Shahabi, and Y.~Liu, ``Diffusion convolutional recurrent
  neural network: Data-driven traffic forecasting,'' \emph{arXiv:1707.01926
  [cs.LG]}, 2018.

\bibitem{6129370}
C.~Kachris and I.~Tomkos, ``A survey on optical interconnects for data
  centers,'' \emph{IEEE Commun. Surv. Tutor.}, vol.~14, no.~4, pp. 1021--1036,
  2012.

\bibitem{candela2020model}
R.~Candela, P.~Michiardi, M.~Filippone, and M.~A. Zuluaga, ``Model monitoring
  and dynamic model selection in travel time-series forecasting,''
  \emph{arXiv:2003.07268 [stat.AP]}, 2020.

\bibitem{dell22}
D.~Gaganam, M.~Cummins, A.~Oliveira Da~Silva, and W.~Biester, ``Integrated edge
  management in smart manufacturing. {A} model-based approach for edge
  computing,'' White paper, Dell Technologies Solutions, 2022.

\bibitem{9060868}
W.~Y.~B. Lim, N.~C. Luong, D.~T. Hoang, Y.~Jiao, Y.-C. Liang, Q.~Yang,
  D.~Niyato, and C.~Miao, ``Federated learning in mobile edge networks: {A}
  comprehensive survey,'' \emph{IEEE Commun. Surv. Tutor.}, vol.~22, no.~3, pp.
  2031--2063, 2020.

\bibitem{Burkart_2021}
N.~Burkart and M.~F. Huber, ``A survey on the explainability of supervised
  machine learning,'' \emph{J. Artif. Intell. Res.}, vol.~70, pp. 245--317,
  2021.

\bibitem{Monarch21}
R.~M. Monarch, \emph{Human-in-the-Loop Machine Learning: {A}ctive {L}earning
  and {A}nnotation for {H}uman-{C}entered {AI}}.\hskip 1em plus 0.5em minus
  0.4em\relax Manning Publications, 2021.

\bibitem{https://doi.org/10.48550/arxiv.1905.10497}
T.~Li, M.~Sanjabi, A.~Beirami, and V.~Smith, ``Fair resource allocation in
  federated learning,'' \emph{arXiv:1905.10497 [cs.LG]}, 2019.

\bibitem{8696290}
J.~{Suárez-Varela}, A.~{Mestres}, J.~{Yu}, L.~{Kuang}, H.~{Feng},
  P.~{Barlet-Ros}, and A.~{Cabellos-Aparicio}, ``Routing based on deep
  reinforcement learning in optical transport networks,'' in \emph{Proc.
  IEEE/OSA Opt. Fiber Commun. Conf. (OFC)}, 2019, pp. 1--3.

\bibitem{HONG2022100677}
Y.~Hong, X.~Hong, and J.~Chen, ``Neural network-assisted decision-making for
  adaptive routing strategy in optical datacenter networks,'' \emph{Optical
  Switching and Networking}, vol.~45, no. 100677, 2022.

\bibitem{mguni2018decentralised}
D.~Mguni, J.~Jennings, and E.~M. de~Cote, ``Decentralised learning in systems
  with many, many strategic agents,'' \emph{arXiv:1803.05028 [cs.MA]}, 2018.

\bibitem{https://doi.org/10.48550/arxiv.1611.03071}
S.~Jabbari, M.~Joseph, M.~Kearns, J.~Morgenstern, and A.~Roth, ``Fairness in
  reinforcement learning,'' \emph{arXiv:1611.03071 [cs.LG]}, 2016.

\bibitem{9738819}
T.~Li, K.~Zhu, N.~C. Luong, D.~Niyato, Q.~Wu, Y.~Zhang, and B.~Chen,
  ``Applications of multi-agent reinforcement learning in future {I}nternet:
  {A} comprehensive survey,'' \emph{IEEE Commun. Surv. Tutor.}, vol.~24, no.~2,
  pp. 1240--1279, 2022.

\bibitem{https://doi.org/10.48550/arxiv.1810.11748}
R.~Arakawa, S.~Kobayashi, Y.~Unno, Y.~Tsuboi, and S.-i. Maeda, ``{DQN-TAMER:
  H}uman-in-the-loop reinforcement learning with intractable feedback,''
  \emph{arXiv:1810.11748 [cs.HC]}, 2018.

\bibitem{8747378}
E.~Donati, M.~Payvand, N.~Risi, R.~Krause, and G.~Indiveri, ``Discrimination of
  {EMG} signals using a neuromorphic implementation of a spiking neural
  network,'' \emph{IEEE Trans. Biomed. Circuits Syst.}, vol.~13, no.~5, pp.
  795--803, 2019.

\end{thebibliography}

\begin{IEEEbiography}[{\includegraphics[width=1in,height=1.25in,clip,keepaspectratio]{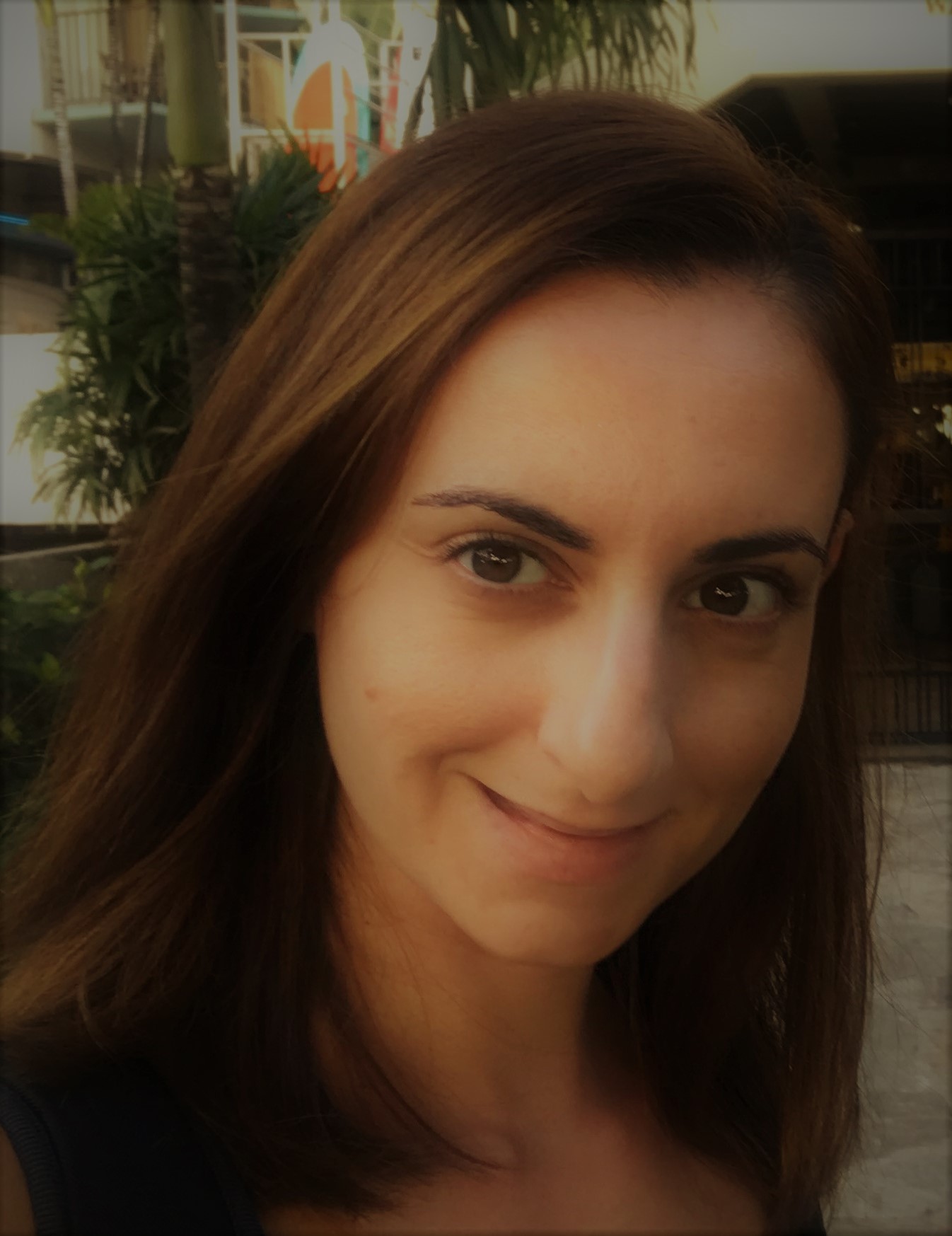}}]{Tania Panayiotou} received the Diploma degree in computer engineering and informatics from the University of Patras, Patras, Greece, in 2005 and the Ph.D. degree in computer engineering from the University of Cyprus (UCY) in 2013. She is currently a Research Associate with the KIOS Research and Innovation Center of Excellence, UCY. She has authored more than 40 articles, conference papers, and book chapters. Her research interests include optical networks and transportation networks.
\end{IEEEbiography}

\begin{IEEEbiography}
[{\includegraphics[width=1in,height=1.25in,clip,keepaspectratio]{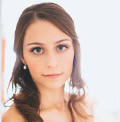}}]{Maria Michalopoulou} holds a B.Sc. in Informatics and Telecommunications from the University of Athens, Greece, and an M.Sc. in Communications Engineering from the RWTH Aachen University, Germany. In 2015 she received her doctoral degree (Dr.-Ing.) on wireless networks from the same university. Since September 2017 she is a Research Associate at the KIOS Research and Innovation Center of Excellence at the University of Cyprus. Her research interests include a variety of topics in the domains of wireless networks and intelligent transportation networks. 
\end{IEEEbiography}

\begin{IEEEbiography}[{\includegraphics[width=1in,height=1.25in,clip,keepaspectratio]{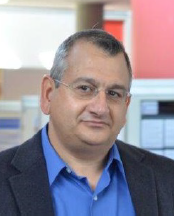}}]{Georgios Ellinas} holds B.Sc., M.Sc., M.Phil., and Ph.D. degrees in Electrical Engineering from Columbia University. He is a Professor at the Department of Electrical and Computer Engineering and a founding member of the KIOS Research and Innovation Center of Excellence at the University of Cyprus. Prior to joining the University of Cyprus, he also served as an Associate Professor of Electrical Engineering at City College of the City University of New York, as a Senior Network Architect at Tellium Inc., and as a Research Scientist/Senior Research Scientist in Telcordia Technologies' (formerly Bell Communications Research (Bellcore)) Optical Networking Research Group. His research interests are in optical/telecommunication networks, intelligent transportation systems, IoT, and autonomous unmanned aerial systems.
\end{IEEEbiography}

\vfill

\end{document}